\documentclass[a4paper,11pt]{article}
\usepackage{jinstpub} 
\usepackage{lineno}

\usepackage{graphicx}
\usepackage{subcaption}

\usepackage[LGRgreek]{mathastext}

\usepackage{textcomp}   

\title{Long-term operation of the screen-printed graphite-based resistive coatings on the HPL electrode for the Resistive Plate Chamber.}

\author[a,b]{Davide Costa,}
\author[a,1]{Francesco Fallavollita,\note{Corresponding author.}}
\author[a,b]{Hubert Kroha,}
\author[a,b]{Oliver Kortner,}
\author[a]{Pavel Maly,}
\author[a]{Giorgia Proto,}
\author[a]{Daniel Soyk,}
\author[a]{Elena Voevodina,}
\author[a]{Jorg Zimmermann}

\affiliation[a]{Max-Planck-Institut für Physik, \\ Werner Heisenberg Institute, \\ Boltzmannstr. 8 - 85748 Garching, \\ Germany}
\affiliation[b]{Physik-Department, \\ Technische Universität München, \\ James-Franck-Str. 1 - 85748 Garching, \\ Germany}

\abstract{The reliability of large-area Resistive Plate Chambers (RPCs) operated under High-Luminosity Large Hadron Collider (HL-LHC) conditions is governed by the long-term stability and radiation tolerance of screen-printed graphite/phenoxy coatings on high-pressure-laminate (HPL) electrodes. This work presents a comprehensive, end-to-end qualification of such coatings that integrates industrial process control and metrology with controlled humidity/temperature campaigns, extended high-voltage stress testing to decade-scale charge levels, and representative neutron and gamma irradiation at CERN facilities. The results establish reproducible industrial coating production, stable performance under sustained operation and irradiation, and practical acceptance criteria with operating and monitoring guidelines. The study provides a transferable quality-assurance framework for graphite-based resistive coatings on HPL electrodes, enabling reproducible production and reliable RPC performance for the HL-LHC upgrades and for future high-rate collider experiments.}


\arxivnumber{1234.56789} 

\begin{document}
\maketitle
\flushbottom

\section{Introduction}
\label{sec:intro}
 The High-Luminosity upgrade of the Large Hadron Collider (HL-LHC) imposes unprecedented demands on the long-term stability, radiation tolerance, and industrial reproducibility of gaseous detectors employed in high-rate muon systems. Resistive Plate Chambers (RPCs), widely adopted for trigger and timing applications in ATLAS \cite{2106380}, and CMS \cite{Hebbeker:2017bix}, rely critically on the electrical stability of their large-area high-pressure-laminate (HPL) electrodes coated with graphite-based resistive layers. These coatings ensure uniform high-voltage distribution across the gas volumes and directly affect detector efficiency, spatial resolution, rate capability, and longevity. While screen-printed carbon-loaded inks have been successfully employed in large-scale RPC production for over two decades, the transition to the HL-LHC environment requires a systematic re-assessment of their reliability under sustained operation. The anticipated integrated charge, mixed-field radiation exposure, and stringent quality-assurance requirements necessitate an end-to-end validation programme that extends beyond conventional acceptance tests. In particular, understanding the interplay between ink composition, environmental factors (temperature, humidity), and bias-induced aging mechanisms is essential to secure reproducible long-term detector performance. This work presents a comprehensive qualification of industrially screen-printed graphite/phenoxy coatings on HPL electrodes. The programme integrates controlled fabrication, extended high-voltage stress testing to decade-equivalent charge accumulation, and irradiation campaigns in representative mixed-field (CHARM) and gamma-dominated (GIF++) facilities at CERN. By correlating macroscopic electrical properties with environmental conditions and cumulative exposure, and by defining robust acceptance and monitoring criteria, the study provides a validated framework for the reliable deployment of resistive coatings in RPC electrodes. Beyond the direct relevance to the HL-LHC upgrades, these results establish a transferable methodology applicable to future high-rate collider experiments.

\section{Silk-Screen Printing of Graphite Resistive Coatings for RPC Electrodes}
\label{sec:silk-screen_printing}

The silk-screen printing, long established as the standard technique for depositing carbon-based resistive inks onto high-pressure-laminate and glass electrodes in RPC detector technology, has been implemented at the Max-Planck-Institute for Physics and its industrial partners as the sole process for both prototyping and large-scale manufacturing of the gas volumes \cite{Fallavollita:2025bns}. The silk-screen printing technique offers:
\begin{enumerate}
    \item  robust scalability, ranging from manual bench-top presses suitable for small pre-series runs of less than $0.5 \, m^2$, up to fully automated screen-printing lines capable of consistently depositing uniform coatings over areas exceeding $10^2 \, m^2 \, / day$.

    \item precise control over the thickness of the carbon-based resistive coating $(\pm 5 \, \mu m)$, resulting in accurate and reproducible control of its surface resistivity.

    \item excellent uniformity and high reproducibility across large-area substrates, ensuring consistent electrical and mechanical properties within stringent production tolerances.

    \item high throughput at relatively low cost; short cycle times on the order of a few minutes per print, combined with inexpensive stainless-steel or polyester mesh screens, make it particularly well suited for mass production of large-area detectors.    
\end{enumerate}

The surface resistivity of a graphite-loaded resistive ink is controlled, in the first instance, by the chemistry of the ink itself, namely the graphite volume fraction, particle size distribution, wetting/dispersing additives, and the choice of polymer binder. Once the formulation is fixed, the surface resistivity can be adjusted over more than an order of magnitude by tailoring the wet film thickness deposited in the screen-printing process. The thickness is set not only by the nominal mass of ink transferred but also by the geometrical and process parameters of the screen: thread count and wire diameter (which together determine the open-area fraction), the emulsion-over-mesh layer, mesh tension, squeegee angle and pressure, and the ink viscosity at the shear rates encountered during printing. These factors jointly govern the amount of material laid down and, consequently, the final surface resistivity of the cured coating.

For the manufacturing campaign conducted at the Max-Planck-Institute for Physics in collaboration with its industrial partners, the carbon-based resistive ink supplied by Winchem\textsuperscript{\textregistered} company (Republic of Korea) has been employed. This formulation, already qualified and deployed for large-scale production of CMS RPC gas volumes at the Korea Detector Laboratory, serves as a proven benchmark for coating uniformity and high process yield \cite{Park:2005ze}.
By jointly tuning the graphite loading in the ink formulation and regulating the wet-film thickness laid down during screen printing, the surface resistivity of the coating can be tuned from $\approx 10 \, k\Omega/\square$ up to >$10 \, M\Omega/\square$, with a reproducibility better than 30\%. The carbon-based resistive ink used in this work is formulated as a carbon-black/phenoxy composite dispersed in an organic solvent. According to the supplier’s safety-data sheet, the mixture contains 5–15 wt \% conductive carbon black, 30–50 wt \% phenoxy resin (poly-hydroxyether binder), and 40–60 wt \% diethylene-glycol monobutyl-ether acetate as the carrying solvent, with no other functional additives declared.

The core technology for large-scale screen printing of resistive layers onto high-pressure-laminate (HPL) plates has been successfully industrialised at the German company Siebdruck Esslinger\textsuperscript{\textregistered}. Before screen-printing the carbon-based resistive ink, the HPL plates have been thoroughly cleaned in two successive steps, first with acetone and then with isopropanol, to ensure a contaminant-free surface and optimal coating adhesion. The target surface resistivity of $350 \, k\Omega/\square \pm 30 \%$ has been consistently achieved with a $15 \pm 5 \, \mu m$ dry-film thickness by employing T90 screens, 90 threads/cm in either stainless steel or polyester, tensioned to 22 N/cm during the screen-printing process. The same screen-printing technique has been used to apply a 5 mm-wide conductive graphite side band, with a surface resistivity of only a few $\times 10 \, k\Omega/\square$, along each long edge of the electrodes. These bands provide an equipotential contact that distributes the high-voltage bias uniformly and thus guarantees a homogeneous electric field throughout the entire active gas volume. A surface resistivity of $20 \, k\Omega/\square \pm 30 \%$ has reproducibly achieved by depositing three consecutive layers of carbon-based resistive ink, resulting in a total dry-film thickness of  $30 \pm 5 \, \mu m$ when printed with a T90 mesh screen.

\begin{figure}[htbp]
  \centering
  \begin{subfigure}[t]{0.49\textwidth}
    \centering
    \includegraphics[width=\textwidth]{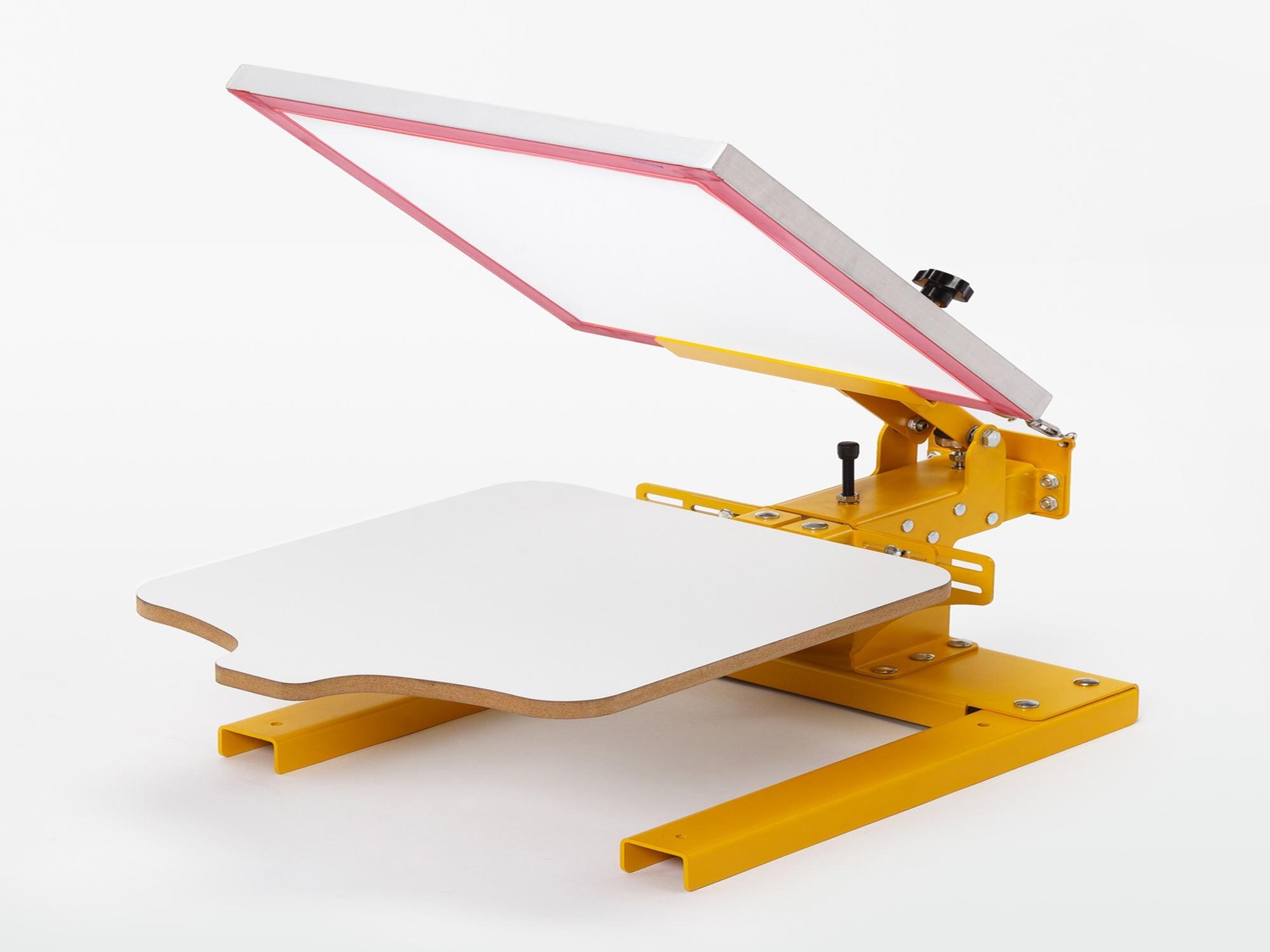}  
    \caption{}             
    \label{fig:screen_printing}
  \end{subfigure}
  \hfill
  \begin{subfigure}[t]{0.49\textwidth}
    \centering
    \includegraphics[width=\textwidth]{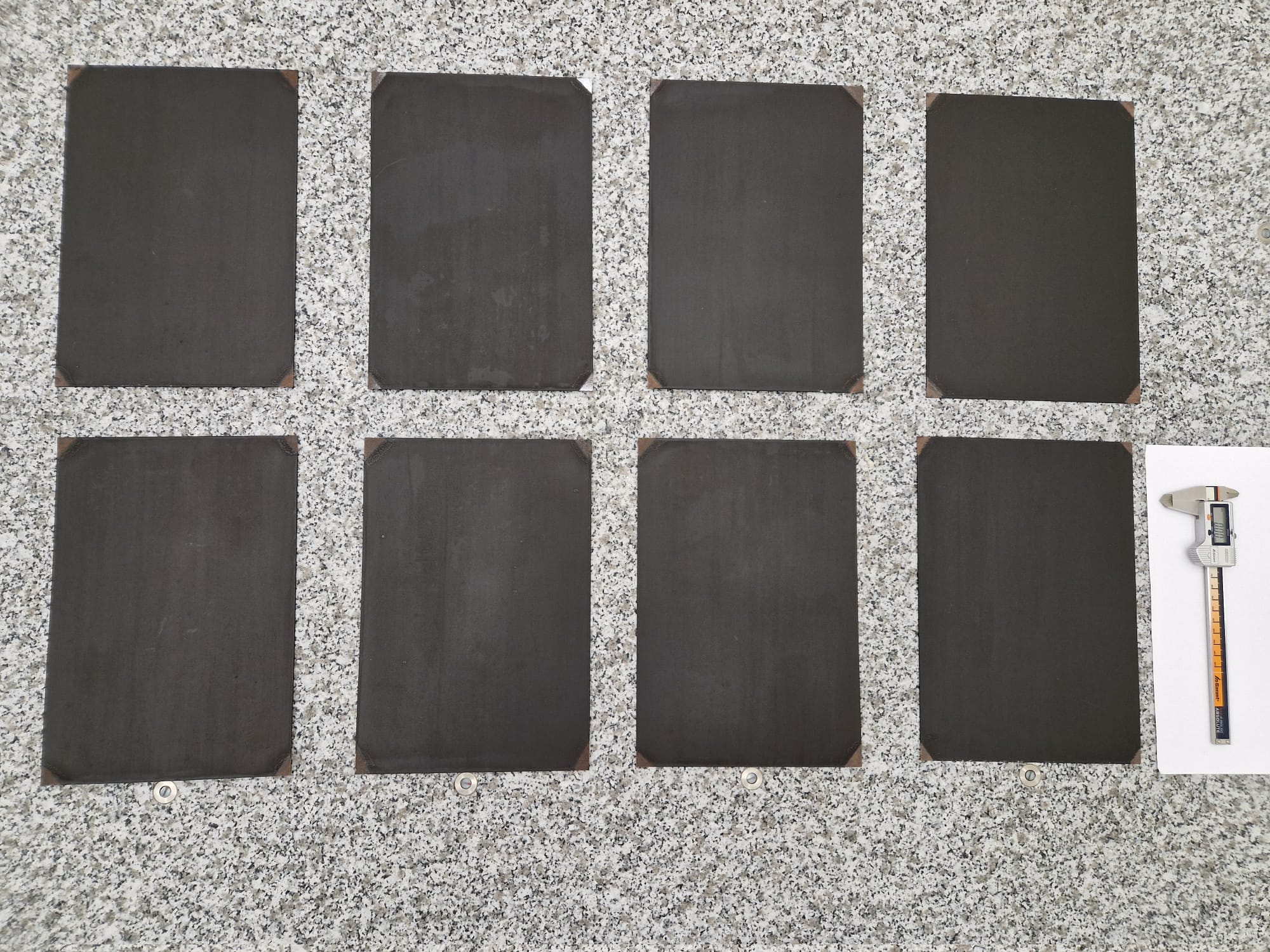}  
    \caption{}             
    \label{fig:HPL_graphite_coating}
  \end{subfigure}
  
  \caption{Silk-screen table and the accessories for the preparation of control samples for screen-printing with a carbon-based resistive ink at the Max Planck Institute for Physics (a). High-pressure-laminate (HPL) substrates after the drying stage, showing a uniform graphite-resistive coating (b).}
  \label{fig:silk_screen_printing}
\end{figure}

\begin{figure}[htbp]
  \centering
  \begin{subfigure}[t]{0.49\textwidth}
    \centering
    \includegraphics[width=\textwidth]{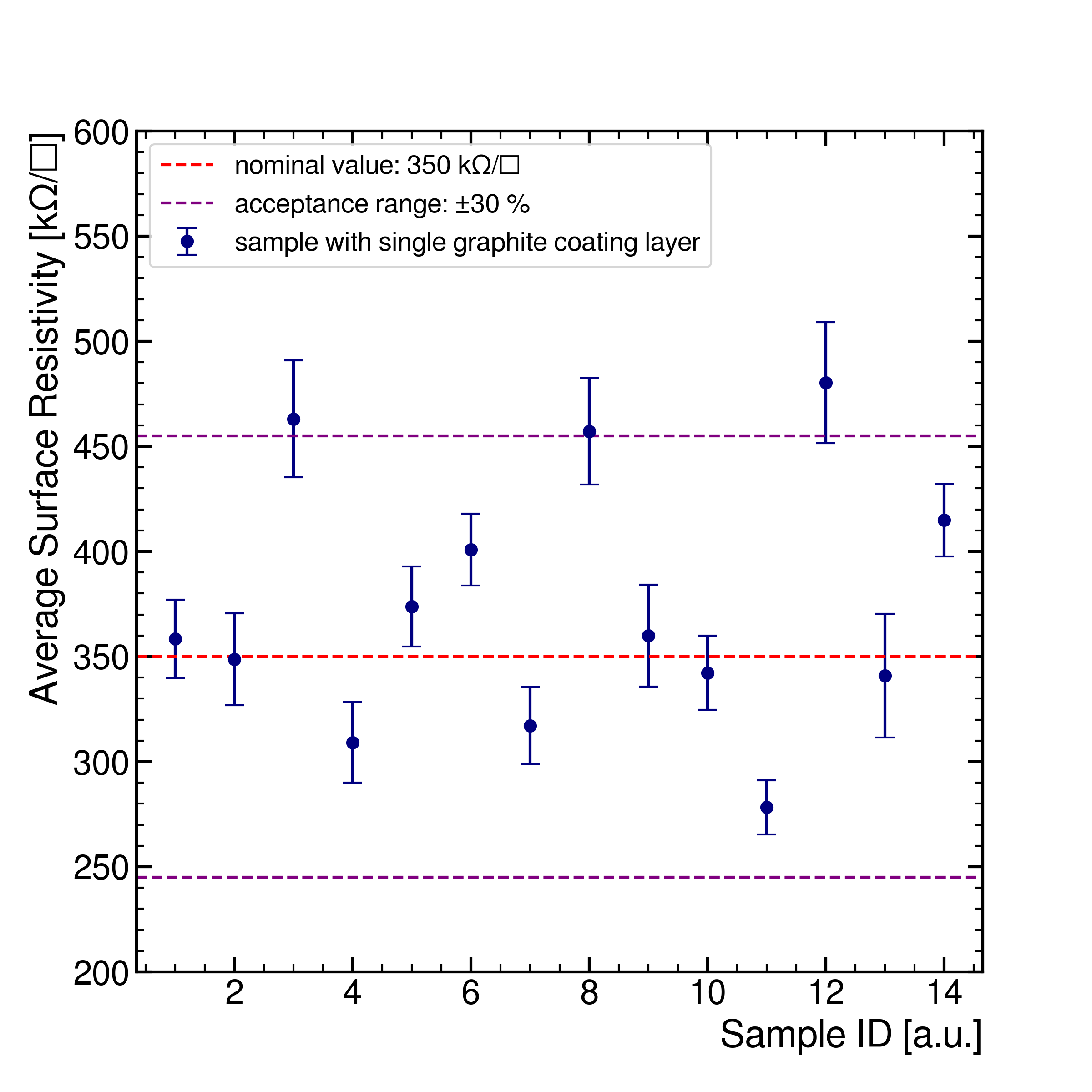}  
    \caption{}             
    \label{fig:triple_a}
  \end{subfigure}
  \hfill
  \begin{subfigure}[t]{0.49\textwidth}
    \centering
    \includegraphics[width=\textwidth]{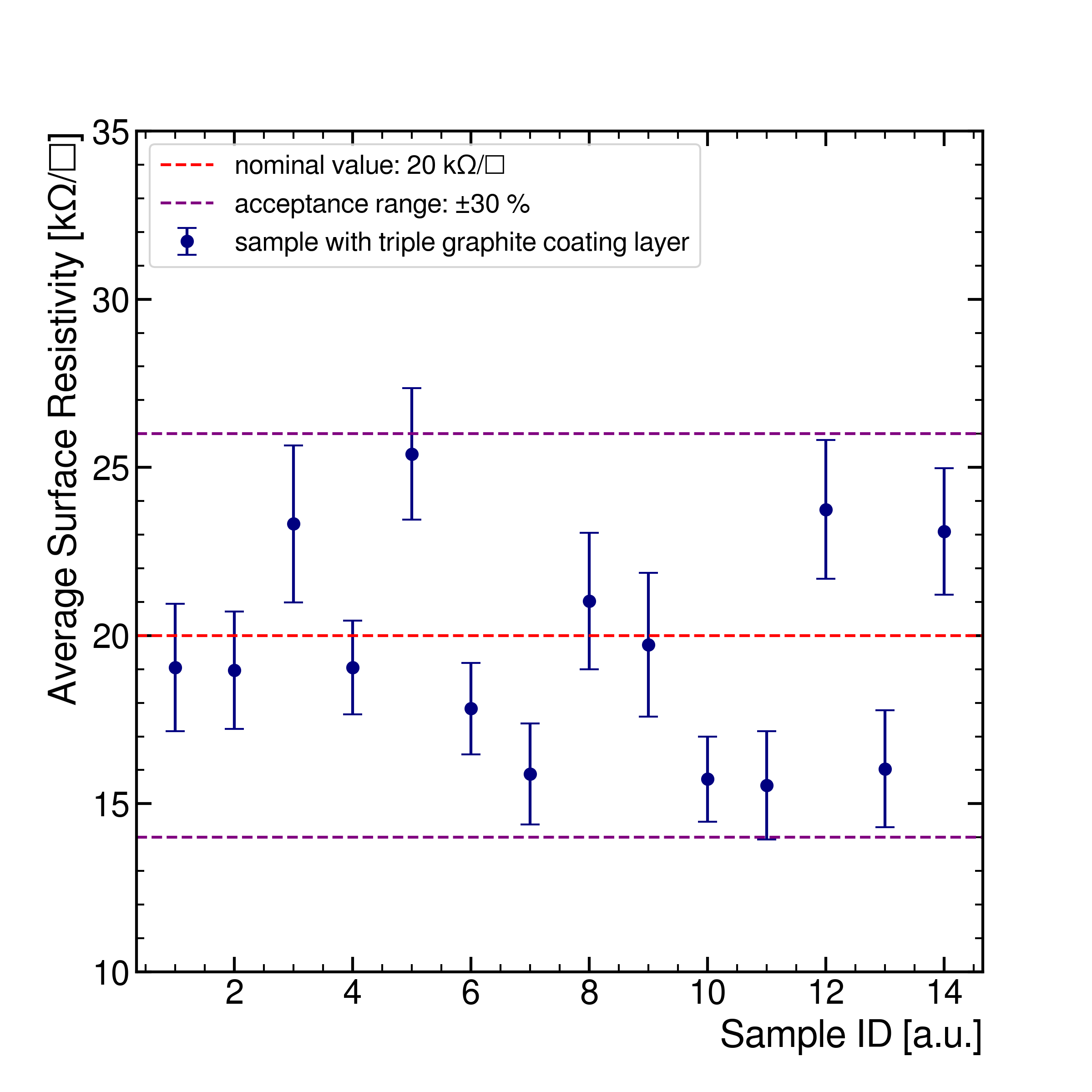}  
    \caption{}             
    \label{fig:triple_b}
  \end{subfigure}
  
  \caption{Average surface resistivity of fourteen HPL plates after silk-screen deposition of a single graphite layer (a) and three consecutive layers (b).}
  \label{fig:triple_layers}
\end{figure}

The reproducibility of the silk-screen-printing process used to deposit a resistive graphite layer on the outer surfaces of HPL electrodes during large-scale fabrication of RPC gas volumes has been investigated by producing fourteen 30 cm $\times$ 20 cm HPL samples, each featuring a screen-printed graphite coating. After printing, the coated electrodes have been conditioned for 1 h in a forced-air oven at 105 $^{\circ}$C to reproduce the thermal load applied during the in-line lamination of the polyethylene-terephthalate (PET) insulating foil in large-scale RPC gas volume production. The surface resistivity of each graphite-coated HPL electrode has been mapped with a METRISO\textsuperscript{\textregistered} 3000 surface resistance meter coupled to a concentric ring probe. The probe uses coaxial inner and outer electrodes plus an intermediate guard ring to realize a guarded two-terminal geometry. Each HPL plate has been overlaid with a virtual 3 $\times$ 4 grid that divided the surface into twelve zones of equal area to ensure uniform spatial coverage. At the center of each zone the probe has been gently lowered until its conductive-rubber electrodes (bulk resistance $\leq \, 5 \, \Omega$) made uniform contact with the coating;  a calibrated 2.27 kg weight was placed on top of the probe to apply a constant normal force during every reading.  A constant DC test voltage V has been applied to the inner ring, the resulting current $I_i$ collected by the outer ring has been measured, 
and the resistance was obtained as $R_i=V \, / \, I_i$. Five replicate measurements have been taken in each zone, yielding sixty resistance values per plate. Individual readings have been converted to surface resistivity $\rho_i$ by taking into account the probe's geometry factor k:
\begin{equation}
\rho_i = k \times R_i , \quad  k = \frac{2\pi}{ln(r_2/r_1)}
\label{eq:surface_resistivity}
\end{equation}

\noindent where $r_1 = 15.3 \, mm$ and $r_2 = 28.6 \, mm$ are the probe radii defined by the standard (ASTM D257 - "Standard Test Methods for DC Resistance or Conductance of Insulating Materials"). For a given plate the set $\left\{ \rho_i \right\}_i^{60}$ has been summarised by its arithmetic mean $\langle \rho_i \rangle$ and by the standard error of the mean, $\sigma \, / \, \sqrt{N}$ with $N = 60$; the former represents the characteristic surface resistivity of the electrode, whereas the latter quantifies intra-plate dispersion. The same characterization procedure has been applied to additional HPL samples carrying a triple-layer, screen-printed graphite coating representative of the two lateral resistive bands on the HPL electrodes, thereby ensuring a uniform distribution of the high voltage along the full length of each electrode. Figure \ref{fig:triple_layers} compares the plate-averaged surface resistivity obtained after one and three silk-screen passes of the same graphite ink. For the single-layer recipe [Figure \ref{fig:triple_a}] the ensemble mean is $\langle \rho_s \rangle_{1\times} = (3.75 \, \pm \, 0.16) \, \times \, 10^5 \, \Omega/\square$ (relative standard deviation, $\sigma/\mu \approx 16 \%$); ten of the fourteen plates fall inside the $\pm \, 30 \%$ specification band around the $350 \, k\Omega/\square$ target, and the remaining plates are compatible within the statistical uncertainty. The triple-layer coating [Figure \ref{fig:triple_b}] yields $\langle \rho_s \rangle_{3\times} = (1.96 \, \pm \, 0.88) \, \times \, 10^4 \, \Omega/\square$ with the same relative spread ($\approx 17 \%$), and all plates comply with the corresponding $\pm \, 30 \%$ tolerance around $20 \, k\Omega/\square$.  The identical coefficients of variation demonstrate that increasing the number of printing passes reduces the surface resistivity by more than an order of magnitude without degrading uniformity: the process therefore provides a robust and fully reproducible means to tailor resistive coatings for both the high- and low-resistivity requirements of large-area RPC electrodes.

\section{Long-Term High-Voltage Stress Test}
\label{sec:Stress_Test_MPI}

To assess the long‑term stability of carbon‑based resistive coatings on HPL electrodes, a dedicated longevity programme  is under way at the at the Max‑Planck‑Institute for Physics. Ten HPL plates measuring $10 \times 10 \, cm^2$ and $1.3 \, mm$ in thickness, with a bulk resistivity of $10^{10} - 10^{11} \, \Omega \cdot cm  $, have been screen‑printed on both sides with a commercial carbon ink (see Section \ref{sec:silk-screen_printing}). On each face the patterned layer covers an active area of $56 \, cm^2$, giving an initial surface resistivity of $350 \, k\Omega/\square \, \pm \, 30 \%$. To homogenize the electric field and mitigate edge effects, 5 mm‑wide low‑resistivity side bands, with surface resistivity below $20 \, k\Omega/\square$, have been printed along two opposite edges. Immediately after printing, all plates underwent a curing cycle in a convection oven at 105 $^{\circ}C$ for 1 hour. Finally, brass contact pads have been affixed to the ends of the side bands at diagonally opposite corners on each face using a silver‑filled conductive epoxy, providing mechanically robust, low‑impedance connections to the high‑voltage supply.

\begin{figure}[htbp]
  \centering
  \begin{subfigure}[t]{0.49\textwidth}
    \centering
    \includegraphics[width=\textwidth]{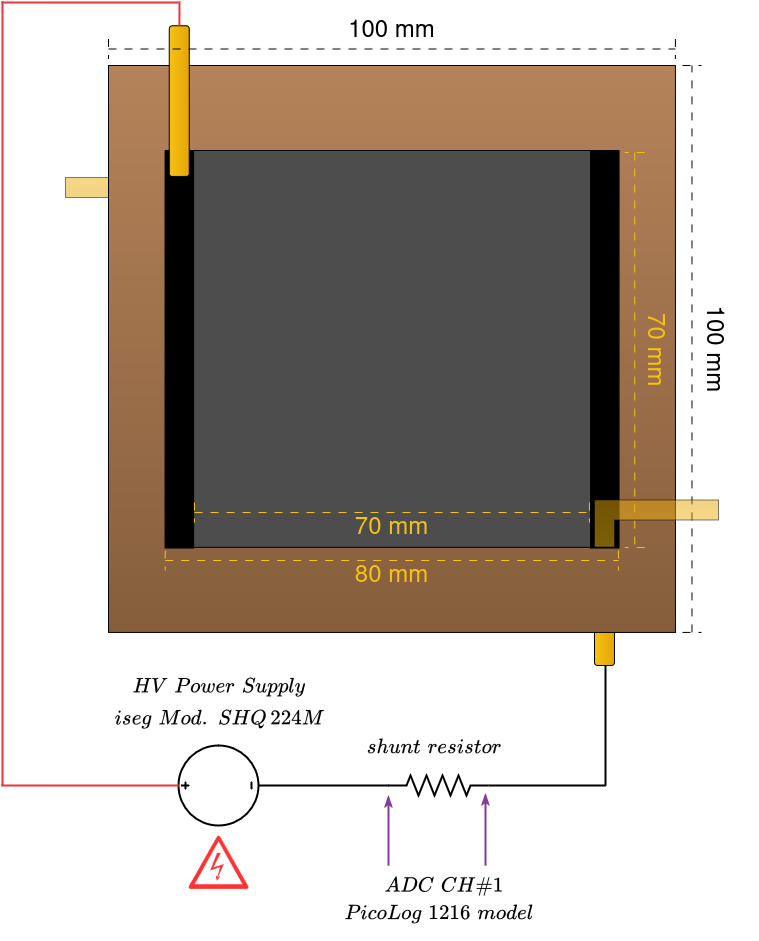}  
    \caption{}             
    \label{fig:current_monitoring_setup}
  \end{subfigure}
  \hfill
  \begin{subfigure}[t]{0.49\textwidth}
    \centering
    \includegraphics[width=\textwidth]{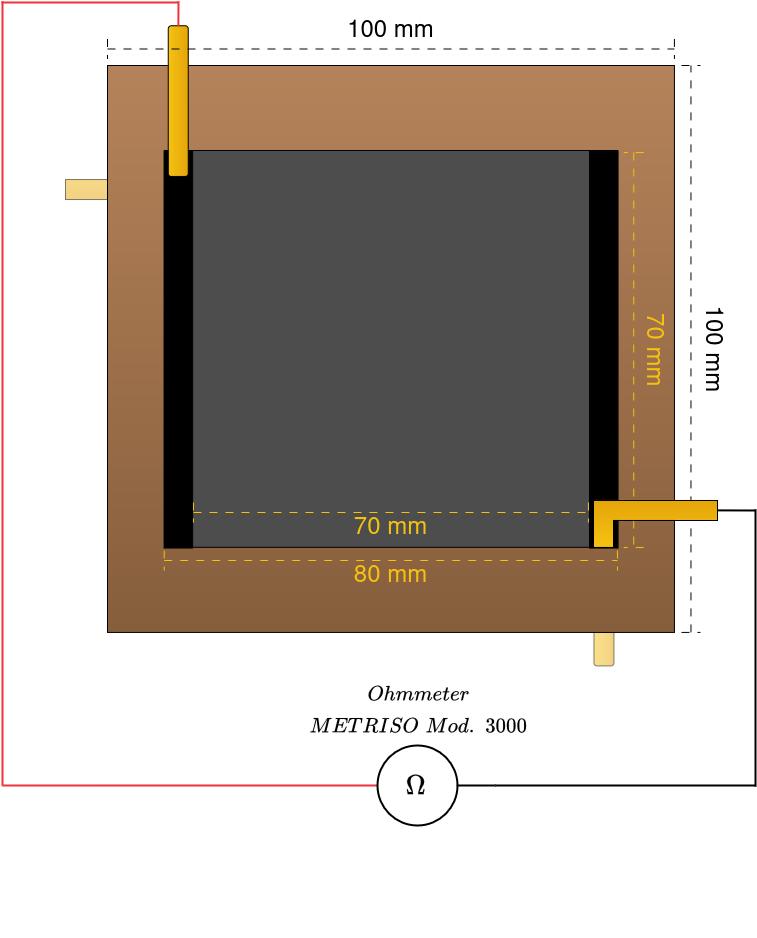}  
    \caption{}             
    \label{fig:surface_resistivity_monitoring_setup}
  \end{subfigure}
  
  \caption{Electrical schematic of the experimental setup: (a) circuit for continuous current monitoring of the test samples; (b) circuit for surface‑resistivity measurement of the carbon‑based resistive ink coatings.
}
  \label{fig:experimental_setup_mpi}
\end{figure}

\begin{figure}[htbp]
  \centering
  \begin{subfigure}[t]{0.49\textwidth}
    \centering
    \includegraphics[width=\textwidth]{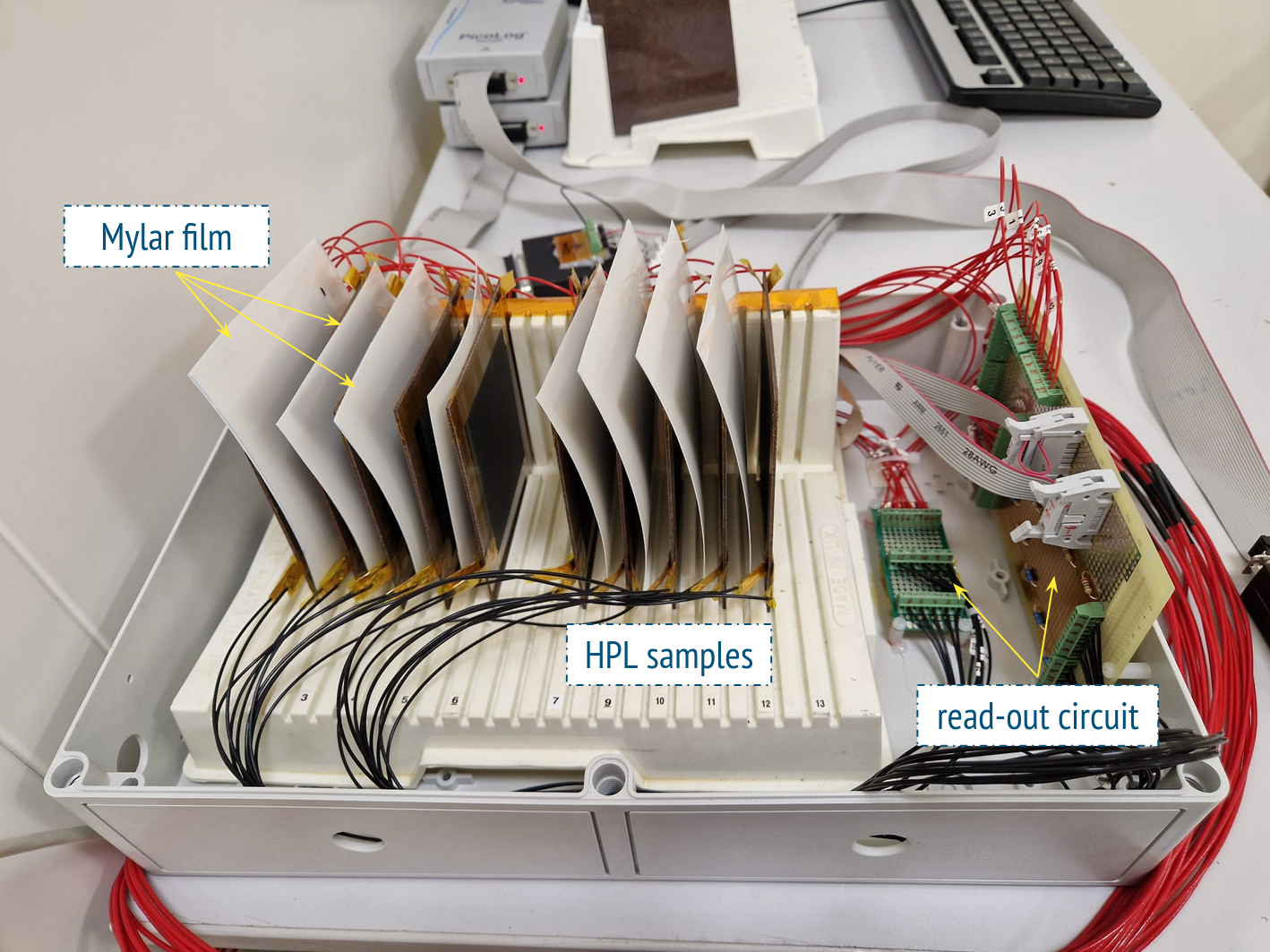}  
    \caption{}             
    \label{fig:bulk_surface_resistivity_connection}
  \end{subfigure}
  \hfill
  \begin{subfigure}[t]{0.49\textwidth}
    \centering
    \includegraphics[width=\textwidth]{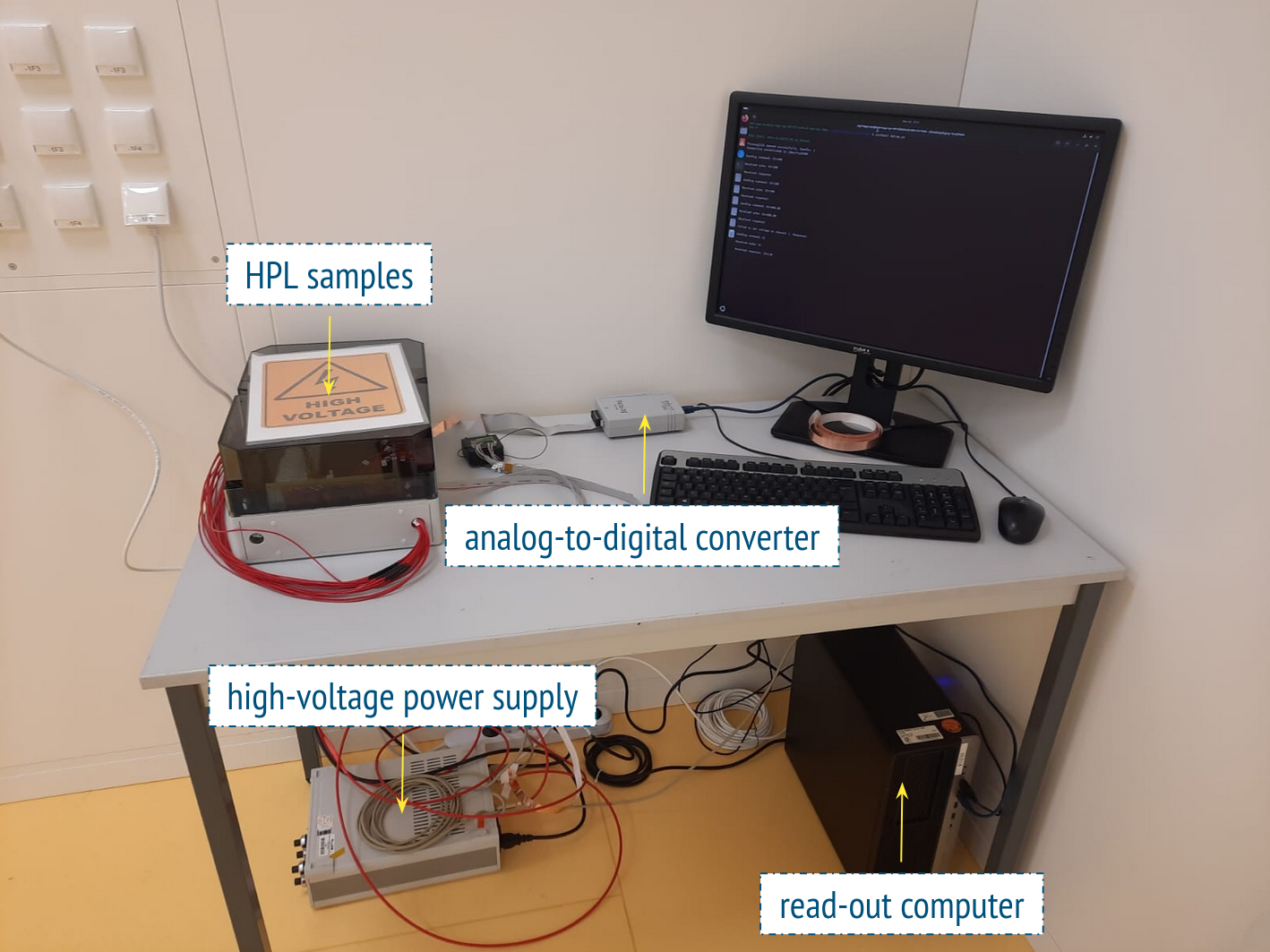}  
    \caption{}             
    \label{fig:setup_installation_clean_room}
  \end{subfigure}
  
  \caption{Experimental setup for the long‑term high‑voltage stress test of HPL samples screen‑printed with a carbon‑based resistive ink. (a) Electrical connections for measuring the current flowing through the HPL plate and the surface resistivity of the carbon-based resistive coating. (b) Complete test assembly installed inside the clean‑room.}
  \label{fig:experimental_setup_clean_room}
\end{figure}

A long-term high-voltage stress test has been carried out by continuously biasing the electrodes coated with carbon-based resistive ink, in order to drive a current across the plate, thereby reproducing realistic operating conditions. To investigate the effects of different integrated charge accumulation rates, five samples have been operated at 800 V and five at 400 V throughout the test period. The current of each electrode has been continuously recorded with an analogue‑to‑digital converter, which digitized the voltage drop across a precision shunt resistor placed in the high‑voltage return path. The surface resistivity of the carbon‑based resistive ink coatings on both faces has been periodically measured, providing a sensitive gauge of grain‑connectivity integrity and enabling a quantitative correlation between resistivity drift and the integrated charge. Figure \ref{fig:experimental_setup_mpi} presents the electrical schematic of the experimental set‑up employed for continuous monitoring of the current drawn by the test samples [Figure \ref{fig:current_monitoring_setup}] and for measuring the surface resistivity of the carbon‑based resistive ink coatings [Figure \ref{fig:surface_resistivity_monitoring_setup}]. The experimental set-up for the long‑term high‑voltage stress test has been installed in a clean-room that provides a controlled environment with real‑time monitoring of atmospheric pressure, ambient temperature, and relative humidity, as shown in Figure \ref{fig:experimental_setup_clean_room}.

Under HL‑LHC operating conditions, the new 1mm‑gap RPC subsystem of the ATLAS Muon Spectrometer is expected to accumulate a total integrated charge of about $284 \, mC/cm^2$, including a safety factor of $\times 3$; taking a single‑count charge of $q_{av} \approx 6 \, pC$, a conservative local interaction rate of $R_{max} \approx 100 \, Hz/cm^2$, and an effective exposure of $t_{eff} \approx 1.58 \times 10^8 \, s$, corresponding to ten years of HL‑LHC running at a 50\% duty cycle, the resulting integrated charge on the electrode surface is therefore estimated to be:
\begin{equation}
Q_{10yr} = q_{av} \times R_{max} \times t_{eff} \approx 94.6 \, mC/cm^2
\label{eq:integrated_charge}
\end{equation}

\noindent Following CERN guidelines, a safety factor of $\times 3$ is adopted to cover systematic uncertainties, operational up-time fluctuations, and local rate enhancements, raising the qualification target to:
\begin{equation}
Q_{target} = 3 \times Q_{10yr} \approx 283.8 \, mC/cm^2
\label{eq:integrated_charge}
\end{equation}

\begin{figure}[htbp]
  \centering
  \begin{subfigure}[t]{0.49\textwidth}
    \centering
    \includegraphics[width=\textwidth]{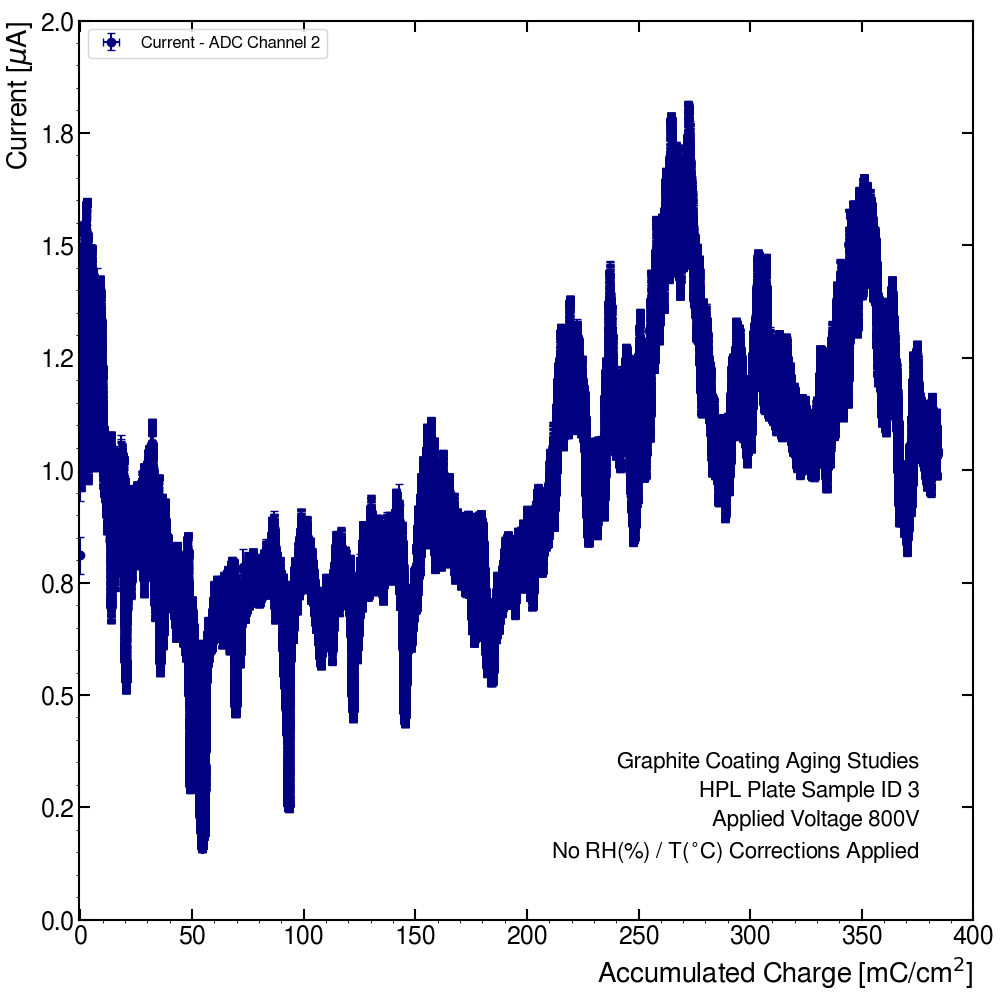}  
    \caption{}             
    \label{fig:current_monitoring_800V}
  \end{subfigure}
  \hfill
  \begin{subfigure}[t]{0.49\textwidth}
    \centering
    \includegraphics[width=\textwidth]{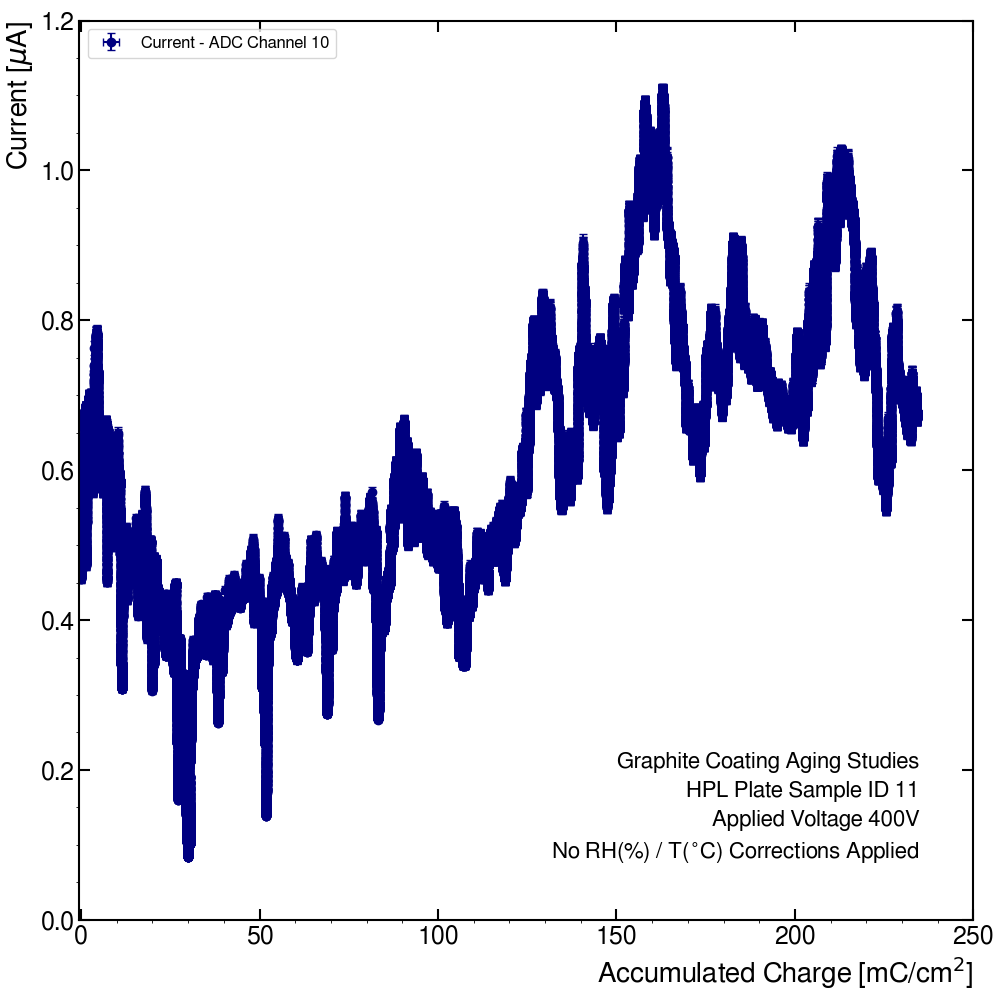}  
    \caption{}             
    \label{fig:current_monitoring_400V}
  \end{subfigure}
  
  \caption{Current drawn by the HPL plate as a function of the integrated charge for two operating voltages: 800 V (a) and 400 V (b). The curves are presented without environmental corrections.}
  \label{fig:summary_currnet_monitoring}
\end{figure}

\begin{figure}[htbp]
  \centering
  \begin{subfigure}[t]{0.49\textwidth}
    \centering
    \includegraphics[width=\textwidth]{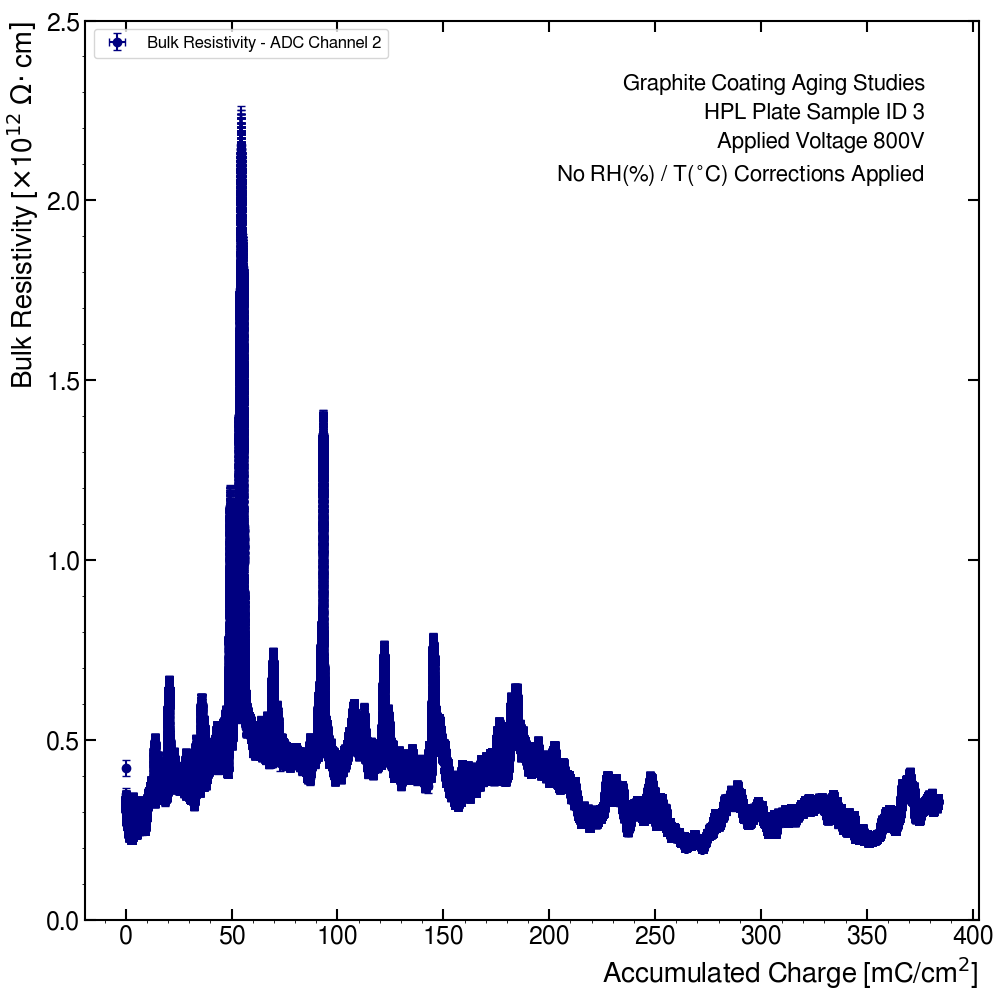}  
    \caption{}             
    \label{fig:bulk_resistivity_monitoring_800V}
  \end{subfigure}
  \hfill
  \begin{subfigure}[t]{0.49\textwidth}
    \centering
    \includegraphics[width=\textwidth]{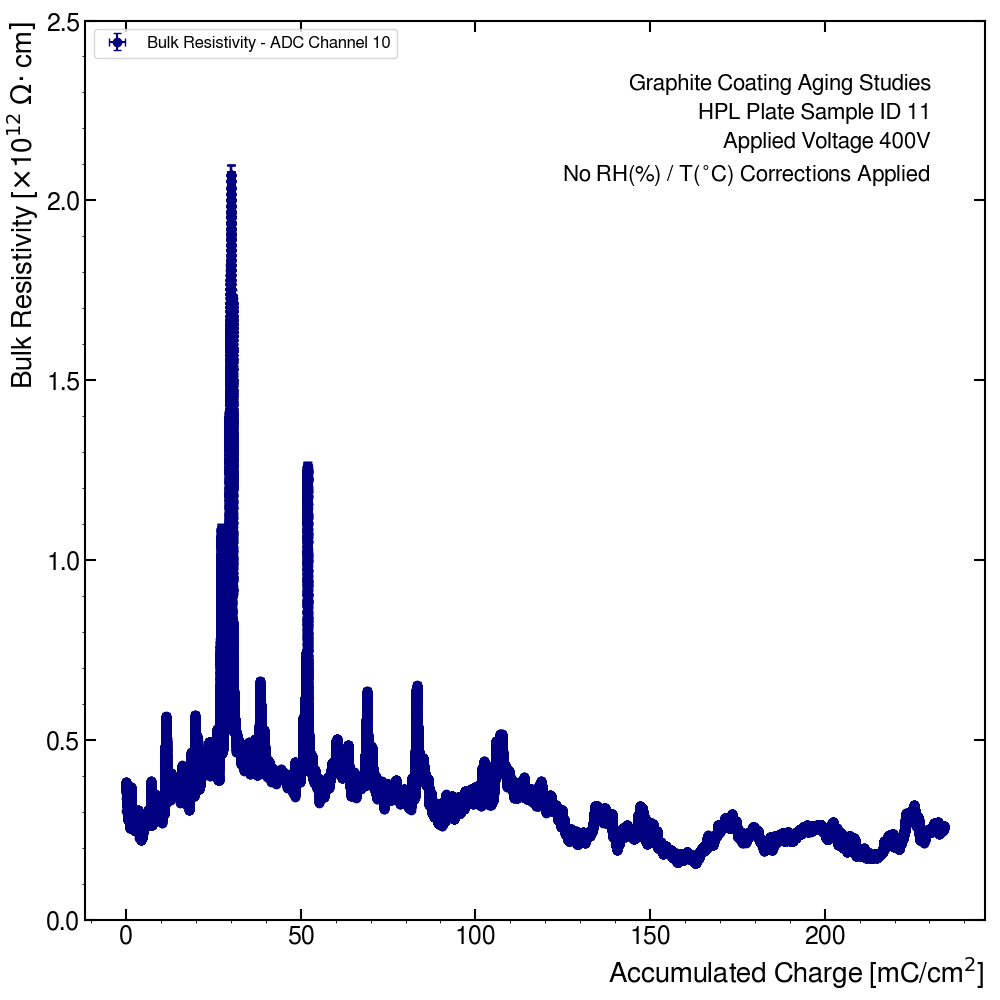}  
    \caption{}             
    \label{fig:bulk_resistivity_monitoring_400V}
  \end{subfigure}
  
  \caption{Bulk resistivity of the HPL plate as a function of the integrated charge for two operating voltages: 800 V (a) and 400 V (b). The curves are presented without environmental corrections.}
  \label{fig:bulk_resistivity_monitoring}
\end{figure}

The measurement of the current flowing through the HPL plate, and the corresponding bulk resistivity, provides an indirect assessment of the quality and long-term stability of the carbon-based resistive ink coating applied to the electrode surfaces throughout the entire high-voltage stress test, as discussed in detail in Section \ref{sec:intro}. Figure \ref{fig:summary_currnet_monitoring} presents a comparison of the current flowing through the HPL plate as a function of the integrated charge for two different bias voltages: 800 V [Figure \ref{fig:current_monitoring_800V}] and 400 V [Figure \ref{fig:current_monitoring_400V}]. Once the sample geometry is defined, the measured current can be directly converted into the bulk resistivity of the material using the following relation:
\begin{equation}
\rho_v = \frac{V}{I} \times \frac{S}{d}
\label{eq:bulk_resistivity}
\end{equation}

\noindent where V is the applied voltage, I the measured current across the HPL sample, S the electrode surface area, and d the sample thickness. Figure \ref{fig:bulk_resistivity_monitoring} presents a comparison of the bulk resistivity of the HPL plate as a function of the integrated charge, evaluated for two different bias voltages: 800 V [Figure \ref{fig:bulk_resistivity_monitoring_800V}] and 400 V [Figure \ref{fig:bulk_resistivity_monitoring_400V}]. The Appendix \ref{sec:appendix_A} tabulates the current flowing through the HPL plate and the bulk resistivity as a function of the integrated charge for the remaining test samples.

\begin{figure}[htbp]
  \centering
  \begin{subfigure}[t]{0.9\textwidth}
    \centering
    \includegraphics[width=\textwidth]{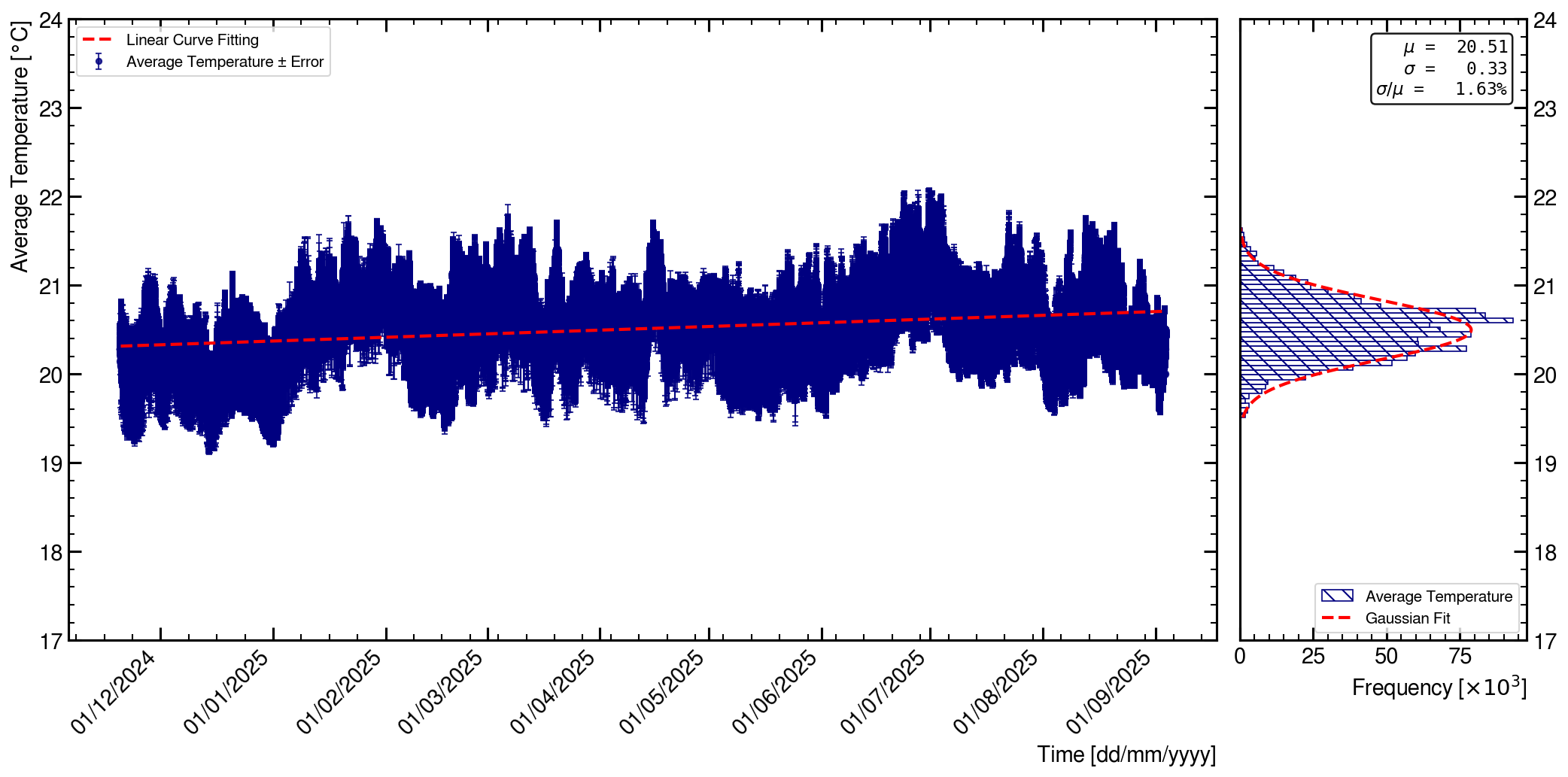}  
    \caption{}             
    \label{fig:temperature_monitoring}
  \end{subfigure}
  \hfill
  \begin{subfigure}[t]{0.9\textwidth}
    \centering
    \includegraphics[width=\textwidth]{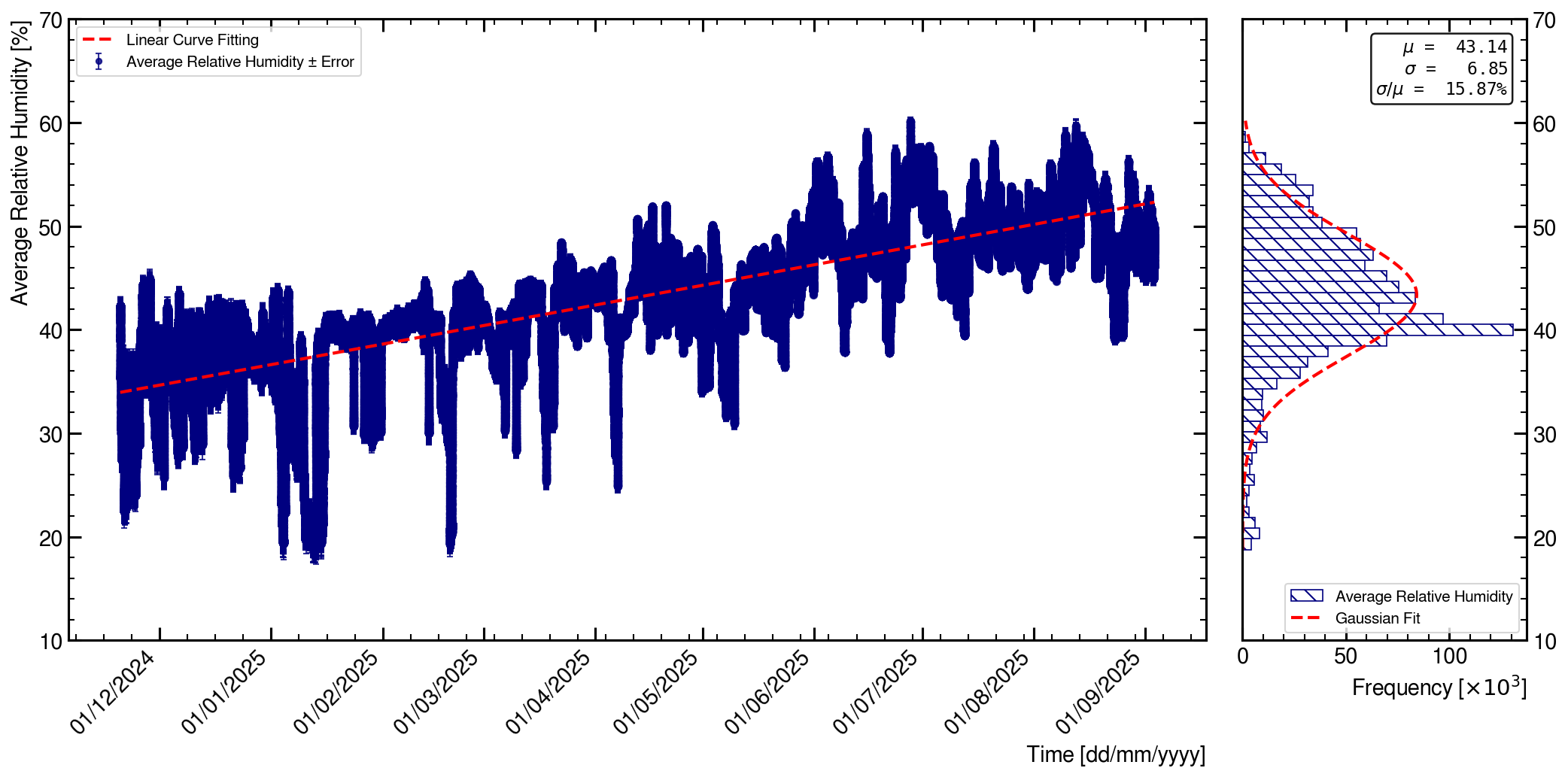}  
    \caption{}             
    \label{fig:humidity_monitoring}
  \end{subfigure}
  
  \caption{Temporal evolution of the average temperature (a) and average relative humidity (b) inside the clean room hosting the long-term high-voltage stress-test setup.}
  \label{fig:environmental_parameter_monitoring}
\end{figure}

\begin{figure}[htbp]
  \centering
  \begin{subfigure}[t]{0.9\textwidth}
    \centering
    \includegraphics[width=\textwidth]{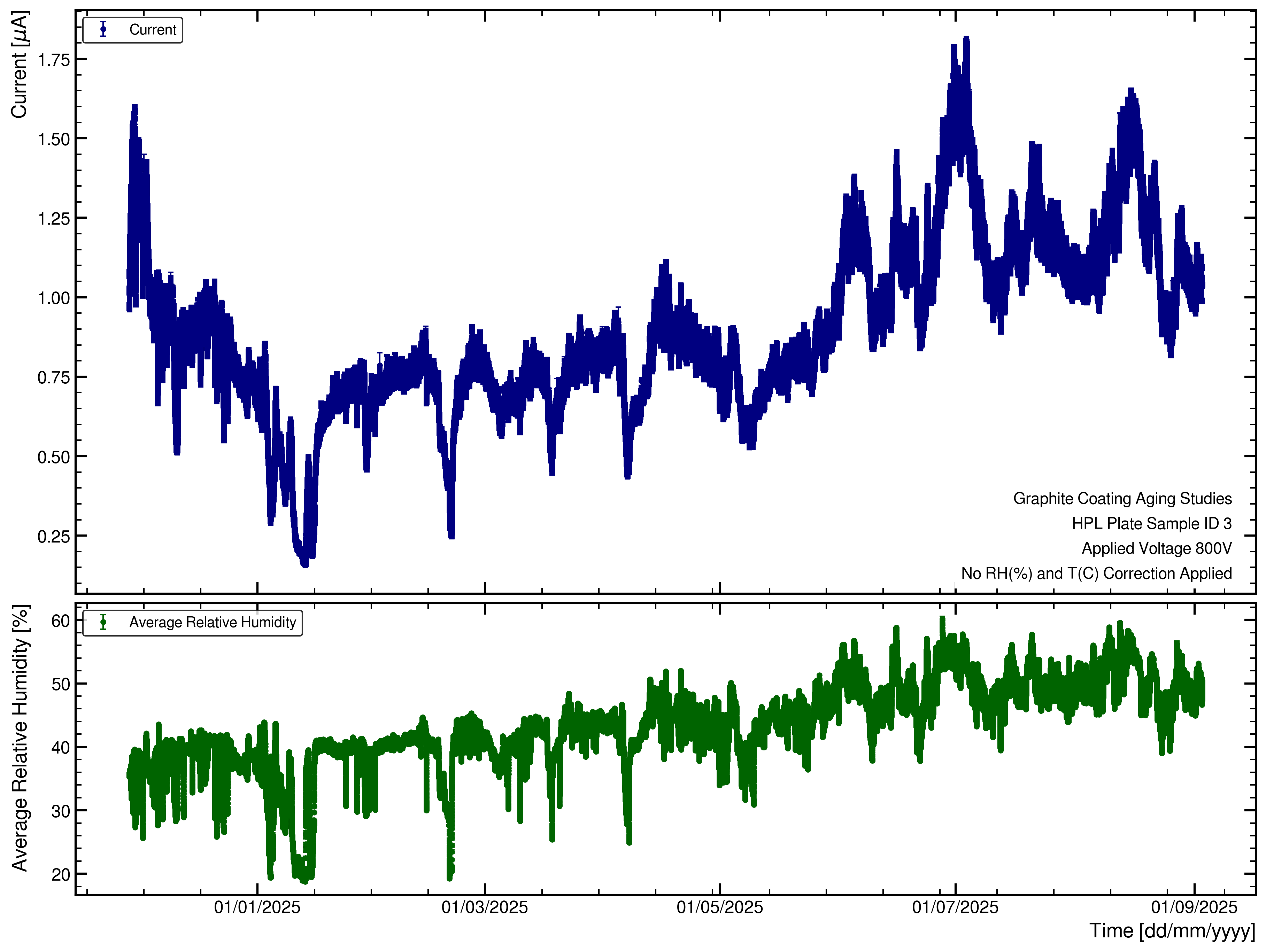}  
    \caption{}             
    \label{fig:current_humidity_correlation_samplt_800V}
  \end{subfigure}
  \hfill
  \begin{subfigure}[t]{0.9\textwidth}
    \centering
    \includegraphics[width=\textwidth]{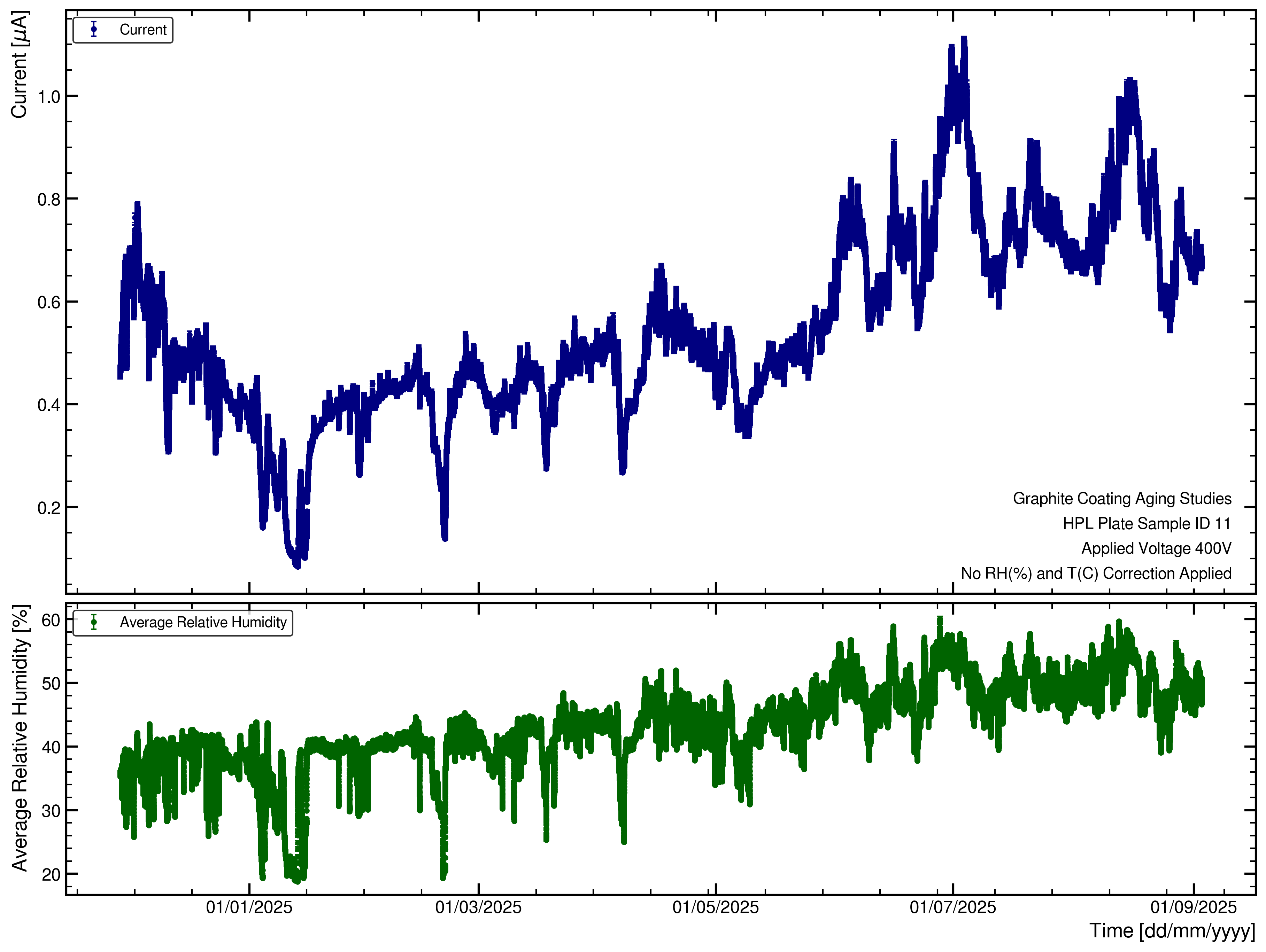}  
    \caption{}             
    \label{fig:current_humidity_correlation_samplt_400V}
  \end{subfigure}
  
  \caption{Temporal evolution of the current and relative humidity measured on the HPL electrode sample under two bias conditions: 800 V (a) and 400 V (b). The comparison highlights the strong dependence of the current on the ambient relative humidity.}
  \label{fig:current_humidity_correlation_summary}
\end{figure}

The significant fluctuations in the current and the corresponding bulk resistivity measured in the HPL samples are largely attributable to changes in ambient relative humidity. The temperature‑ and humidity‑dependence of the bulk resistivity of HPL plates has been the subject of extensive investigation over the past three decades \cite{ALICE:2000mot}, \cite{Ahn:2000tv}, \cite{Meghna:2016its}. Although the measurement campaign is carried out in a clean‑room with relative humidity set‑point fixed at 40\% RH,  excursions of several percentage points are still occasionally observed.  Two primary sources have been identified: (i) temporary shutdowns, whether unscheduled failures or routine maintenance, of the HVAC humidification unit, and (ii) daily and weather‑driven variations in outdoor humidity that propagate through the HVAC fresh‑air (make‑up) stream. Furthermore, the gradual up‑trend in the time-averaged current is driven by the seasonal drift of the clean‑room humidity from $\gtrsim 35\%$ RH in mid‑winter to $\gtrsim 55\%$ RH by early summer. The resulting $\approx 20 \%$ RH increase yields an $\approx 57 \%$ rise in the time‑averaged current, corresponding to an $\approx 72\%$ reduction in bulk resistivity of the HPL plate. Figure \ref{fig:environmental_parameter_monitoring} presents the temporal evolution of the average temperature [Figure \ref{fig:temperature_monitoring}] and the average relative humidity [Figure \ref{fig:humidity_monitoring}] inside the clean room hosting the long-term high-voltage stress-test setup. The average values of the ambient temperature and relative humidity have been recorded by four environmental monitoring stations distributed throughout the clean room. The corresponding histograms show a tight, near‑Gaussian temperature distribution ($\mu \approx 20.5 \, ^{\circ}C$, $\sigma \approx 0.33 \, ^{\circ}C$, $\sigma/\mu \approx 1.6\%$), indicating highly stable thermal conditions. In contrast, the humidity distribution is markedly broader and mildly asymmetric ($\mu \approx 43.1\%$, $\sigma \approx 6.9\%$, $\sigma/\mu \approx 15.9\%$), reflecting larger excursions and a gradual upward drift over the measurement period. Figure \ref{fig:current_humidity_correlation_summary} presents the temporal profiles of current and relative humidity for an HPL sample under two bias conditions: 800 V [Figure \ref{fig:current_humidity_correlation_samplt_800V}] and 400 V [Figure \ref{fig:current_humidity_correlation_samplt_400V}], highlighting the dependence of the current on the relative humidity.

\begin{figure}[htbp]
  \centering
  \begin{subfigure}[t]{0.49\textwidth}
    \centering
    \includegraphics[width=\textwidth]{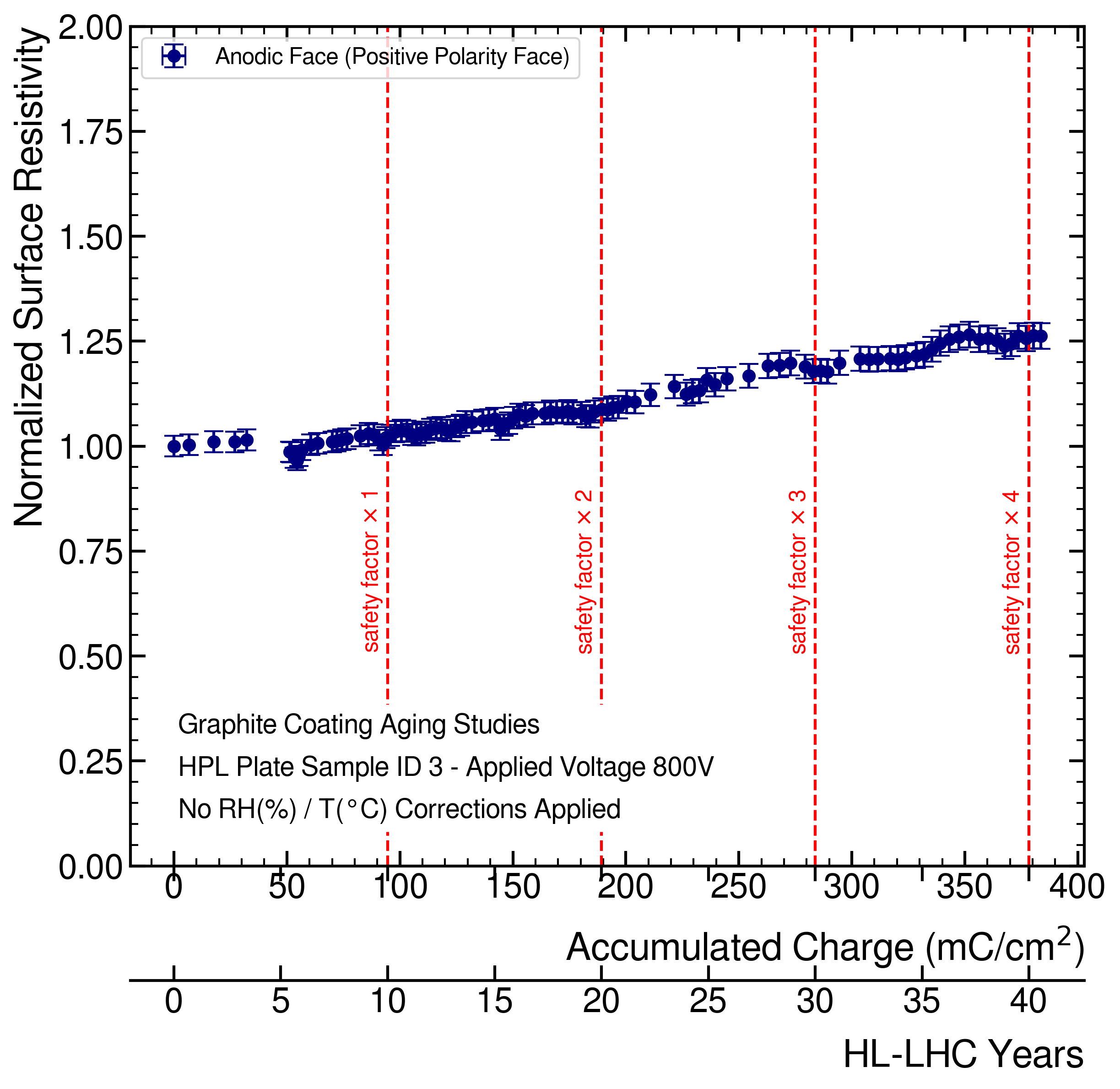}  
    \caption{}             
    \label{fig:surface_resistivity_anode_face_sample_800V}
  \end{subfigure}
  \hfill
  \begin{subfigure}[t]{0.49\textwidth}
    \centering
    \includegraphics[width=\textwidth]{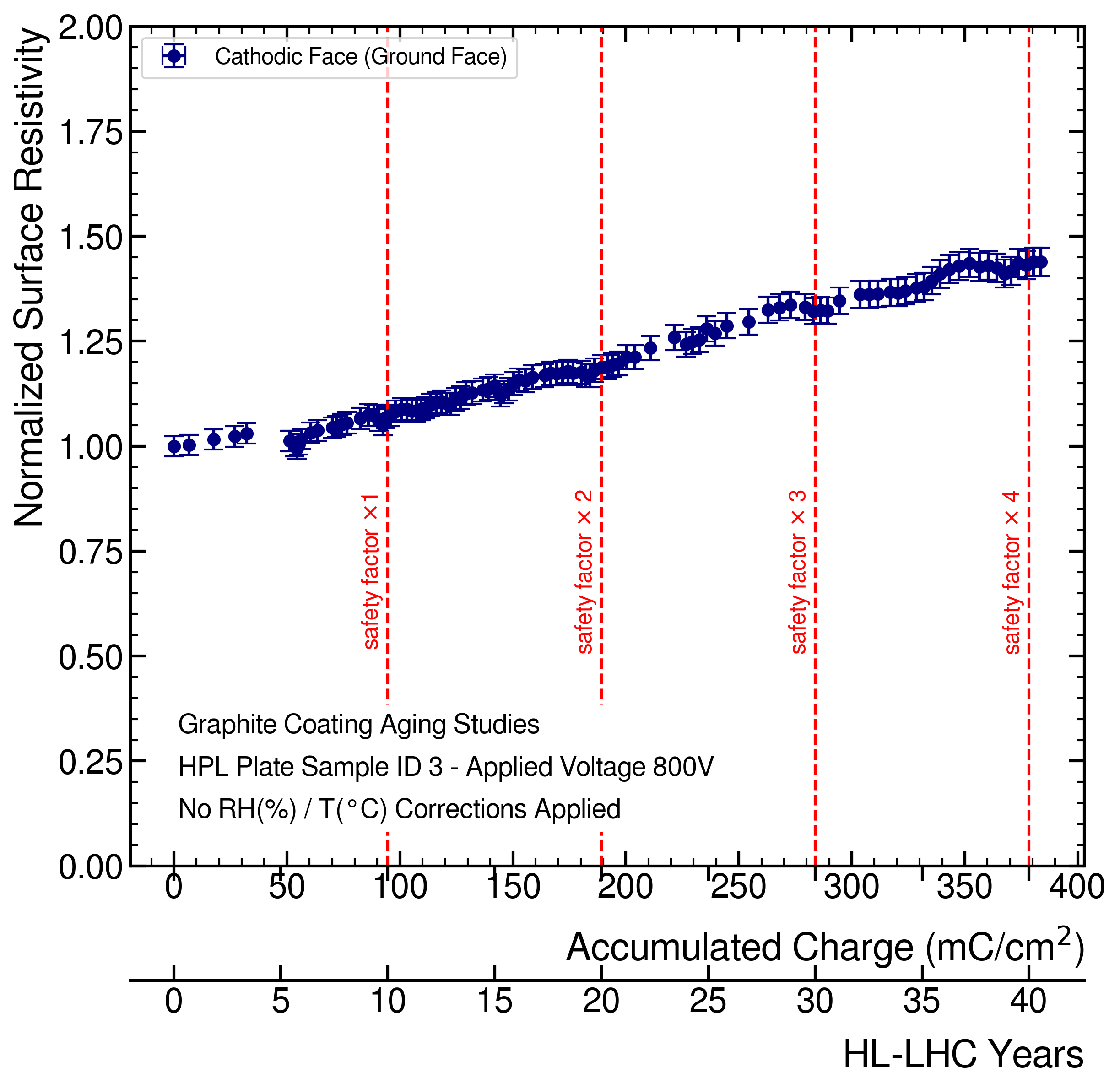}  
    \caption{}             
    \label{fig:surface_resistivity_cathode_face_sample_800V}
  \end{subfigure}
  
  \caption{Surface resistivity of the carbon-based resistive-ink coating measured on the anode (a) and cathode (b) faces of the HPL electrode sample at 800 V as a function of the integrated charge.}
  \label{fig:surface_resistivity_result_sample_800V}
\end{figure}

\begin{figure}[htbp]
  \centering
  \begin{subfigure}[t]{0.49\textwidth}
    \centering
    \includegraphics[width=\textwidth]{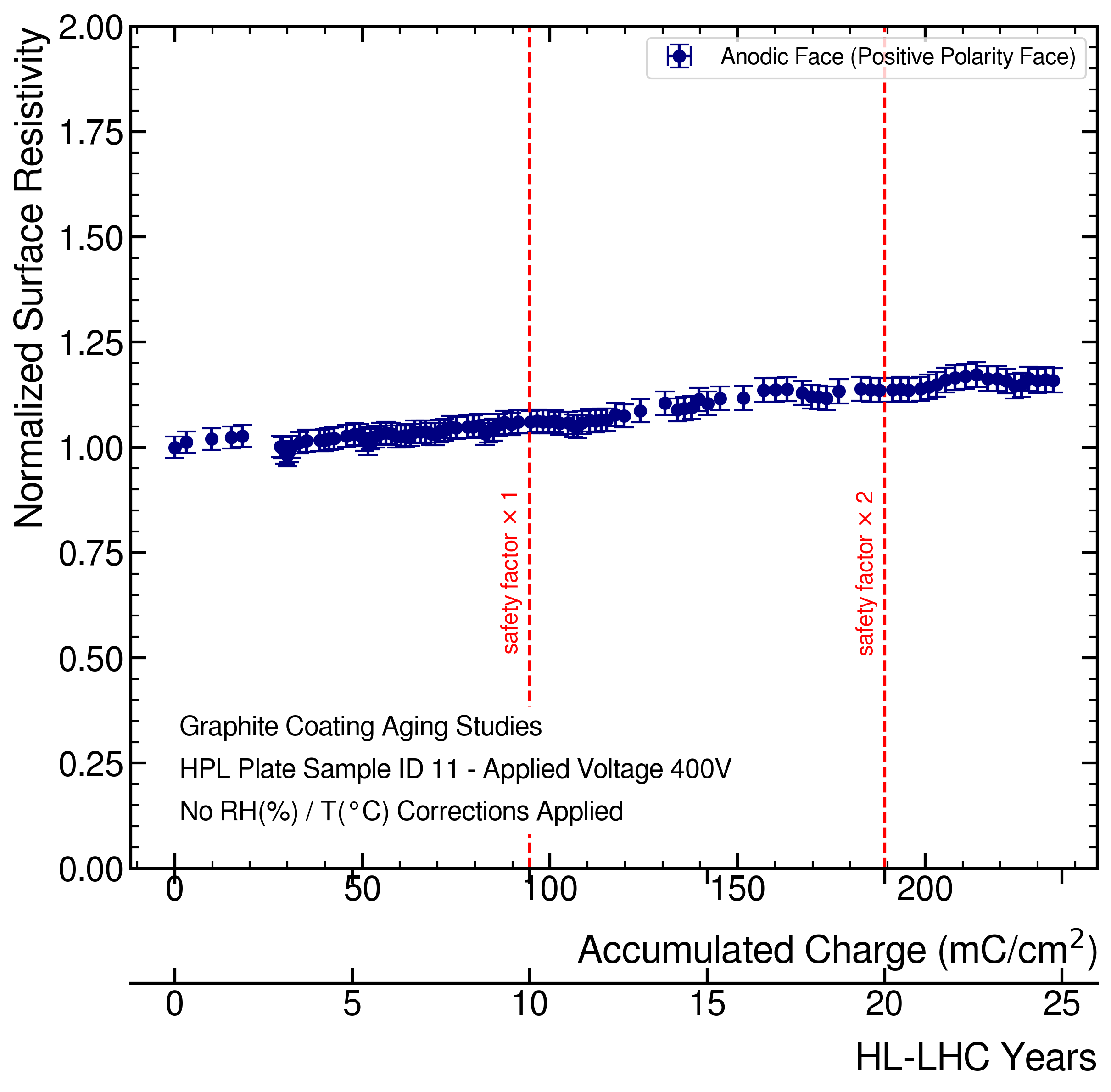}  
    \caption{}             
    \label{fig:surface_resistivity_anode_face_sample_400V}
  \end{subfigure}
  \hfill
  \begin{subfigure}[t]{0.49\textwidth}
    \centering
    \includegraphics[width=\textwidth]{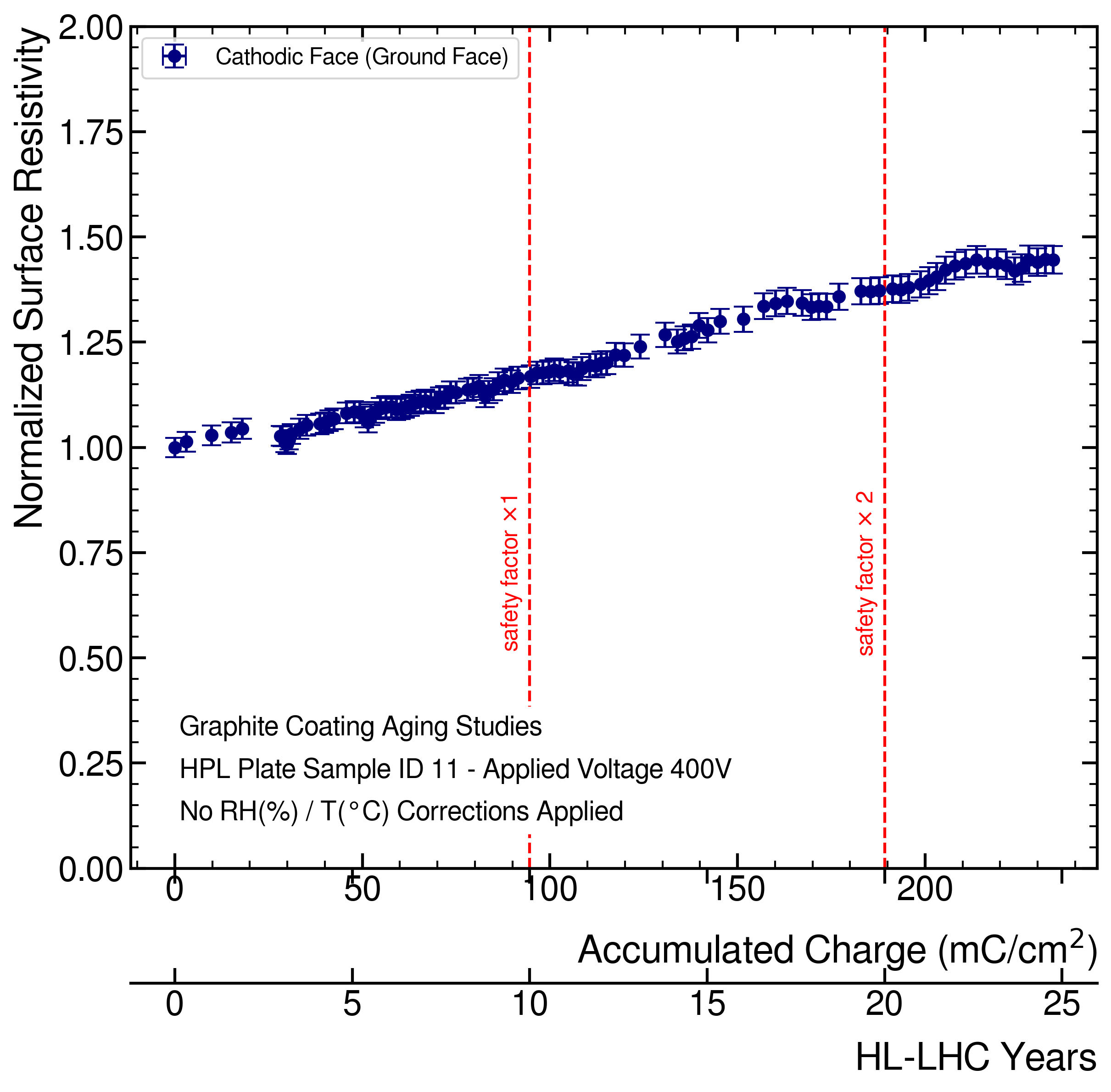}  
    \caption{}             
    \label{fig:surface_resistivity_cathode_face_400V}
  \end{subfigure}
  
  \caption{Surface resistivity of the carbon-based resistive-ink coating measured on the anode (a) and cathode (b) faces of the HPL electrode sample at 400 V as a function of the integrated charge.}
  \label{fig:surface_resistivity_result_sample_400V}
\end{figure}

Figure \ref{fig:surface_resistivity_result_sample_800V} compares the surface resistivity of the carbon‑based resistive ink coating on the anode [Figure \ref{fig:surface_resistivity_anode_face_sample_800V}] and cathode [Figure \ref{fig:surface_resistivity_cathode_face_sample_800V}] faces of the HPL electrode sample at 800 V as a function of the integrated charge. Figure \ref{fig:surface_resistivity_result_sample_400V} presents the corresponding measurements at 400 V. The Appendix \ref{sec:appendix_A} tabulates the surface resistivity as a function of the integrated charge for the remaining HPL electrode samples. 

Throughout the measurement campaign, the surface resistivity of the HPL electrode samples has exhibited a polarity-independent, strictly monotonic response to ambient relative humidity: increases in humidity have yielded higher surface resistivity on both the anode‑ and cathode‑facing sides, whereas humidity reductions have produced reversible decreases. Considering the chemical composition of the carbon‑based resistive ink, 5–15 wt\% conductive carbon black, 30–50 wt\% phenoxy resin (poly‑hydroxyether binder), and 40–60 wt\% diethylene‑glycol monobutyl‑ether acetate, an increase in relative humidity promotes water uptake by the hydroxyl‑rich phenoxy matrix, inducing slight volumetric swelling. This hygroscopic expansion increases the separation between adjacent carbon-black aggregates within the percolative network, wherein electron transport is governed primarily by contact resistance and tunneling; even $\mathring{A}$-scale increases in separation yield an exponential increase in electrical resistance. Consequently, both the number and effective cross-sectional area of conductive interparticle junctions between carbon-black aggregates diminish, thereby fragmenting the percolative pathways and increasing the overall surface resistivity. In contrast, a decrease in relative humidity drives water desorption from the phenoxy matrix, leading to polymer densification. This densification reduces the tunneling distances between adjacent carbon-black aggregates, restores conductive pathways within the percolative network, and consequently lowers the measured surface resistivity. Moreover, the cathode face of the HPL electrodes systematically exhibits a higher initial surface resistivity than the anode face and, consequently, a larger fractional variation during relative‑humidity excursions. This trend is consistent with the percolation/tunneling framework for carbon‑black filled polymers: as the effective carbon‑black volume fraction approaches the percolation threshold, small humidity‑induced increases in interparticle spacing, caused by uptake and swelling of the phenoxy binder, produce large increases in surface resistivity. Because the cathode surface is already closer to the threshold, the same volumetric strain yields a much larger surface resistivity increase compared to the anode surface.

Besides the ambient relative humidity‑dependent, reversible variations in surface resistivity, bias‑induced aging mechanism has been hypothesized on both the anode‑ and cathode‑facing sides of the HPL electrode samples, mediated by electric‑field‑driven ion transport and interfacial oxidation-reduction. During prolonged high‑voltage biasing, field‑driven migration could cause mobile ionic contaminants ($Na^+$,  $K^+$, $Ca^{2+}$, …) and in‑situ generated $OH^-/HCO_3^-$ to accumulate at the counter‑electrode, where they precipitate or react at the carbon-black/phenoxy interface, thereby increasing interparticle contact resistance. At the cathode, reduction of dissolved $O_2/H_2O$ generates $OH^-$, promoting alkaline cleavage of ether linkages in the binder and de-wetting of carbon-black aggregates; at the anode, oxidative processes convert the carbon-black surface sites into $\mathrm{C{-}O}$/$\mathrm{C{=}O}$ functionalities and could oxidize the polymer backbone, both effects diminishing the number and quality of conductive junctions. Furthermore, repetitive hygroscopic swelling-contraction cycles, and occasional micro-discharges further open micro‑gaps within the conductive percolation network. These aging processes induce an irreversible upward shift in baseline surface resistivity, which superimposes on and biases the reversible ambient relative humidity‑driven variations.

To confirm the presence of bias-induced aging mechanisms on both the anode and cathode surfaces of the HPL electrodes subjected to high-voltage stress tests, a comprehensive morphological and elemental chemical composition analyses have been planned at the CERN Materials and Metrology Laboratory and the Technical University of Munich. The analysis strategy will combine Scanning Electron Microscopy (SEM) and Energy‑Dispersive X‑ray Spectroscopy (EDS) with Focused Ion Beam (FIB) cross‑sectioning to achieve high‑resolution morphological characterization and spatially‑resolved elemental mapping. In parallel, Secondary Ion Mass Spectrometry (SIMS) will be employed for qualitative and quantitative depth profiling of ionic and molecular contaminants on both electrode surfaces.

\section{Irradiation Test Campaign at CERN CHARM Facility}
\label{sec:Irradiation_Test_CHARM}
\begin{figure}[htbp]
  \centering
  \begin{subfigure}[t]{0.49\textwidth}
    \centering
    \includegraphics[width=\textwidth]{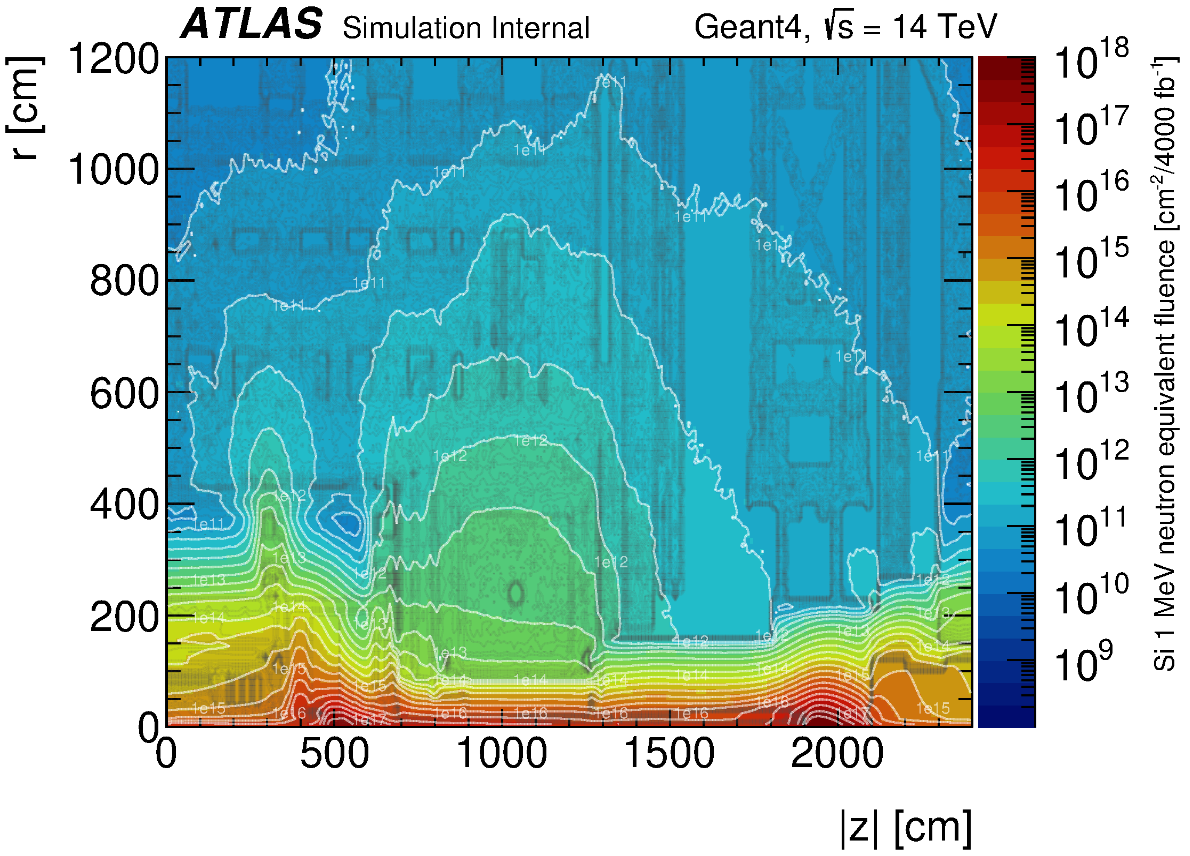}  
    \caption{}             
    \label{fig:NIEL}
  \end{subfigure}
  \hfill
  \begin{subfigure}[t]{0.49\textwidth}
    \centering
    \includegraphics[width=\textwidth]{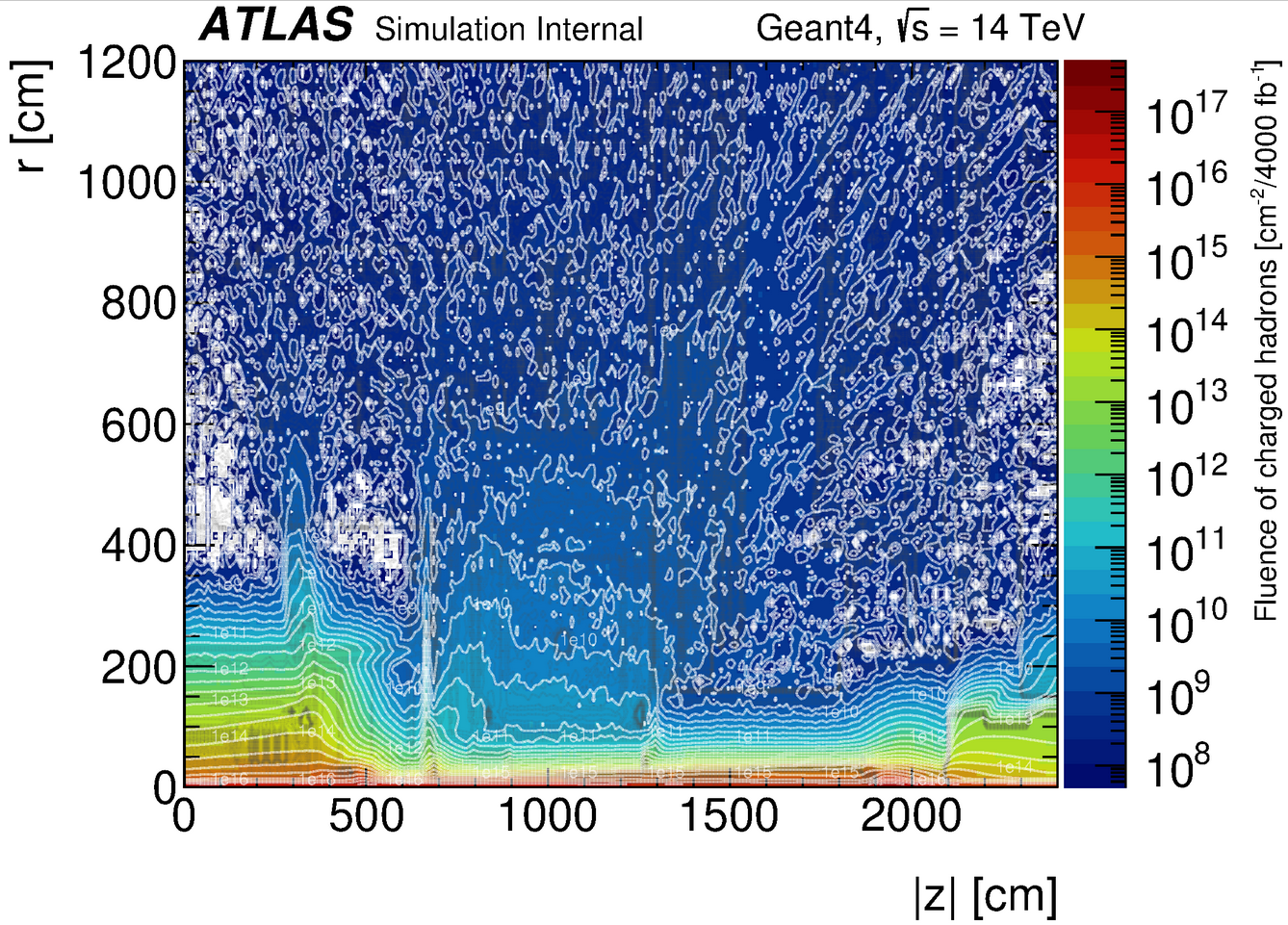}  
    \caption{}             
    \label{fig:HEH}
  \end{subfigure}
  
  \caption{GEANT4 simulation for the 1 MeV Si‑equivalent neutron fluence (a) and charge hadron fluence (b) irradiation map in the ATLAS cavern for $pp$ collisions at $\sqrt{s}=14;\text{TeV}$.}
  \label{fig:ATLAS_NIEL_HEH_ATLAS_Cavern}
\end{figure}

In order to evaluate the long-term stability and radiation tolerance of the carbon-based resistive coating applied to the electrodes of the RPC detector technology, an accelerated irradiation campaign has been carried out at the CHARM facility at CERN. The irradiation test has been designed to reproduce, in a compressed timescale, the mixed-field neutron radiation environment expected in the CMS and ATLAS experiments over ten years of operation under HL-LHC conditions. The  CERN’s High‑energy AcceleRator Mixed‑field (CHARM), located on the T8 beam‑line in the PS East Area, converts the 24 GeV/c proton beam into a broad‑spectrum mixed field by intercepting the beam with a Cu/Al target. Movable concrete‑and‑iron shielding blocks modulate the flux and spectral composition, dominated by neutrons, charged hadrons and photons \cite{Mekki:2016tbj}. Under HL‑LHC operating conditions, the new 1 mm‑gap RPC subsystem of the ATLAS Muon Spectrometer is expected to receive a 1 MeV Si‑equivalent neutron fluence of approximately $1.2 \times 10^{12} \, n \, / \, cm^2$. Figure \ref{fig:ATLAS_NIEL_HEH_ATLAS_Cavern} shows the GEANT4‑simulated map of the 1 MeV Si‑equivalent neutron fluence [Figure \ref{fig:NIEL}] and charge hadron fluence [Figure \ref{fig:HEH}] in the ATLAS cavern for pp collisions at $\sqrt{s}=14 \, \text{TeV}$, normalized to an integrated luminosity of $4000 \, fb^{-1}$ \cite{ATLASUpgradeTWiki}. 

\begin{figure}[htbp]
  \centering
  \begin{subfigure}[t]{0.49\textwidth}
    \centering
    \includegraphics[width=\textwidth]{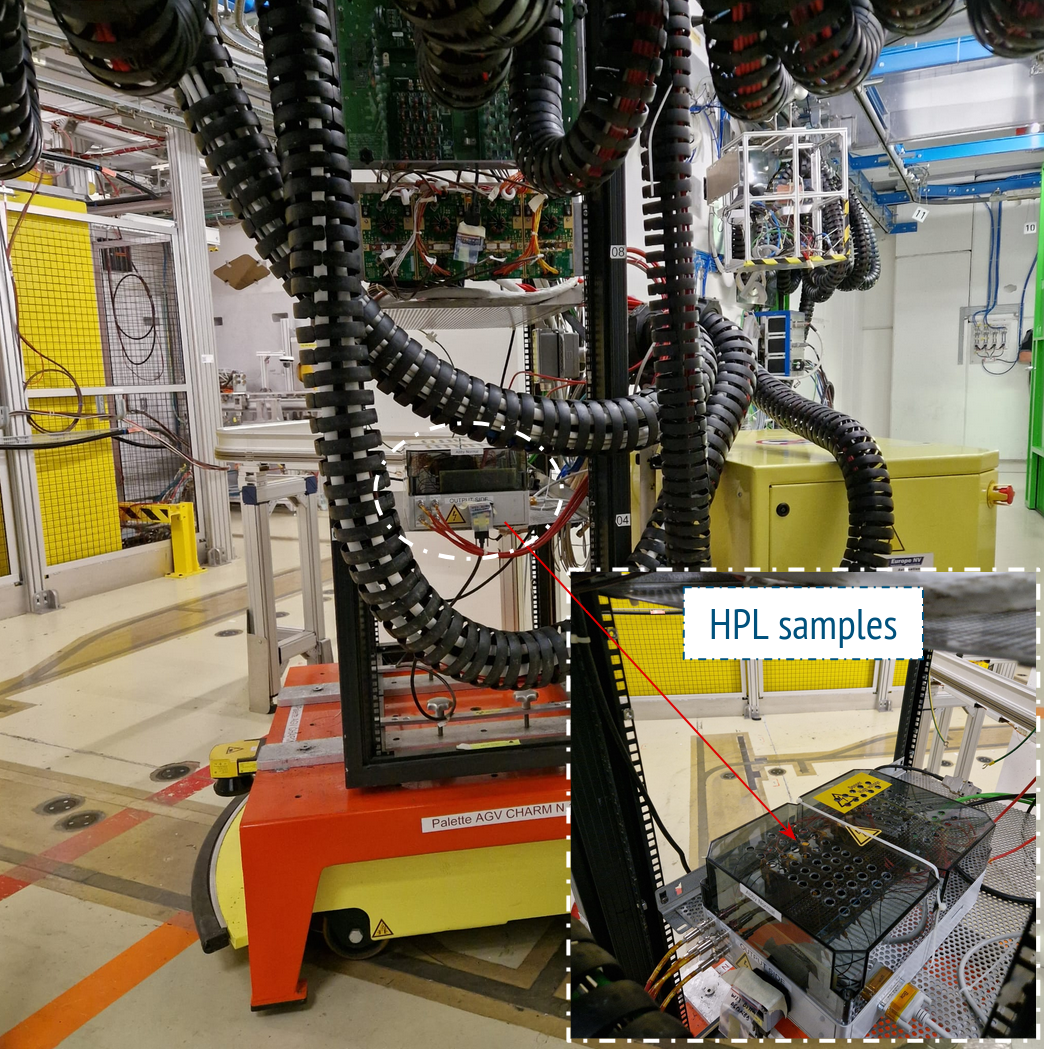}  
    \caption{}             
    \label{fig:CERN_CHARM_Bunker}
  \end{subfigure}
  \hfill
  \begin{subfigure}[t]{0.49\textwidth}
    \centering
    \includegraphics[width=\textwidth]{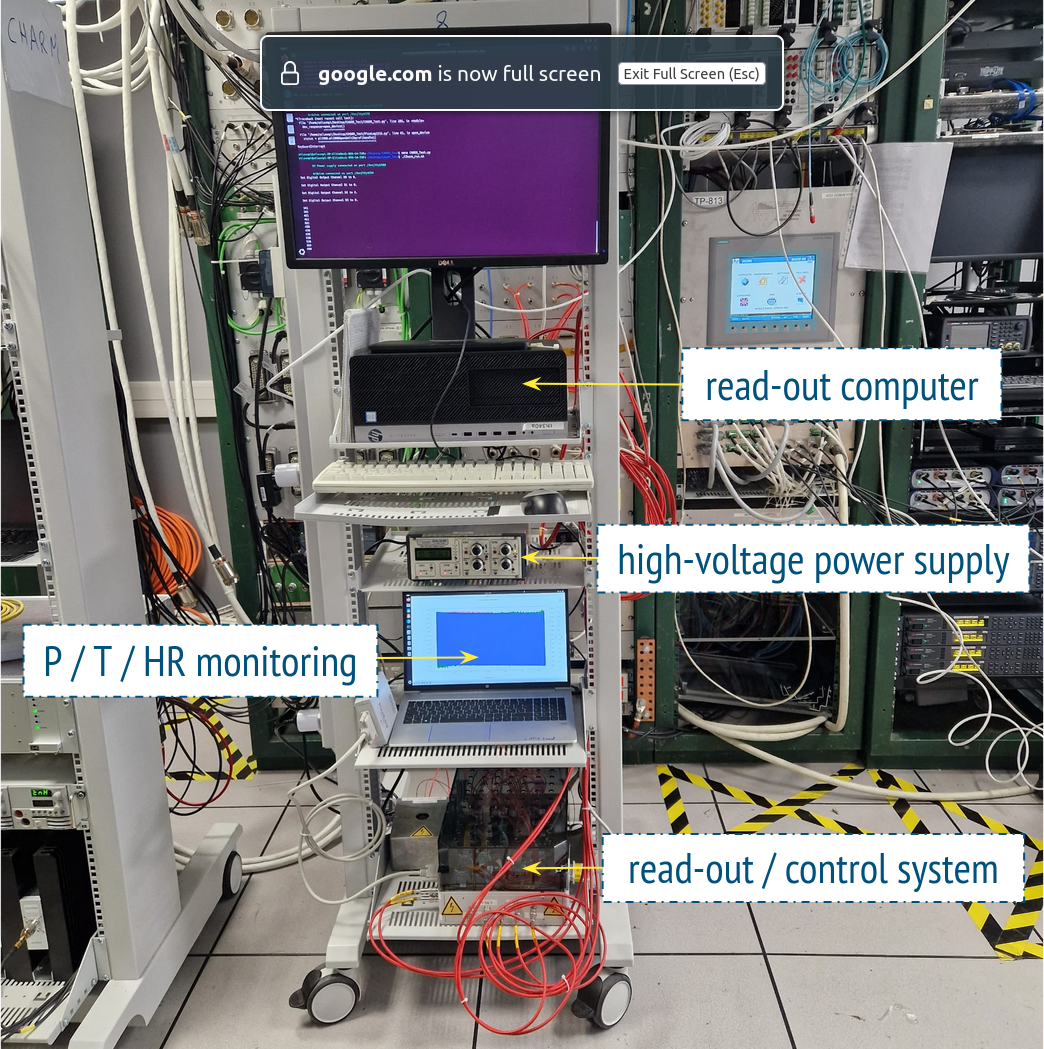}  
    \caption{}             
    \label{fig:CERN_CHARM_Control_Room}
  \end{subfigure}
  
  \caption{Experimental setup at the CERN CHARM facility. Installation of the HPL electrode samples within the irradiation area (a). Deployment of the remote readout and control system in the control room of the facility (b).}
  \label{fig:CERN_CHARM_Experimental_Setup}
\end{figure}

A total of five $10 \, cm \times 10 \, cm$ HPL samples have been coated with carbon-based resistive ink using the screen-printing technique detailed in Sections \ref{sec:silk-screen_printing} and \ref{sec:Stress_Test_MPI}. Due to restricted access to the irradiation area at the CHARM facility resulting from neutron activation, the data acquisition system has been fully automated, enabling uninterrupted remote measurement of both bulk and surface resistivity of the HPL electrode samples over the full duration of the irradiation campaign. Figure \ref{fig:CERN_CHARM_Experimental_Setup} illustrates the installation of the experimental setup at the CERN CHARM facility: the positioning of the HPL electrode samples inside the irradiation zone [Figure \ref{fig:CERN_CHARM_Bunker}], and the deployment of the readout and control system within the control room [Figure \ref{fig:CERN_CHARM_Control_Room}].

\begin{figure}[htbp]
  \centering
  \begin{subfigure}[t]{0.49\textwidth}
    \centering
    \includegraphics[width=\textwidth]{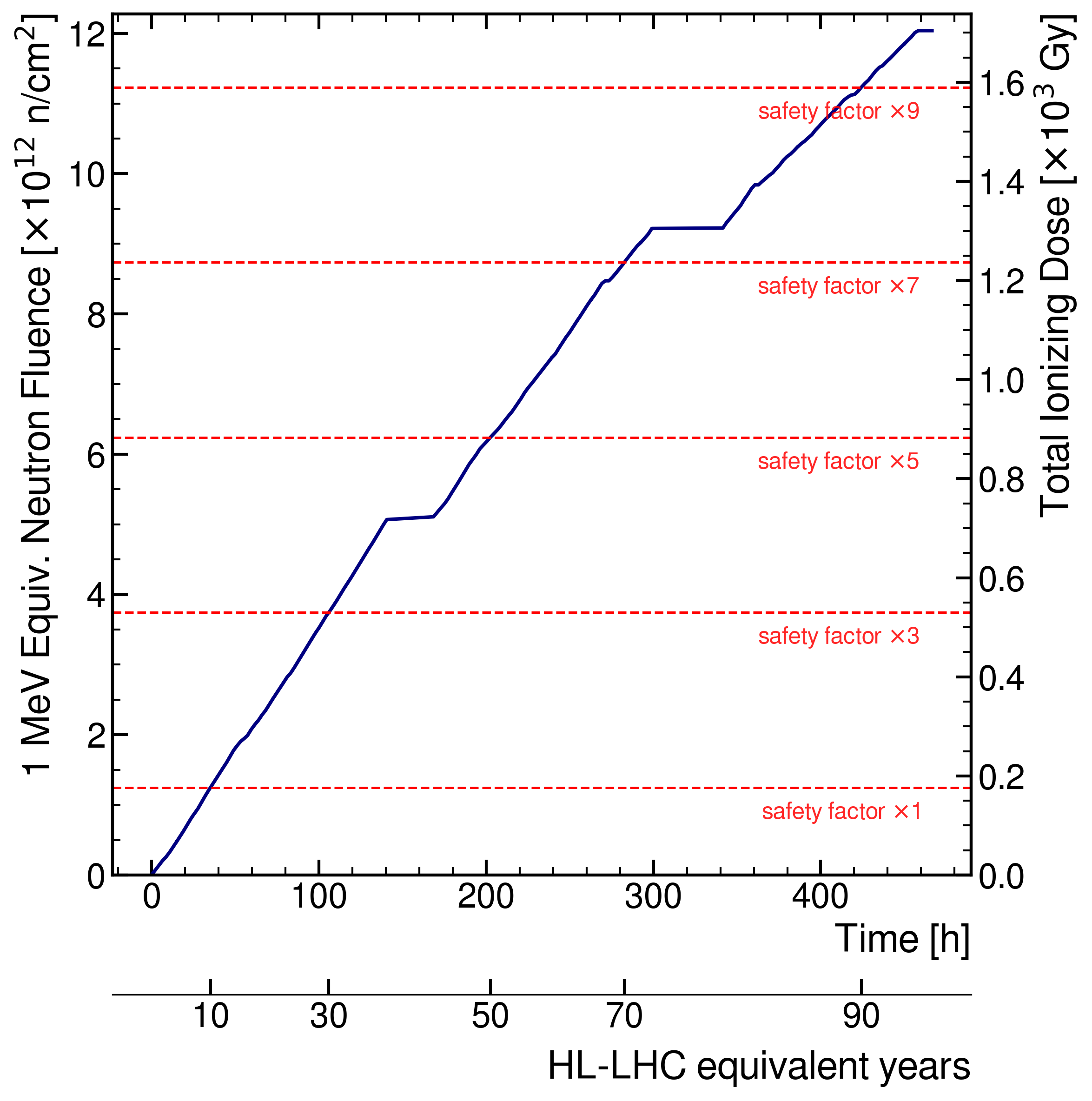}  
    \caption{}             
    \label{fig:Irradiation_Dose_Results}
  \end{subfigure}
  \hfill
  \begin{subfigure}[t]{0.49\textwidth}
    \centering
    \includegraphics[width=\textwidth]{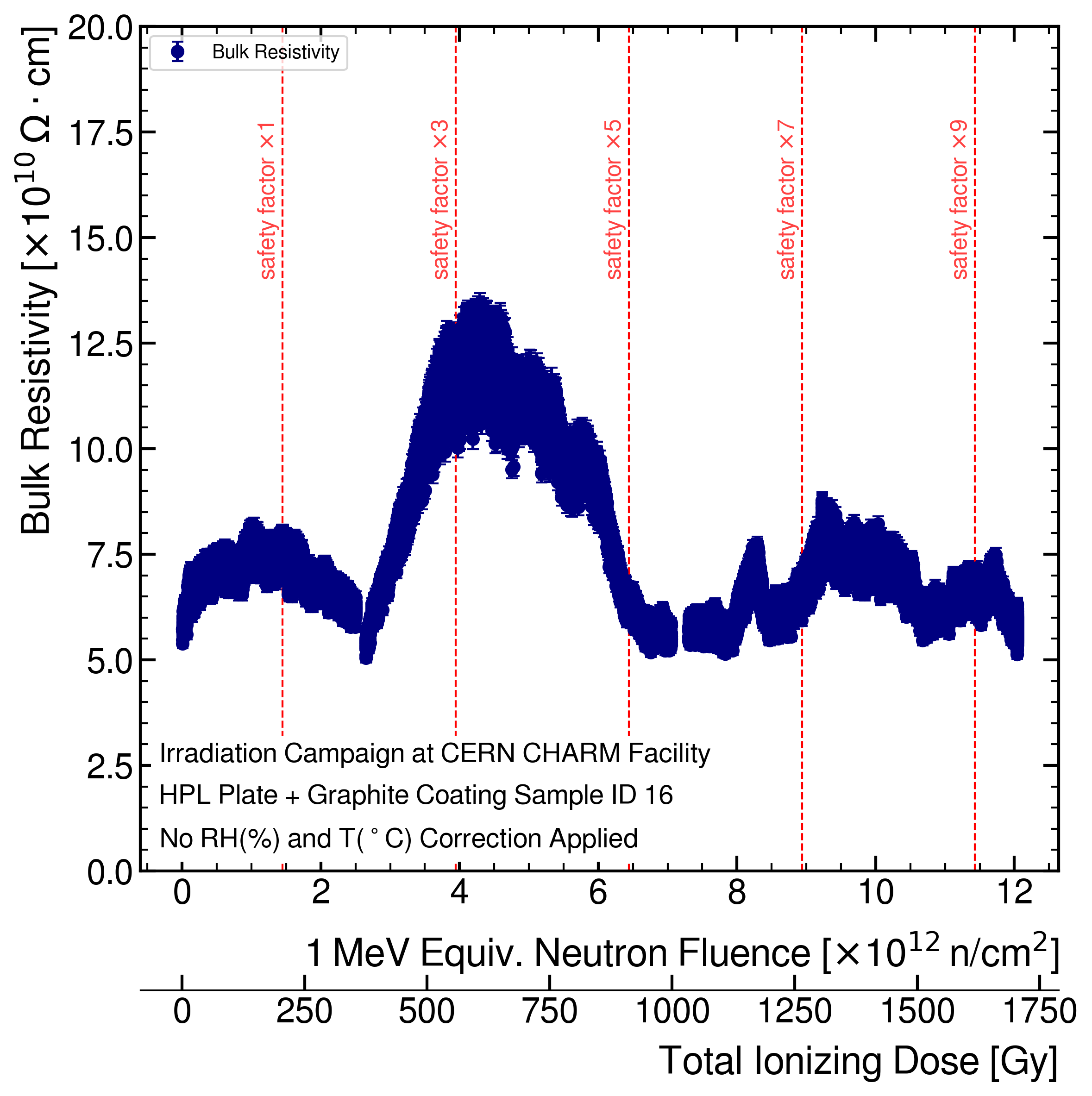}  
    \caption{}             
    \label{fig:Bulk_Resistivity_Results}
  \end{subfigure}
  
  \caption{(a) Time evolution of the cumulative 1 MeV Si-equivalent neutron fluence and total ionizing dose (TID) measured with the RADMON system during the irradiation campaign. (b) Bulk resistivity of the high-pressure-laminate (HPL) electrode sample as a function of the accumulated 1 MeV Si-eq. neutron fluence and TID.}
  \label{fig:CERN_CHARM_Bulk_Resistivity_Results}
\end{figure}

\begin{figure}[htbp]
  \centering
  \begin{subfigure}[t]{0.49\textwidth}
    \centering
    \includegraphics[width=\textwidth]{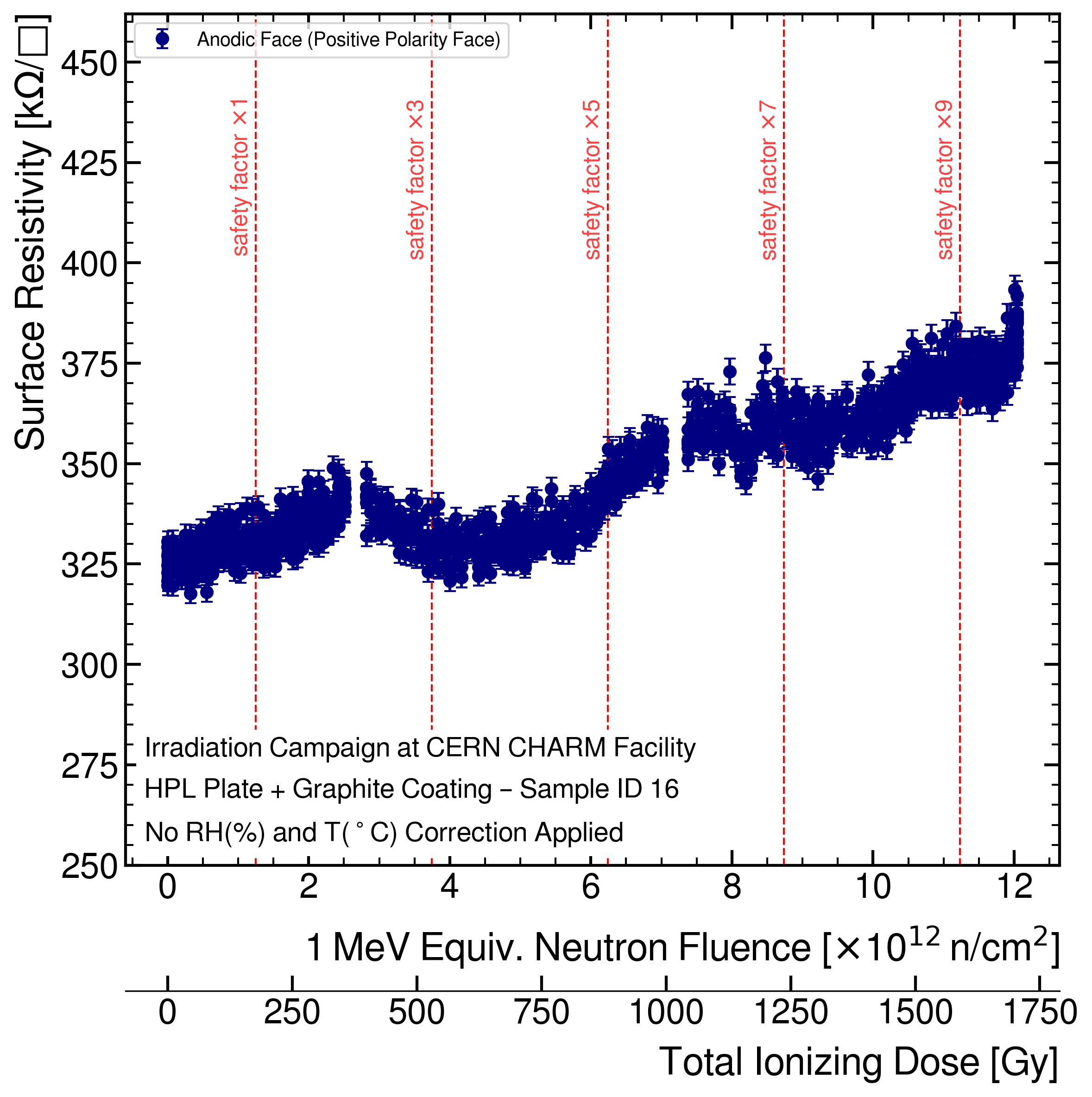}  
    \caption{}             
    \label{fig:CERN_CHARM_Surface_Resistivity_Results_Anode}
  \end{subfigure}
  \hfill
  \begin{subfigure}[t]{0.49\textwidth}
    \centering
    \includegraphics[width=\textwidth]{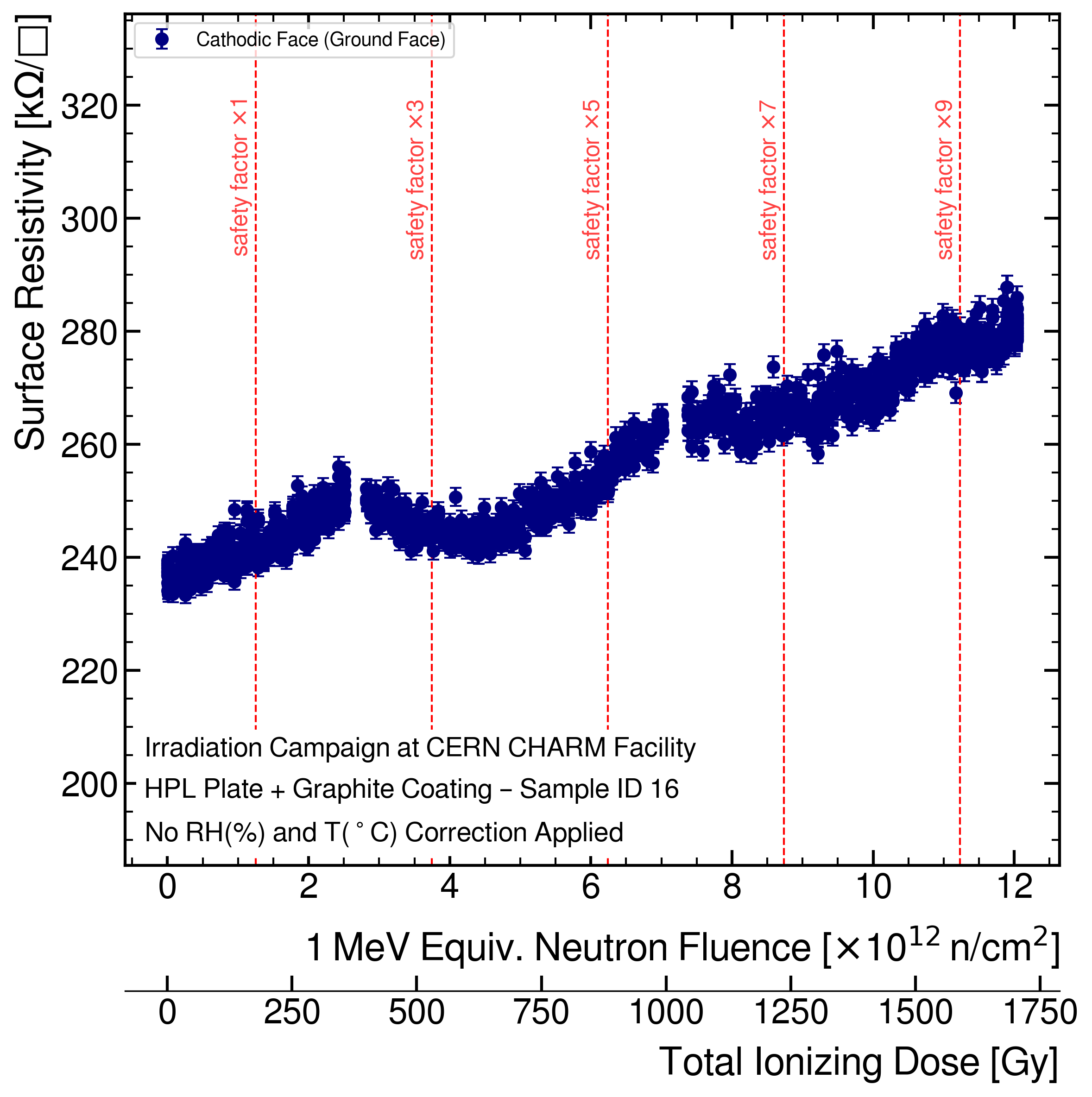}  
    \caption{}             
    \label{fig:CERN_CHARM_Surface_Resistivity_Results_Cathode}
  \end{subfigure}
  
  \caption{Surface resistivity of the carbon-based resistive-ink coating on the high-pressure-laminate (HPL) electrode sample as a function of accumulated 1 MeV Si-equivalent neutron fluence and total ionizing dose (TID): (a) anode face; (b) cathode face.}
  \label{fig:CERN_CHARM_Surface_Resistivity_Results}
\end{figure}

\begin{figure}[htbp]
    \centering
    \includegraphics[width=0.95\textwidth]{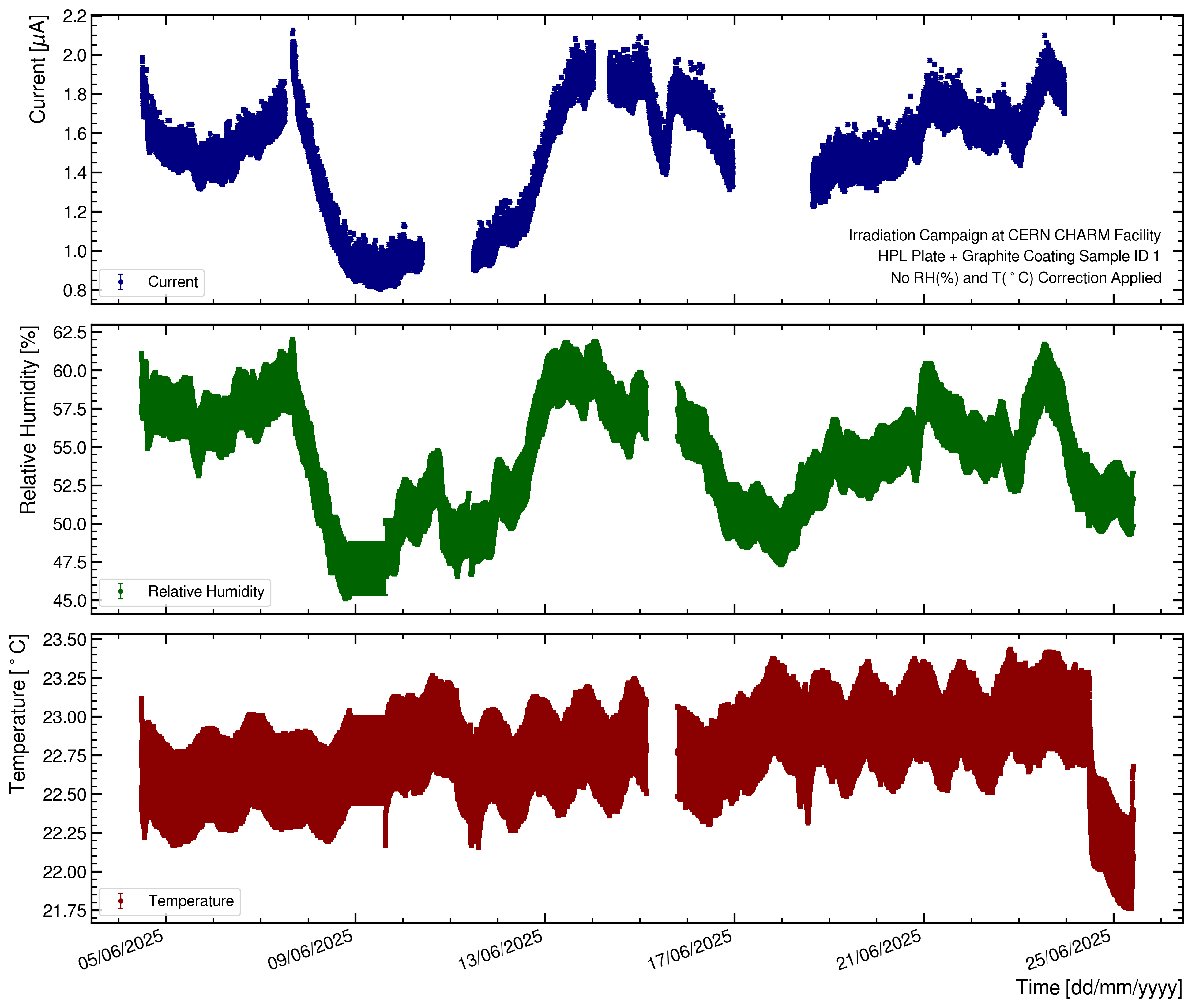}
    \caption{Temporal evolution of the current, relative humidity and temperature measured on the HPL electrode sample under irradiation campaign at CERN CHARM facility. The comparison highlights the strong dependence of the current on the ambient relative humidity.}
    \label{fig:current_relative_humidity_charm_test}
\end{figure}

The neutron fluence and the absorbed ionizing dose have been continuously monitored throughout the irradiation campaign using the CERN Radiation Monitoring (RADMON) system \cite{Spiezia:2014iua}. Simultaneously, environmental parameters, including atmospheric pressure, ambient temperature, and relative humidity, have been systematically recorded within the irradiation area to ensure accurate control and interpretation of the exposure conditions. Figure \ref{fig:Irradiation_Dose_Results} presents the cumulative 1 MeV Si‑equivalent neutron fluence and total ionizing dose as a function of the time, as recorded by the RADMON system throughout the irradiation test. Figure \ref{fig:Bulk_Resistivity_Results} illustrates the evolution of the bulk resistivity of the HPL electrode sample as a function of both the 1 MeV Si‑equivalent neutron fluence and the accumulated total ionizing dose. Figure \ref{fig:CERN_CHARM_Surface_Resistivity_Results} compares the surface resistivity of the carbon‑based resistive ink coating on the anode [Figure \ref{fig:CERN_CHARM_Surface_Resistivity_Results_Anode}] and cathode [Figure \ref{fig:CERN_CHARM_Surface_Resistivity_Results_Cathode}] faces of the HPL electrode sample as a function of both the 1 MeV Si‑equivalent neutron fluence and the accumulated total ionizing dose. As detailed in Section \ref{sec:Stress_Test_MPI}, the fluctuations observed in both the current through the HPL electrode samples, the corresponding bulk resistivity, and the surface resistivity of the carbon‑based resistive ink coating on the anode and cathode faces of the HPL electrode sample are primarily attributable to variations in relative humidity within the irradiation area of the CHARM facility. 
Figure \ref{fig:current_relative_humidity_charm_test} presents the temporal profiles of current, relative humidity and temperature for an HPL electrode sample under irradiation campaign at CERN CHARM facility, highlighting the dependence of the current on the relative humidity. The Appendix \ref{sec:appendix_B} tabulates both the bulk resistivity and surface resistivity as a function of the 1 MeV Si‑equivalent neutron fluence and the accumulated total ionizing dose for the remaining HPL electrode samples. 

The irradiation campaign comprised $\approx400 \, h$ of uninterrupted exposure. Throughout this period, both the bulk resistivity of the HPL electrode samples and the surface resistivity of the carbon-based resistive ink coating remained stable, while the samples accumulated a 1 MeV Si‑equivalent neutron fluence of $\approx 10 \times 10^{12}\, n/cm^2$ and high energy hadrons ($> 20 \, MeV$) fluence of $\approx 10^6 \, p/cm^2$, corresponding to the maximum 1 MeV Si‑equivalent neutron fluence and high energy hadron fluence expected in the RPC subsystem of the ATLAS Muon Spectrometer over ten years of operation at the HL-LHC, including a safety factor of $\times 9$ and $\times 4$, respectively. As outlined in Section \ref{sec:Stress_Test_MPI}, a comprehensive series of post-irradiation morphological and elemental chemical composition  analyses have been scheduled at the CERN Materials and Morphology Laboratory and at the Technical University of Munich, with the aim of identifying potential aging mechanisms in the carbon-based resistive ink layer.

\section{Irradiation Test Campaign at CERN GIF++ Facility}
\label{sec:Irradiation_Test_GIFpp}
\begin{figure}[htbp]
  \centering
  \begin{subfigure}[t]{0.49\textwidth}
    \centering
    \includegraphics[width=\textwidth]{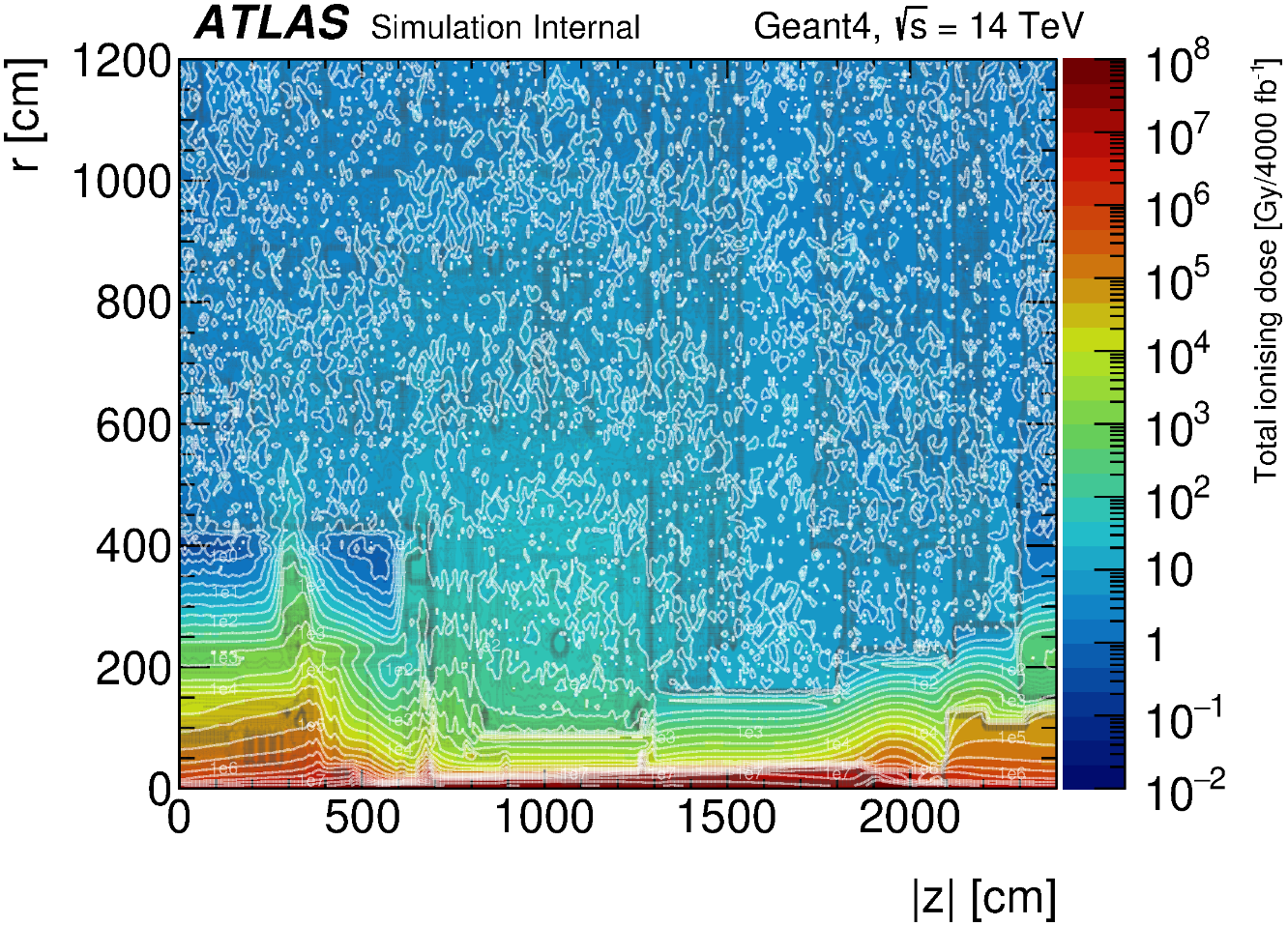}  
    \caption{}             
    \label{fig:TID}
  \end{subfigure}
  \hfill
  \begin{subfigure}[t]{0.49\textwidth}
    \centering
    \includegraphics[width=\textwidth]{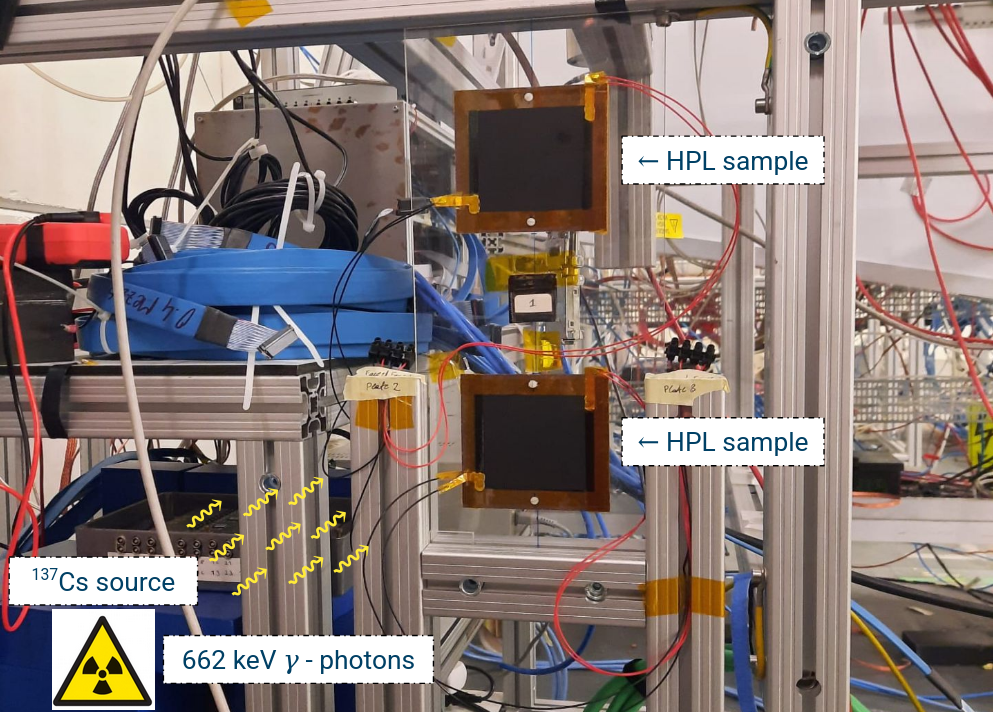}  
    \caption{}             
    \label{fig:GIFpp_setup}
  \end{subfigure}
  
  \caption{GEANT4 simulation for the TID irradiation map in the ATLAS cavern for $pp$ collisions at $\sqrt{s}=14;\text{TeV}$ (a). Installation of the HPL samples coated with the carbon-based resistive ink layer at the CERN GIF++ facility (b).}
  \label{fig:ATLAS_TID_GIFpp}
\end{figure}

A further irradiation campaign is currently under way at the GIF++ gamma-irradiation facility at CERN. The test facility uniquely combines a $^{137}Cs$ source ($\approx$ 11 TBq in 2025) with two sets of adjustable filters to vary the intensity, and a high-energy muon beam (100 GeV/c) delivered by the SPS secondary line H4 in EHN1 during dedicated running periods. The photon spectrum of the $^{137}Cs$ source, dominated by its 662 keV primary line and Compton-scattered low-energy tail, closely reproduces the characteristic energies of neutron-induced backgrounds observed in LHC experiments such as ATLAS and CMS \cite{Pfeiffer:2016hnl}. Under HL-LHC operating conditions, the new 1 mm-gap RPC subsystem of the ATLAS Muon Spectrometer is expected to accumulate a maximum total-ionizing dose (TID) of about 43 Gy. Figure \ref{fig:TID} shows the GEANT4 simulation for the total-ionizing-dose map in the ATLAS cavern for $pp$ collisions at $\sqrt{s}=14 \, \text{TeV}$, normalized to an integrated luminosity of $4000 \, fb^{-1}$ \cite{ATLASUpgradeTWiki}. 

\begin{figure}[h!]
    \centering
    \begin{subfigure}[b]{0.49\textwidth}
        \centering
        \includegraphics[width=\textwidth]{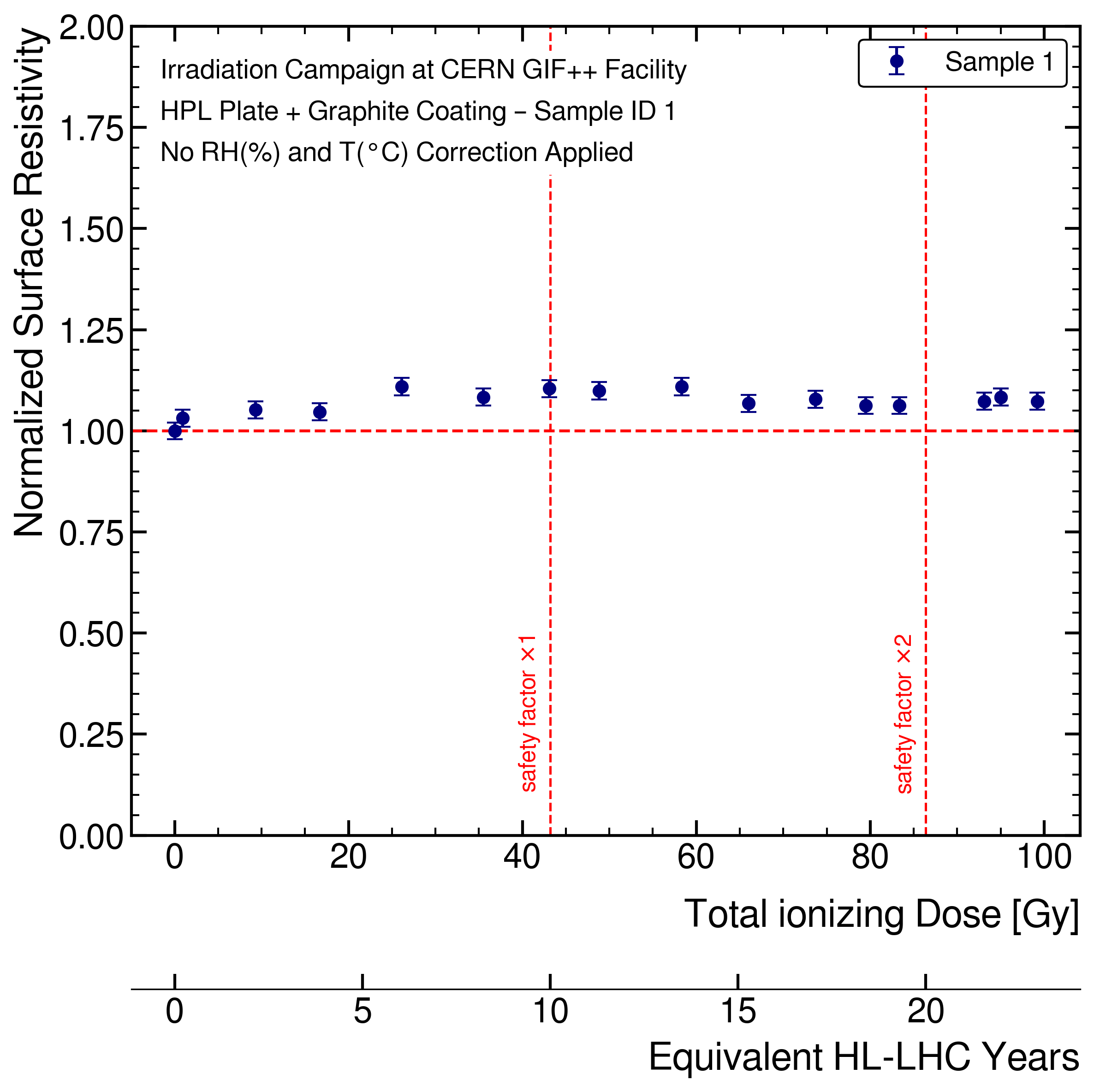}
        \caption{ }
        \label{fig:Sample1}
    \end{subfigure}
    \hfill
    \begin{subfigure}[b]{0.49\textwidth}
        \centering
        \includegraphics[width=\textwidth]{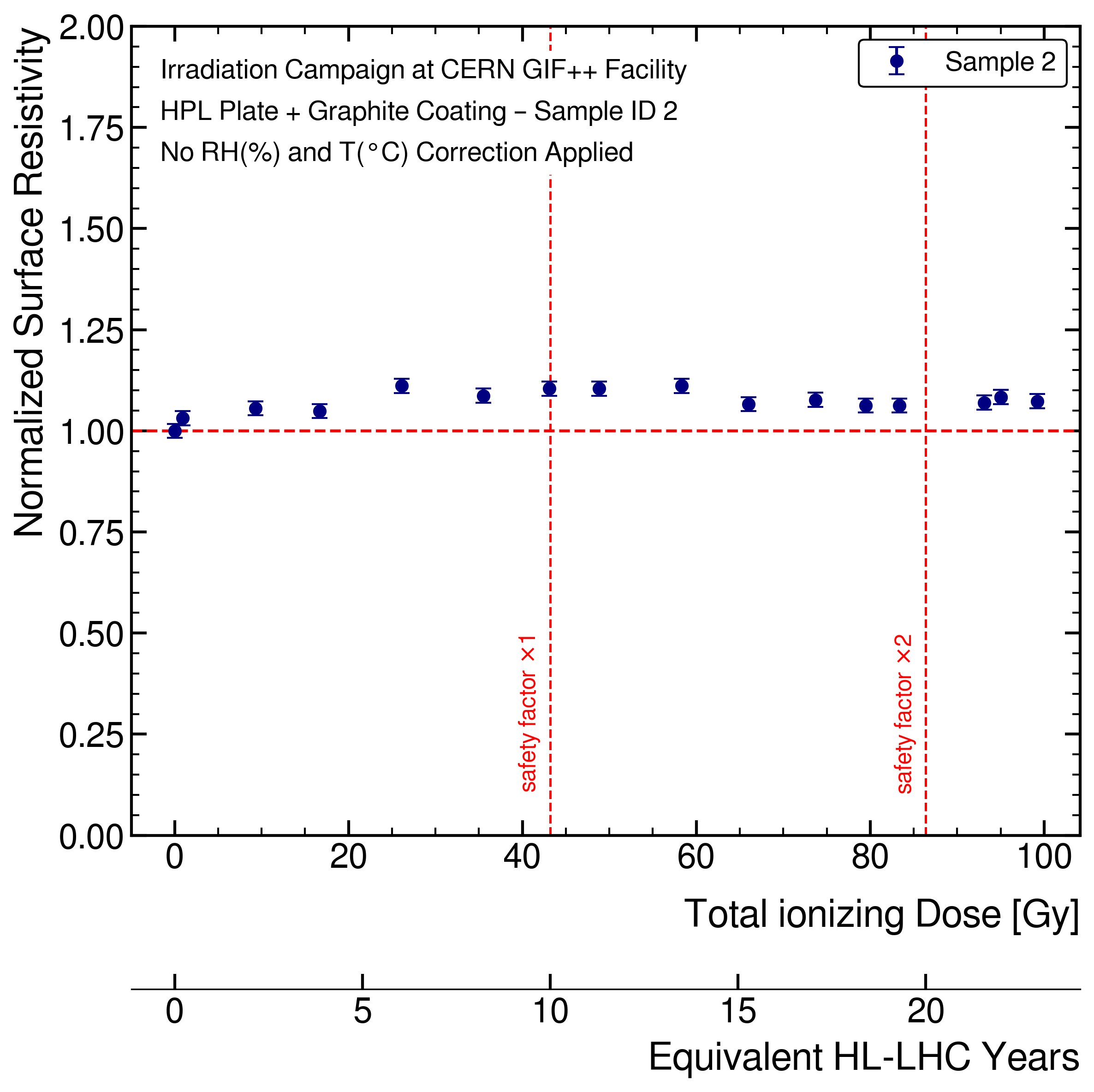}
        \caption{ }
        \label{fig:Sample2}
    \end{subfigure}

    \vspace{0.5cm} 

    \begin{subfigure}[b]{0.49\textwidth}
        \centering
        \includegraphics[width=\textwidth]{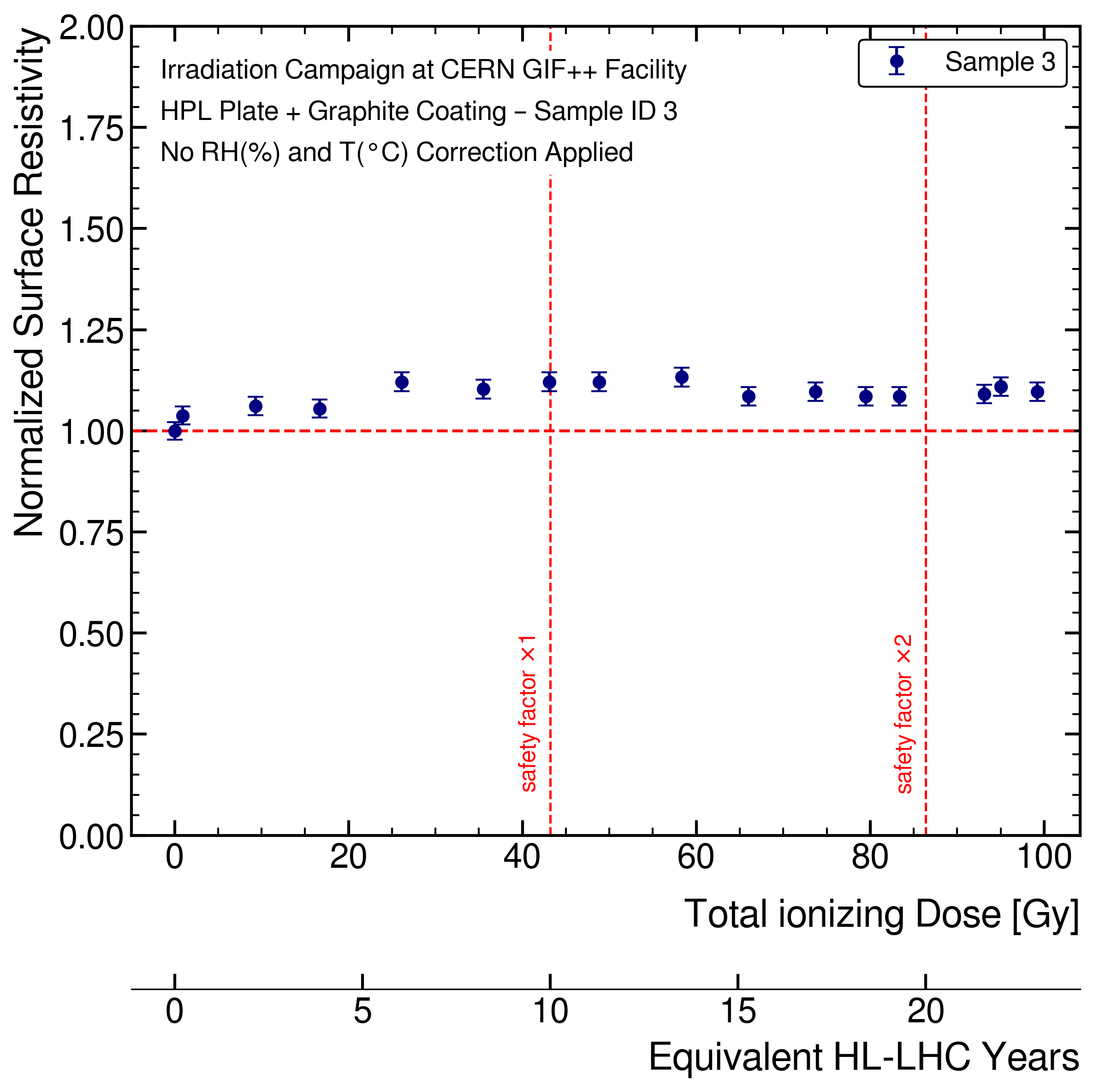}
        \caption{ }
        \label{fig:Sample3}
    \end{subfigure}
    \hfill
    \begin{subfigure}[b]{0.49\textwidth}
        \centering
        \includegraphics[width=\textwidth]{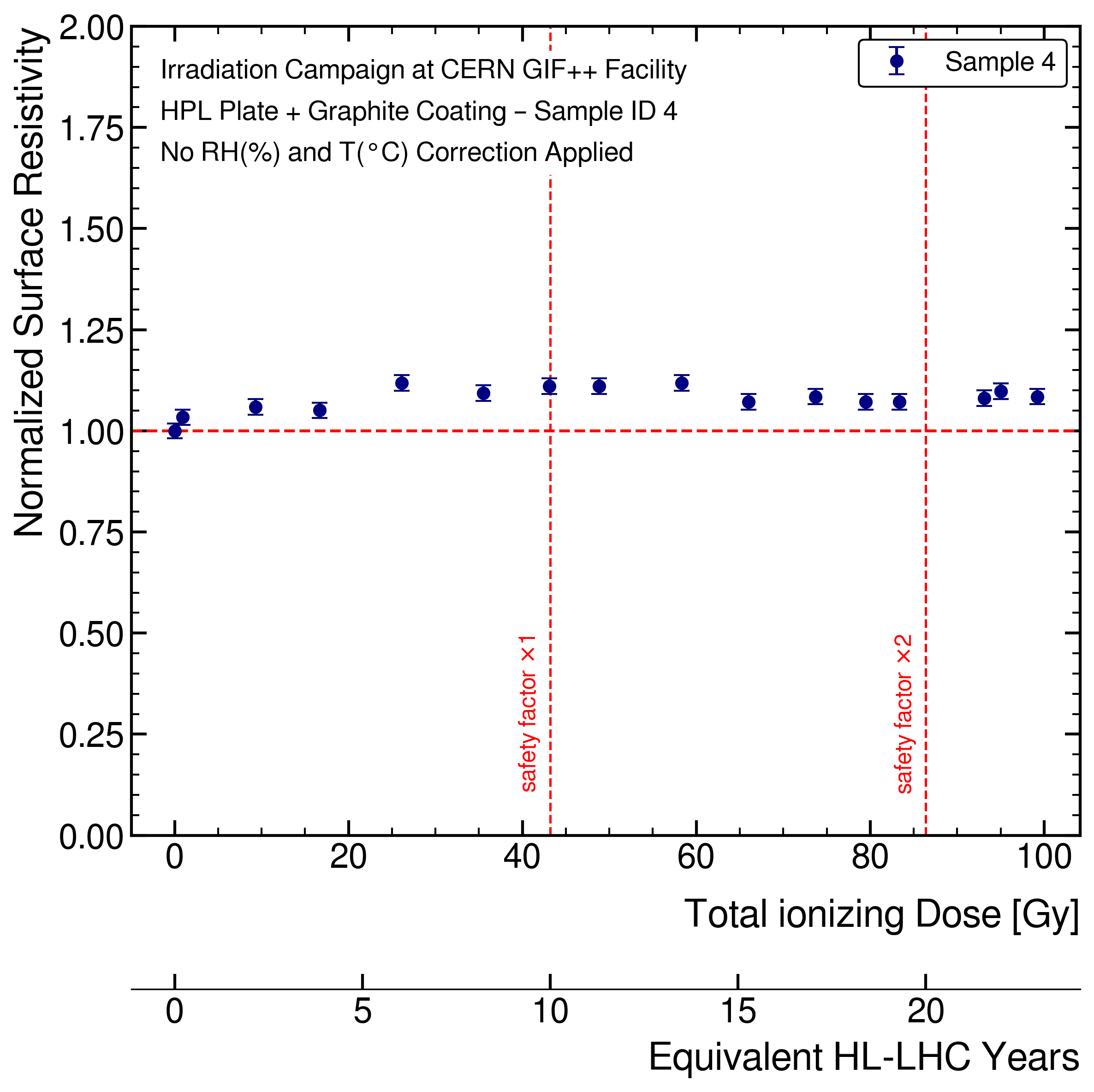}
        \caption{ }
        \label{fig:Sample4}
    \end{subfigure}
    
    \caption{Irradiation test campaign at CERN GIF++ facility: normalized surface resistivity of the carbon-based resistive ink coating on the HPL plates as a function of the accumulated TID.}
    \label{fig:GIFpp_Irradiation_Test_Results}
\end{figure}

A total of four $10 \, cm \times 10 \, cm$ HPL samples have been coated with carbon-based resistive ink using the screen-printing technique detailed in Sections \ref{sec:silk-screen_printing}. The samples have been subsequently installed at the CERN GIF++ facility, positioned approximately 1 meter from the $^{137}Cs$ source, as shown in Figure \ref{fig:GIFpp_setup}. Under these conditions, the samples have been exposed to an average total ionizing dose of approximately 8 Gy per week. The irradiation campaign has been conducted in passive mode, meaning that no high voltage has been applied to the samples during exposure. The surface resistivity of the carbon-based resistive ink coating has been periodically monitored throughout the irradiation period to assess potential degradation effects induced by ionizing radiation. The absorbed ionizing dose has been continuously monitored throughout the irradiation campaign using the CERN Radiation Monitoring (RADMON) system. In parallel, environmental parameters, including atmospheric pressure, ambient temperature, and relative humidity, have been recorded within the irradiation area to ensure proper control and interpretation of the exposure conditions. Figures \ref{fig:Sample1} - \ref{fig:Sample4} present the evolution of the surface resistivity, normalized to its pre-irradiation value, as a function of the accumulated total ionizing dose. The surface resistivity remains stable up to an integrated dose of 130 Gy, corresponding to the maximum TID expected in the RPC subsystem of the ATLAS Muon Spectrometer over ten years of operation at the HL-LHC, including a safety factor of $\times 3$. Observed variations remain below 10\% and are primarily attributed to fluctuations in relative humidity within the irradiation area, as discussed in detail in Section \ref{sec:Stress_Test_MPI}. The irradiation campaign will continue with the aim of identifying clear signs of aging in the carbon-based resistive ink coating, in order to better understand both the damage mechanisms and the operational limits of the material.

\section{Conclusions}

The qualification programme presented in this work establishes the long-term reliability of screen-printed graphite/phenoxy resistive coatings on high-pressure-laminate electrodes for Resistive Plate Chambers under HL-LHC conditions. By combining controlled industrial fabrication, extended high-voltage stress testing, and irradiation campaigns in mixed-field and gamma environments at CERN, the study demonstrates that the adopted technology fulfils the stringent requirements of next-generation high-rate muon detectors. The silk-screen process, when operated within well-defined industrial tolerances, yields reproducible coatings with the target surface resistivity for the electrode surfaces, ensuring uniform electric-field distribution across large-area gas volumes. Long-term stress tests up to integrated charges equivalent to - and exceeding by a safety factor of four - ten years of HL-LHC operation confirm that the electrical response of the coatings remains stable, with only humidity-driven reversible fluctuations and a modest irreversible upward drift consistent with microscopic percolation and aging mechanisms. Complementary irradiation tests at the CERN CHARM and GIF++ facilities further confirm the robustness of the coatings against cumulative damage. The coatings retained their functional performance up to neutron fluences, hadron fluxes, and total ionising doses that reach or surpass the maximum levels expected over the lifetime of the ATLAS and CMS RPC subsystems at the HL-LHC. Within the explored dose ranges, no significant degradation of surface or bulk resistivity was observed beyond environmental fluctuations, underscoring the intrinsic radiation hardness of the carbon-black/phenoxy composite. Together, these findings provide a solid experimental basis for the deployment of silk-screen printed graphite coatings in the large-scale production of RPC electrodes for HL-LHC upgrades. Beyond their immediate relevance for ATLAS and CMS, the results establish a transferable quality-assurance framework for resistive coatings in gaseous detectors, offering a validated pathway for future collider experiments operating at high rates and extreme radiation levels.

\appendix
\section{High-Voltage Stress Test Summary Results}
\label{sec:appendix_A}
\begin{figure}[htbp]
  \centering
  \begin{subfigure}[t]{0.49\textwidth}
    \centering
    \includegraphics[width=\textwidth]{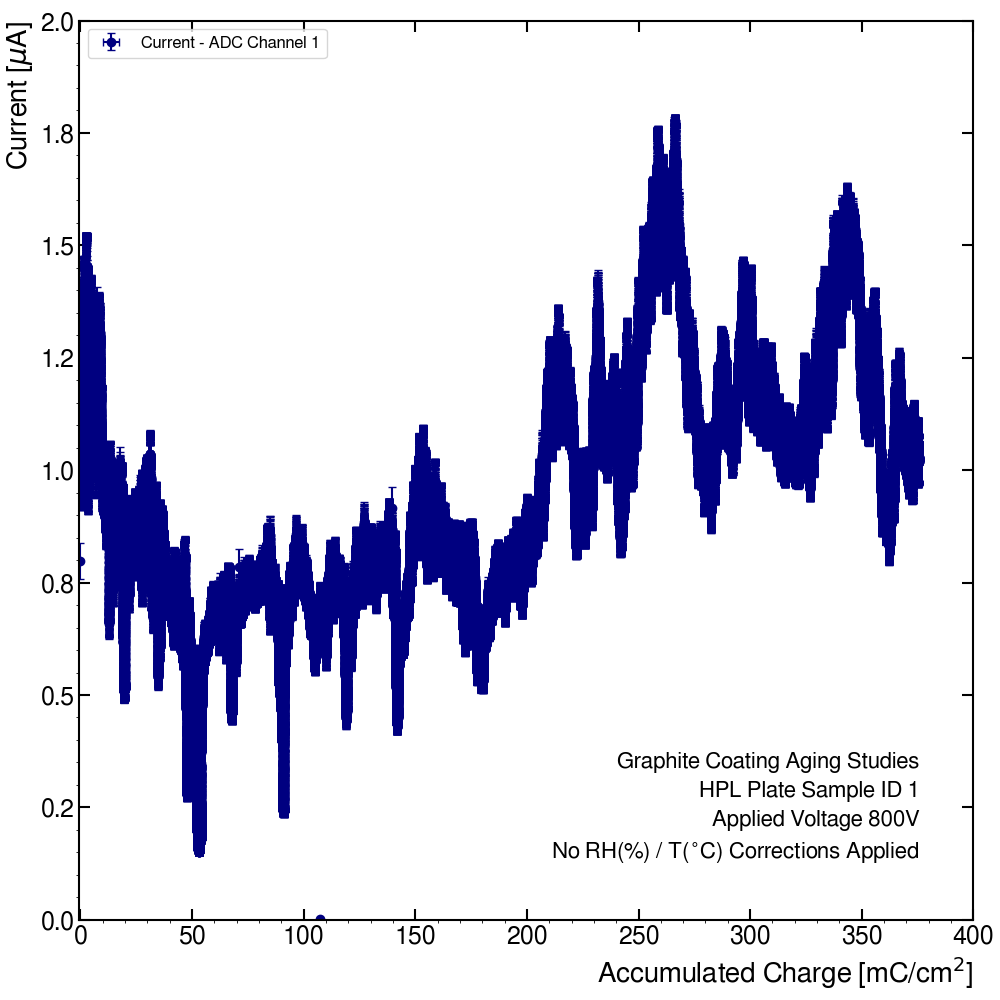}  
    \caption{}             
    \label{fig:current_800V_a}
  \end{subfigure}
  \hfill
  \begin{subfigure}[t]{0.49\textwidth}
    \centering
    \includegraphics[width=\textwidth]{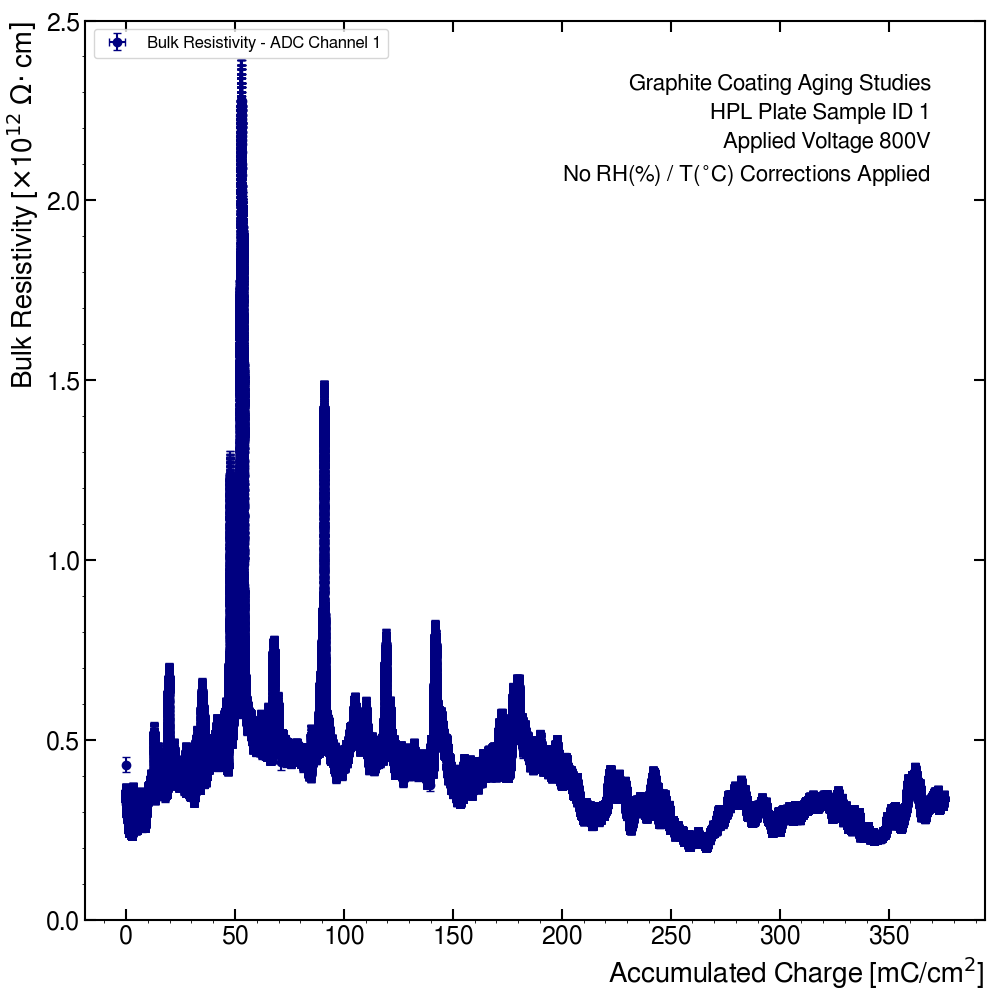}  
    \caption{}             
  \end{subfigure}
  \hfill
  \begin{subfigure}[t]{0.49\textwidth}
    \centering
    \includegraphics[width=\textwidth]{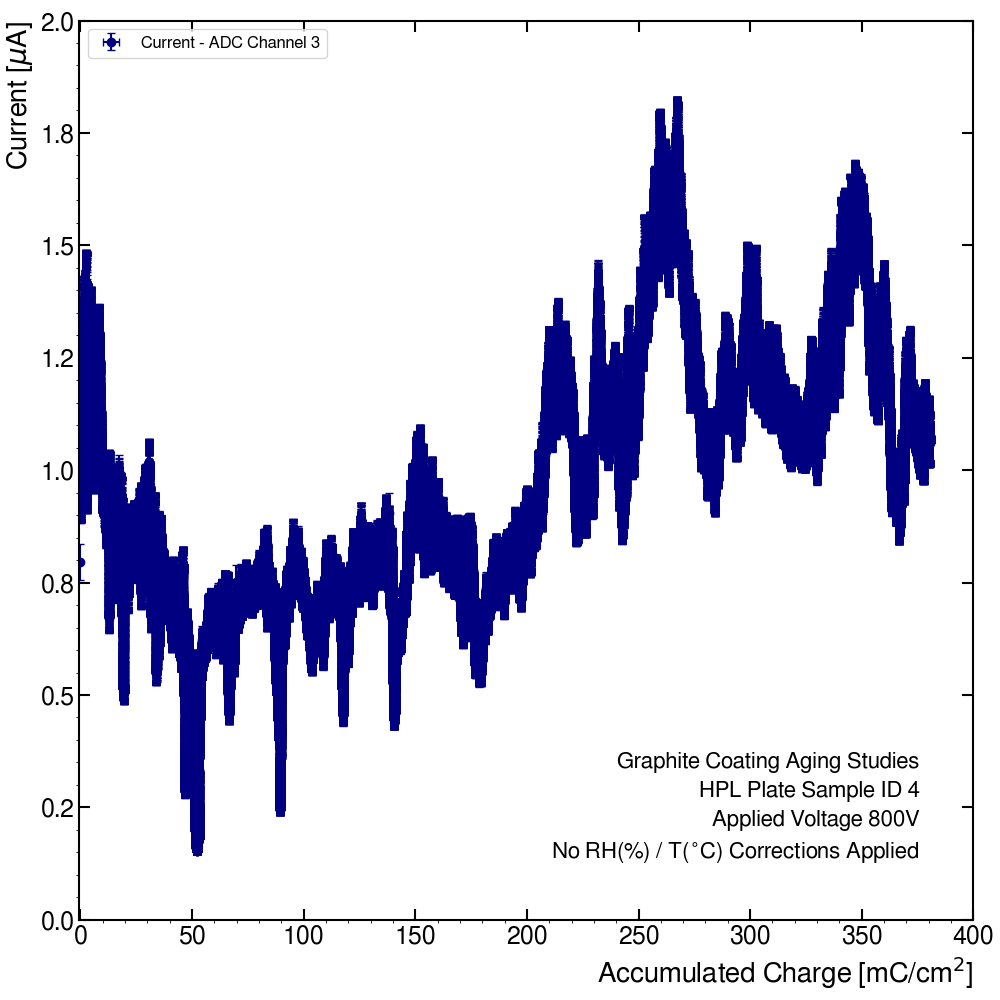}  
    \caption{}             
  \end{subfigure}
  \hfill
  \begin{subfigure}[t]{0.49\textwidth}
    \centering
    \includegraphics[width=\textwidth]{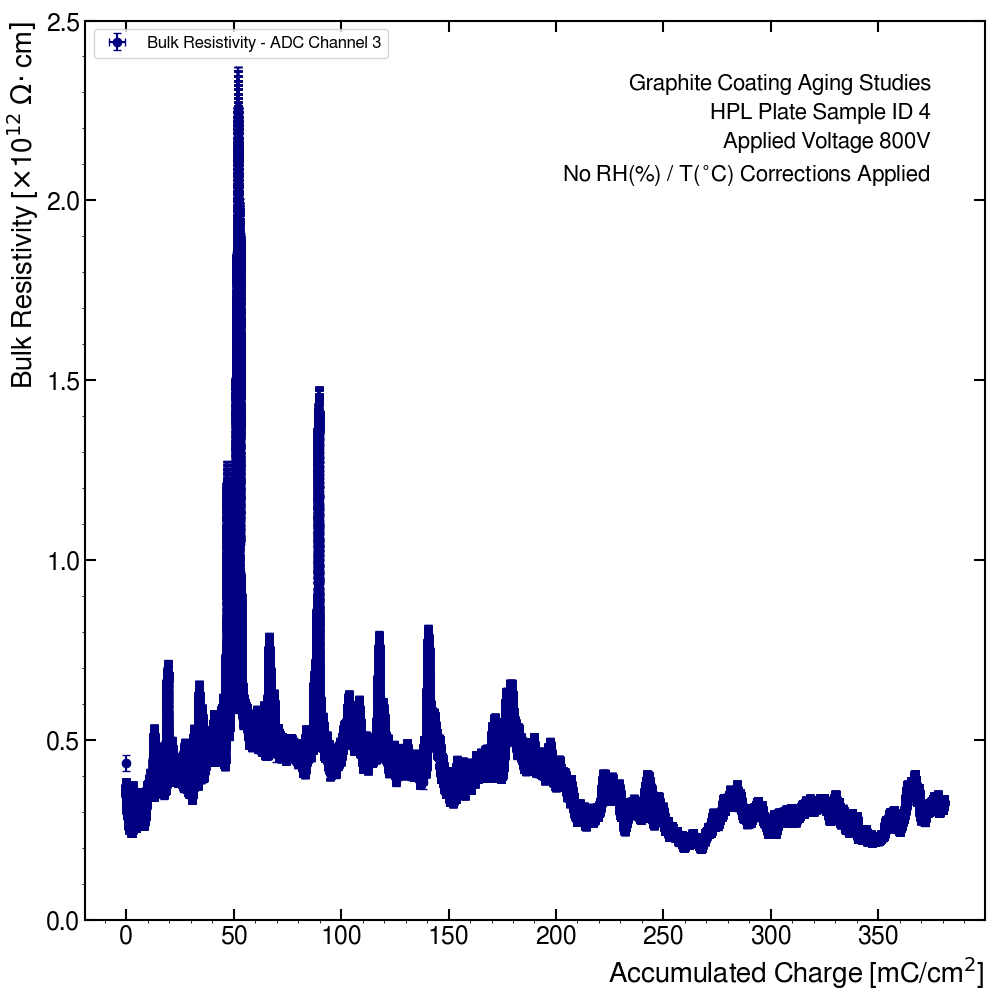}  
    \caption{}             
  \end{subfigure}
    
  \caption{Current (left) and the corresponding bulk resistivity (right) of the HPL test samples as functions of integrated charge at an operating voltage of 800 V. No corrections for temperature or relative‑humidity variations have been applied.}
  \label{fig:appendix_test_sample_800V_abcd}
\end{figure}

\begin{figure}[htbp]
  \centering
  \begin{subfigure}[t]{0.49\textwidth}
    \centering
    \includegraphics[width=\textwidth]{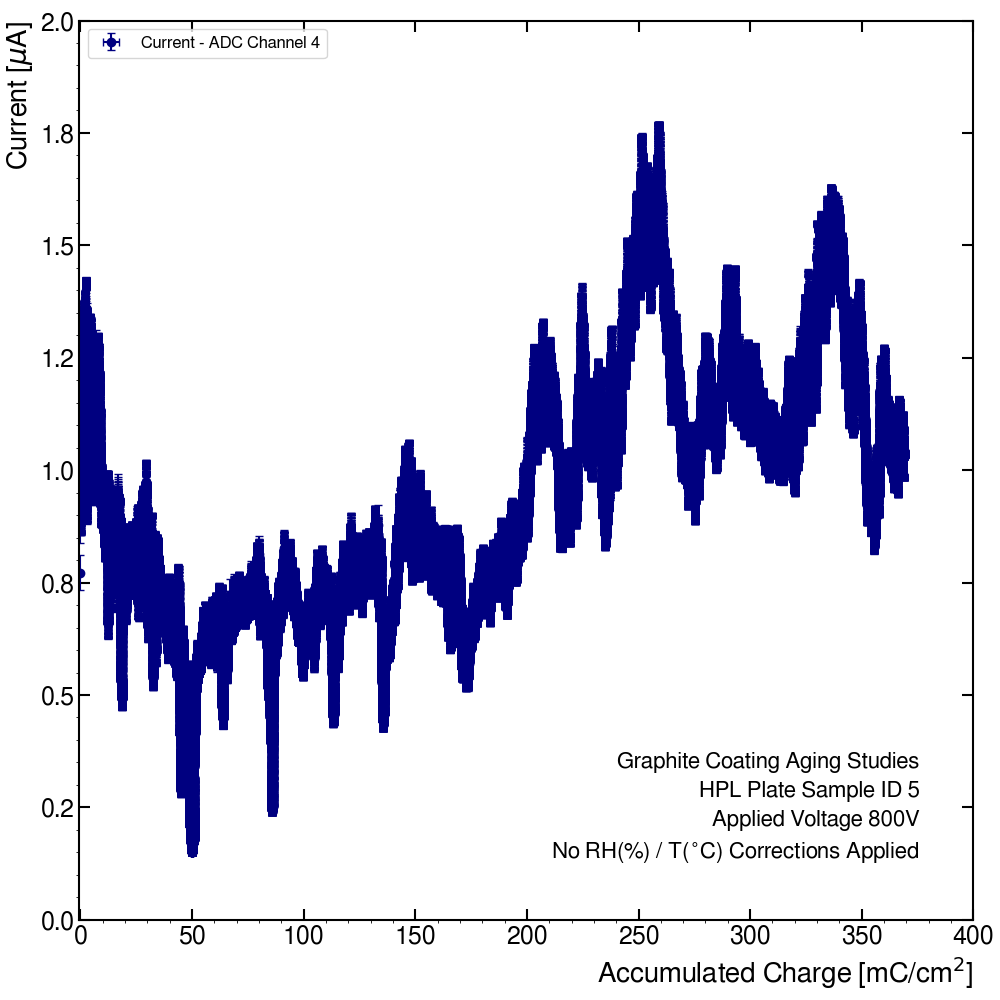}  
    \caption{}             
    \label{fig:current_800V_aa}
  \end{subfigure}
  \hfill
  \begin{subfigure}[t]{0.49\textwidth}
    \centering
    \includegraphics[width=\textwidth]{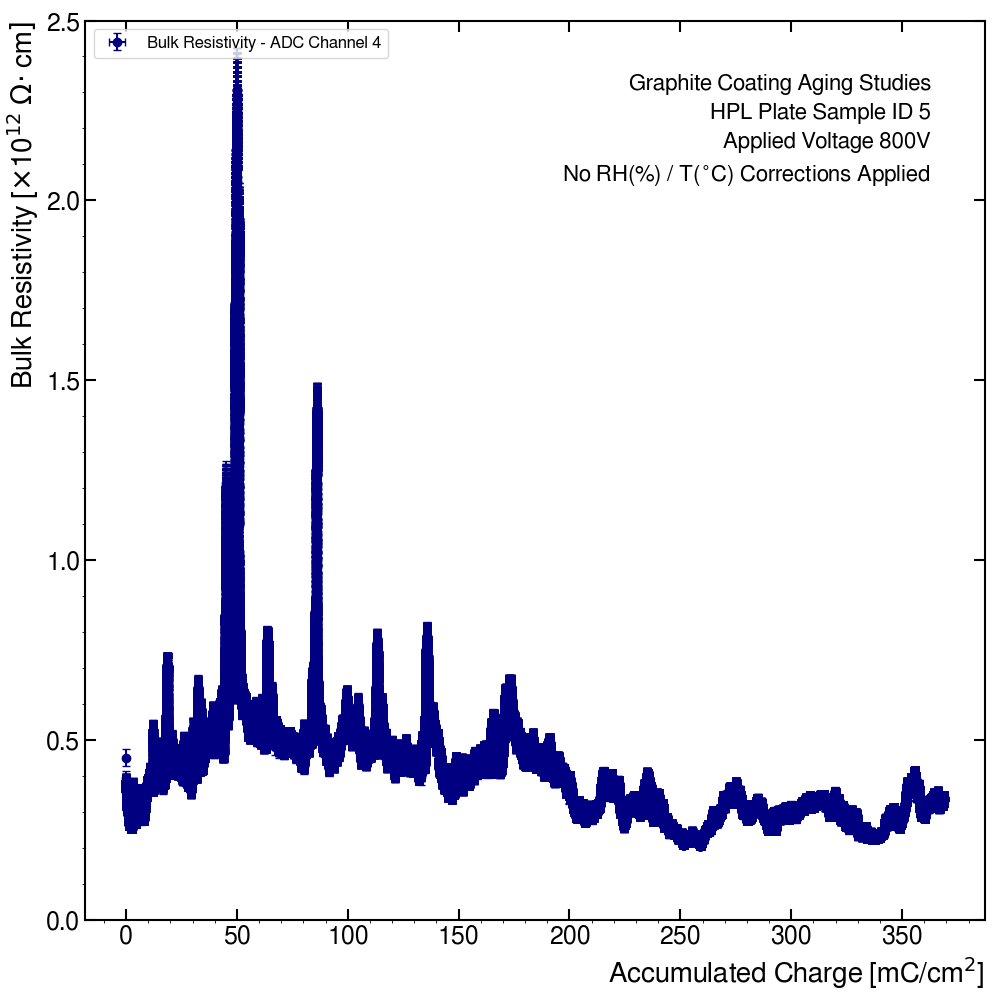}  
    \caption{}             
  \end{subfigure}
  \hfill
  \begin{subfigure}[t]{0.49\textwidth}
    \centering
    \includegraphics[width=\textwidth]{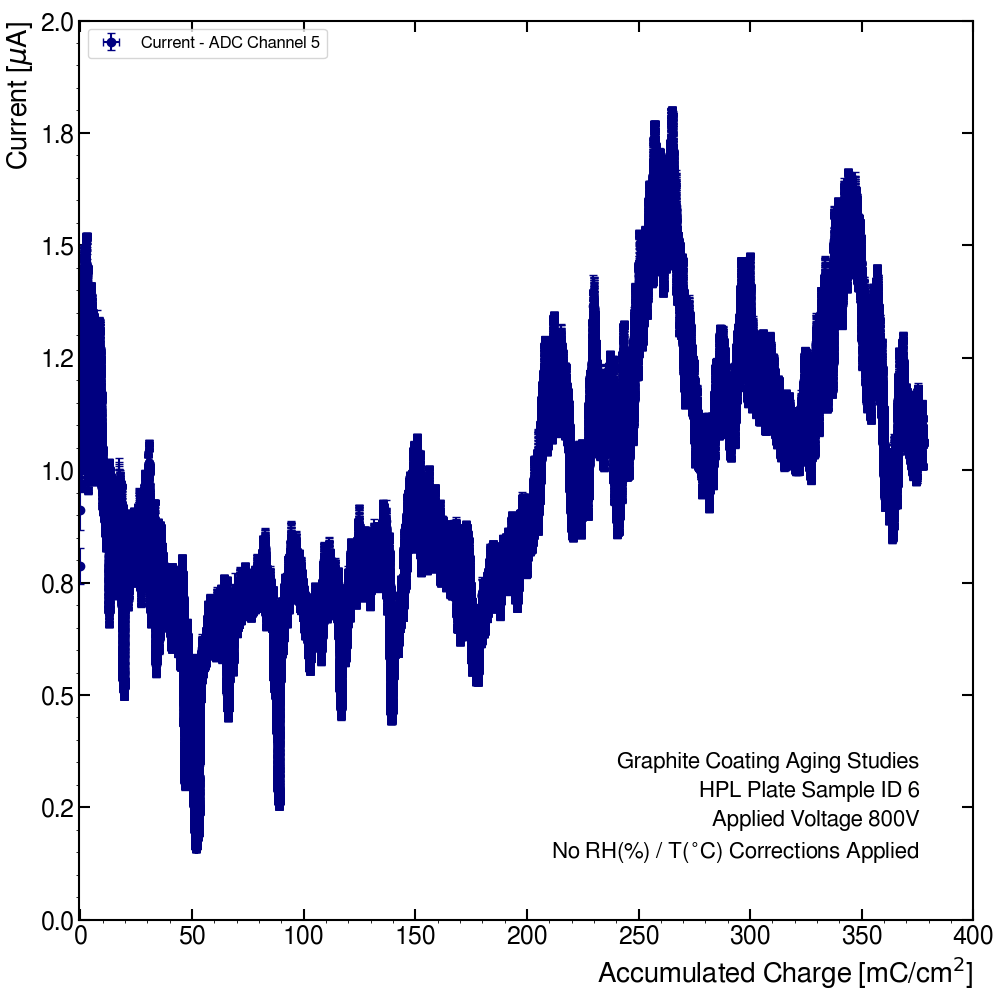}  
    \caption{}             
  \end{subfigure}
  \hfill
  \begin{subfigure}[t]{0.49\textwidth}
    \centering
    \includegraphics[width=\textwidth]{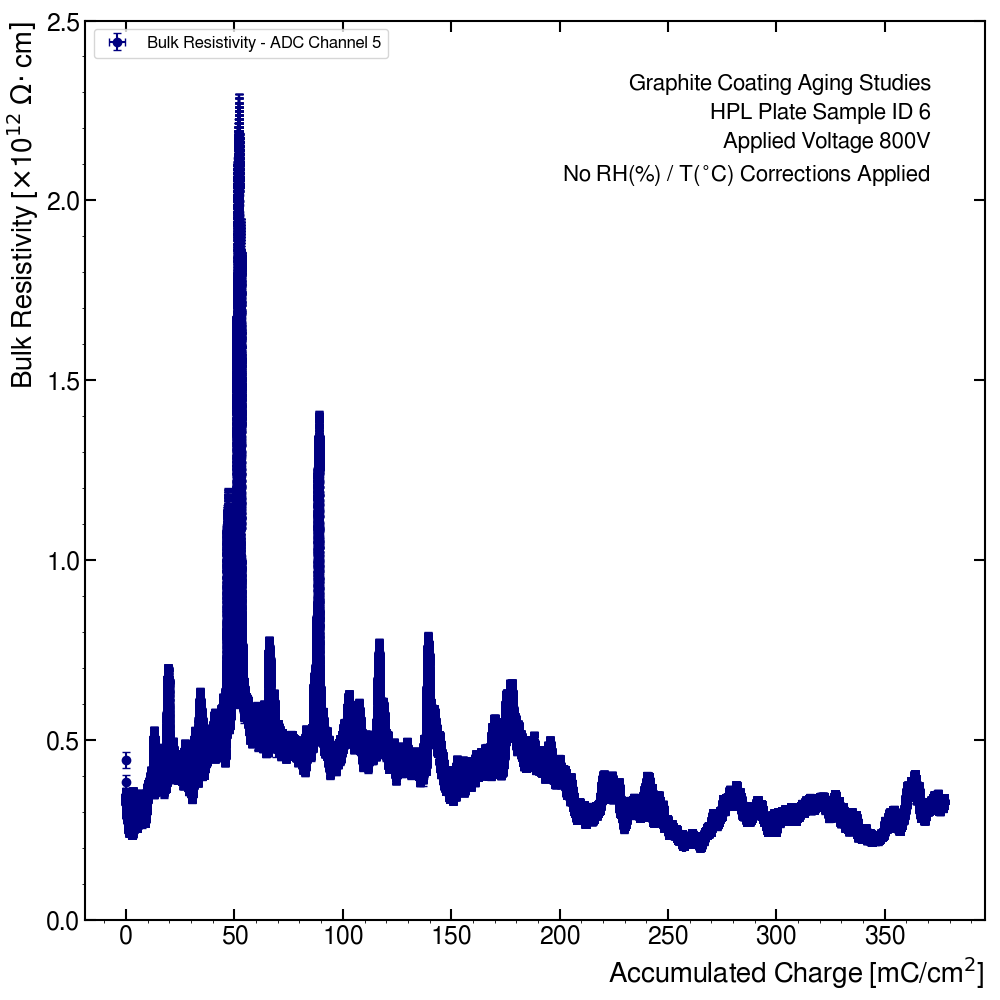}  
    \caption{}             
  \end{subfigure}
    
  \caption{Current (left) and the corresponding bulk resistivity (right) of the HPL test samples as functions of integrated charge at an operating voltage of 800 V. No corrections for temperature or relative‑humidity variations have been applied.}
  \label{fig:appendix_test_sample_800V_aabbccdd}
\end{figure}

\begin{figure}[htbp]
  \centering
  \begin{subfigure}[t]{0.49\textwidth}
    \centering
    \includegraphics[width=\textwidth]{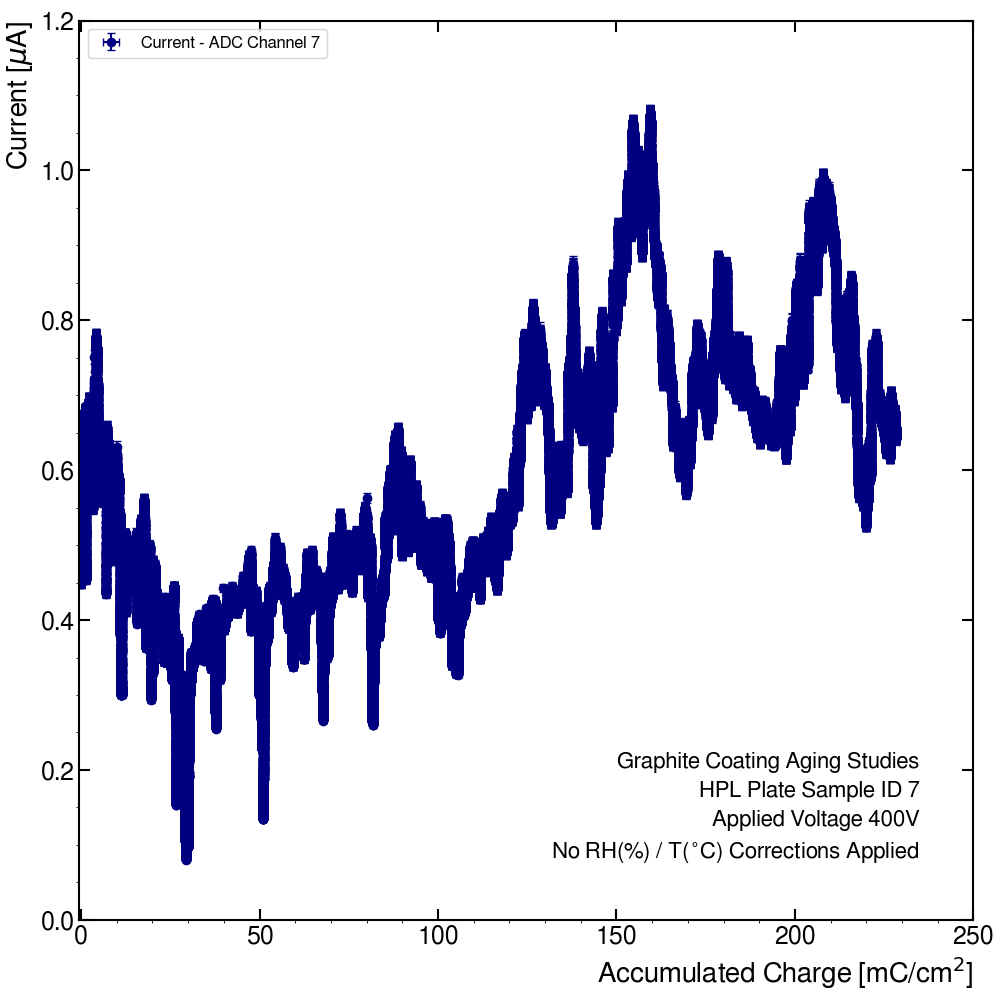}  
    \caption{}             
    \label{fig:current_800V_a}
  \end{subfigure}
  \hfill
  \begin{subfigure}[t]{0.49\textwidth}
    \centering
    \includegraphics[width=\textwidth]{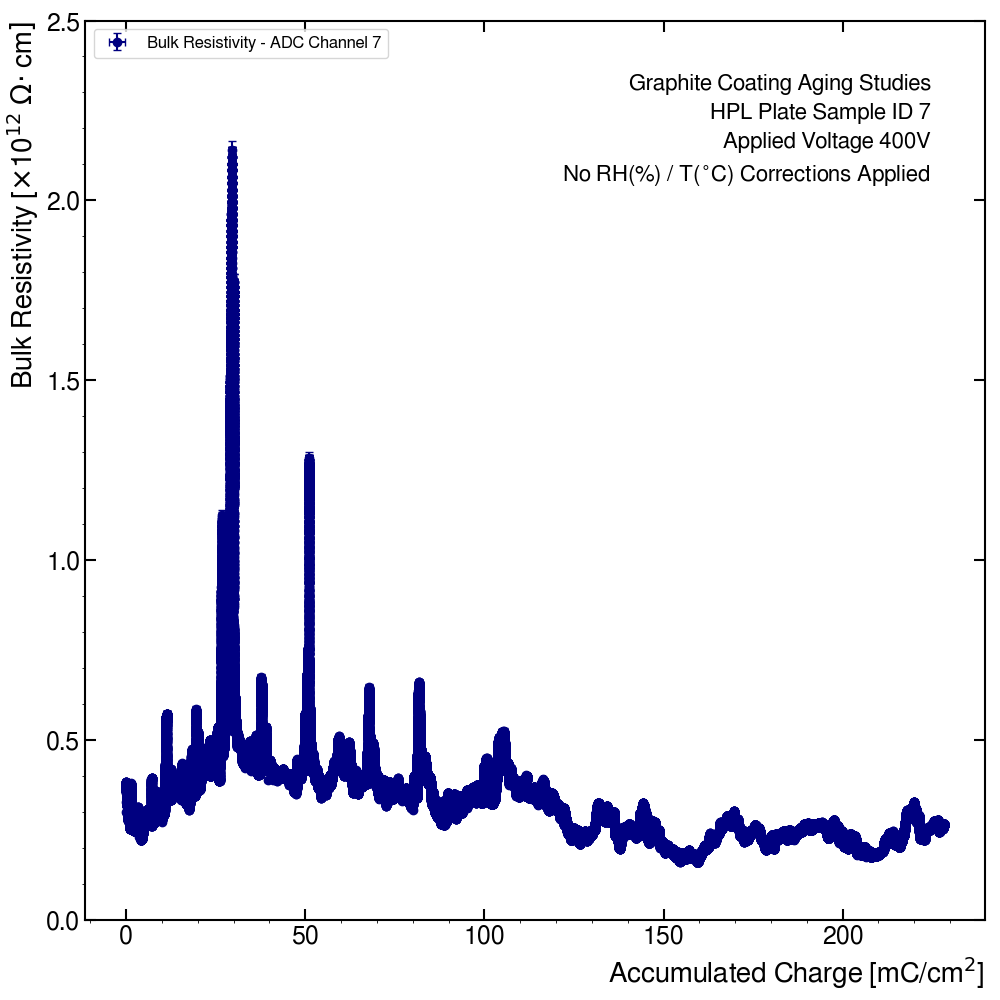}  
    \caption{}             
  \end{subfigure}
  \hfill
  \begin{subfigure}[t]{0.49\textwidth}
    \centering
    \includegraphics[width=\textwidth]{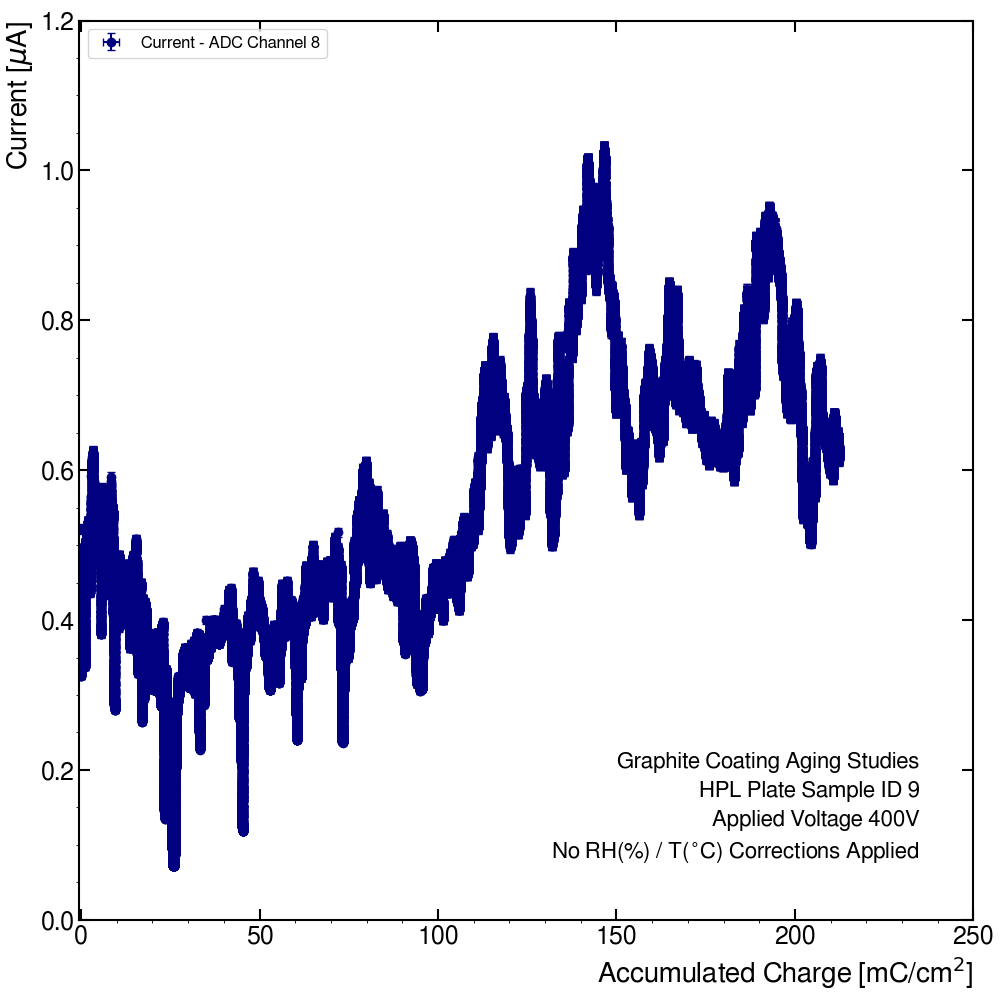}  
    \caption{}             
  \end{subfigure}
  \hfill
  \begin{subfigure}[t]{0.49\textwidth}
    \centering
    \includegraphics[width=\textwidth]{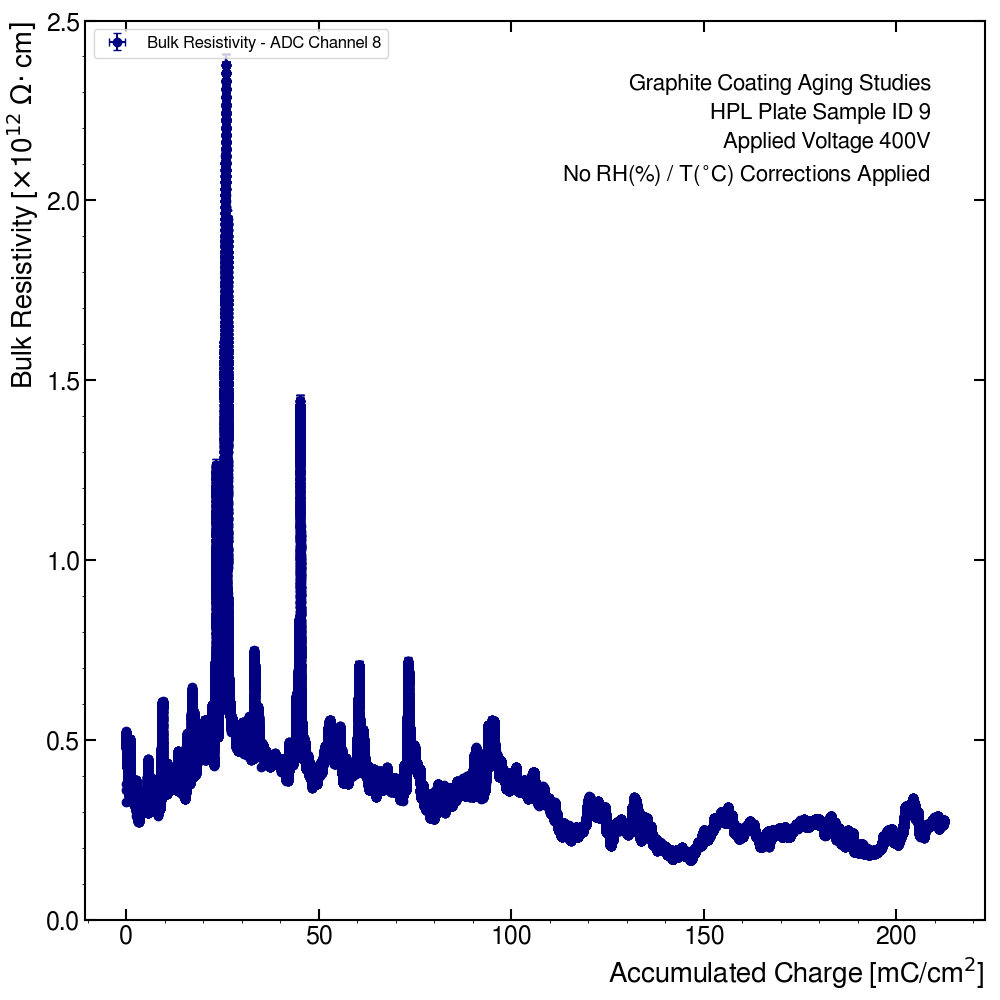}  
    \caption{}             
  \end{subfigure}
    
  \caption{Current (left) and the corresponding bulk resistivity (right) of the HPL test samples as functions of integrated charge at an operating voltage of 400 V. No corrections for temperature or relative‑humidity variations have been applied.}
  \label{fig:appendix_test_sample_400V_abcd}
\end{figure}

\begin{figure}[htbp]
  \centering
  \begin{subfigure}[t]{0.49\textwidth}
    \centering
    \includegraphics[width=\textwidth]{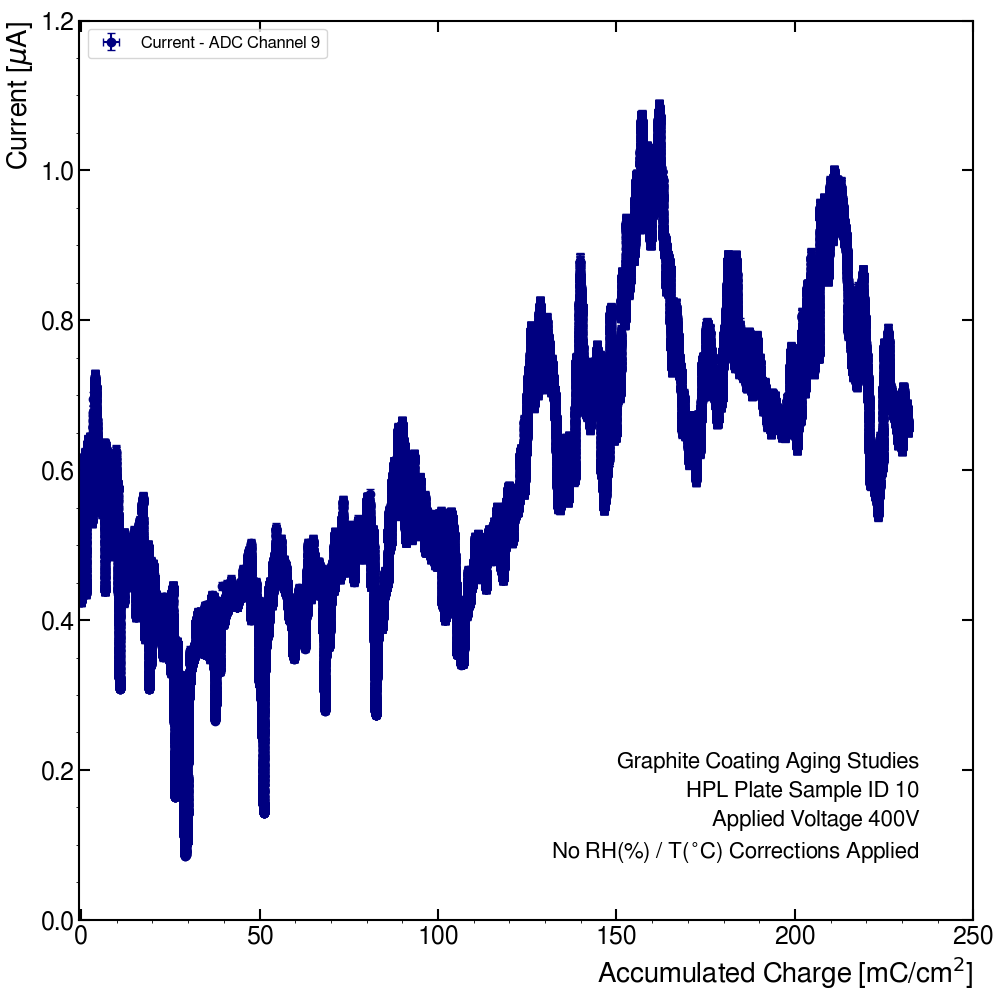}  
    \caption{}             
    \label{fig:current_400V_aa}
  \end{subfigure}
  \hfill
  \begin{subfigure}[t]{0.49\textwidth}
    \centering
    \includegraphics[width=\textwidth]{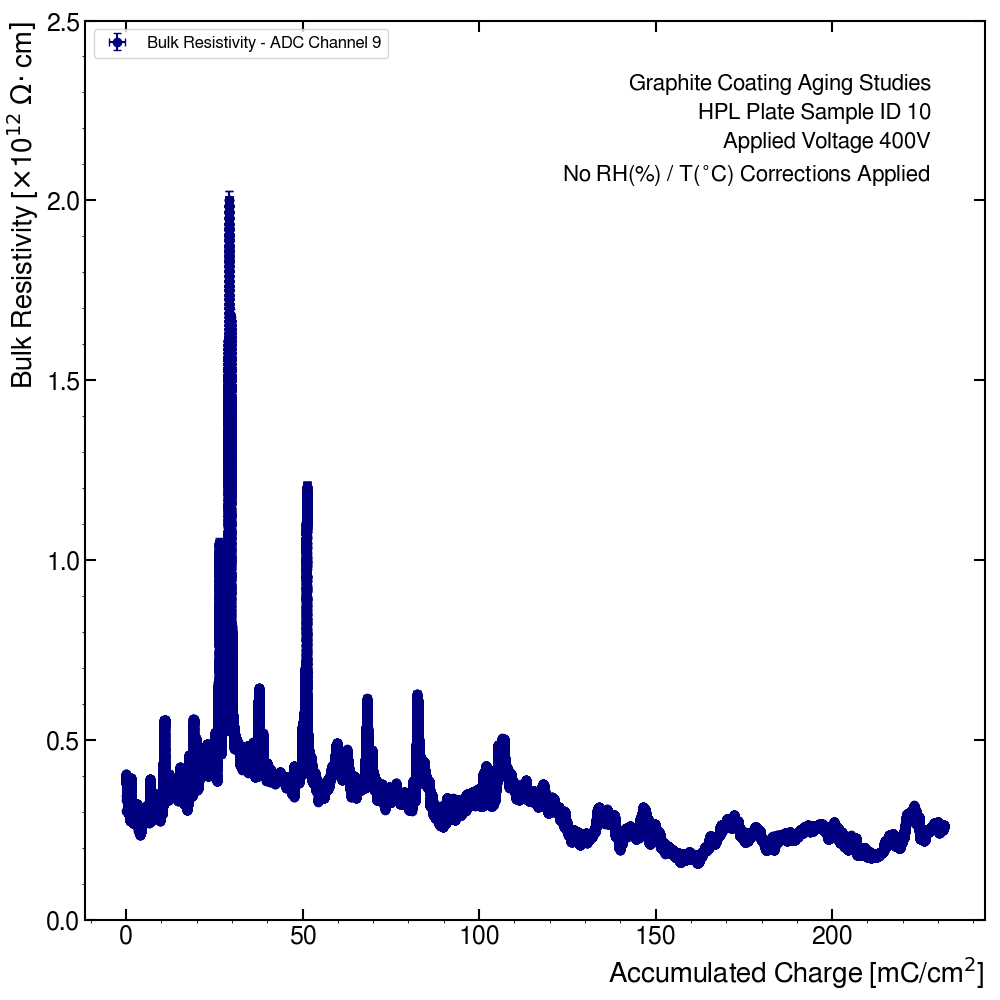}  
    \caption{}             
  \end{subfigure}
  \hfill
  \begin{subfigure}[t]{0.49\textwidth}
    \centering
    \includegraphics[width=\textwidth]{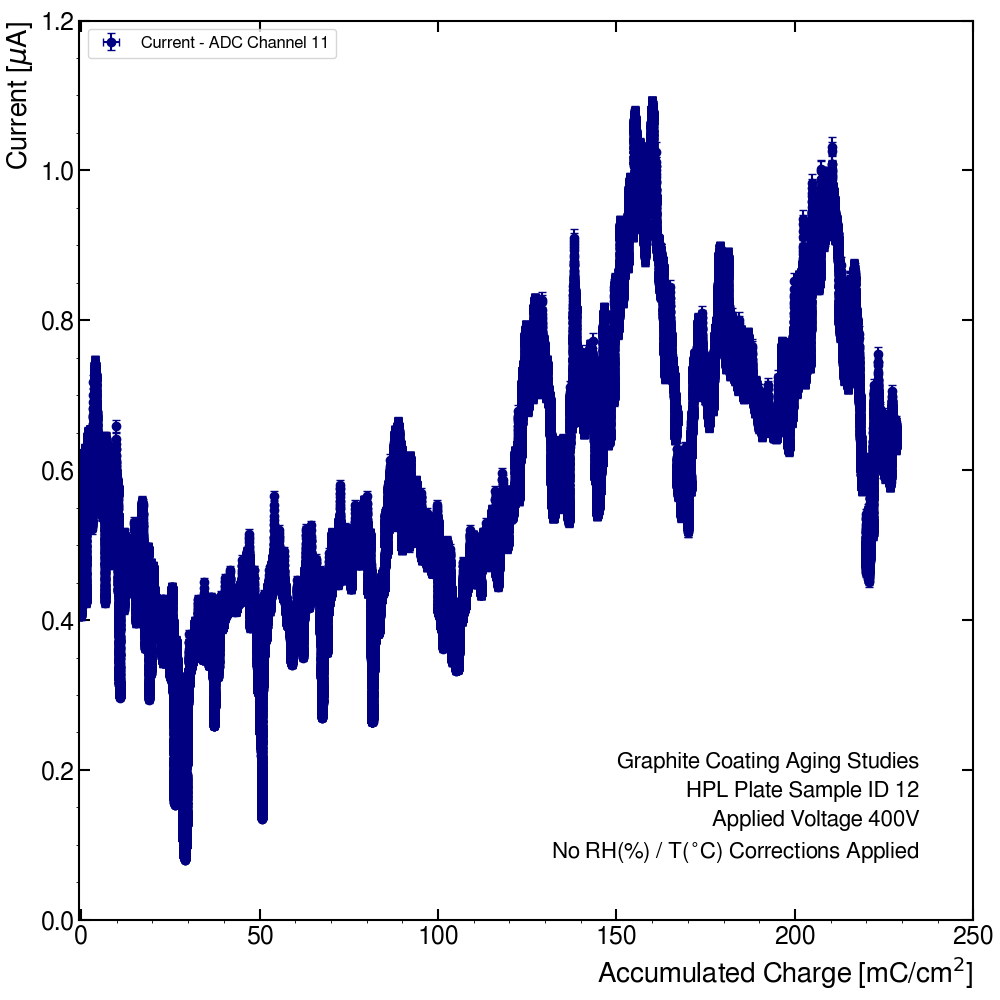}  
    \caption{}             
  \end{subfigure}
  \hfill
  \begin{subfigure}[t]{0.49\textwidth}
    \centering
    \includegraphics[width=\textwidth]{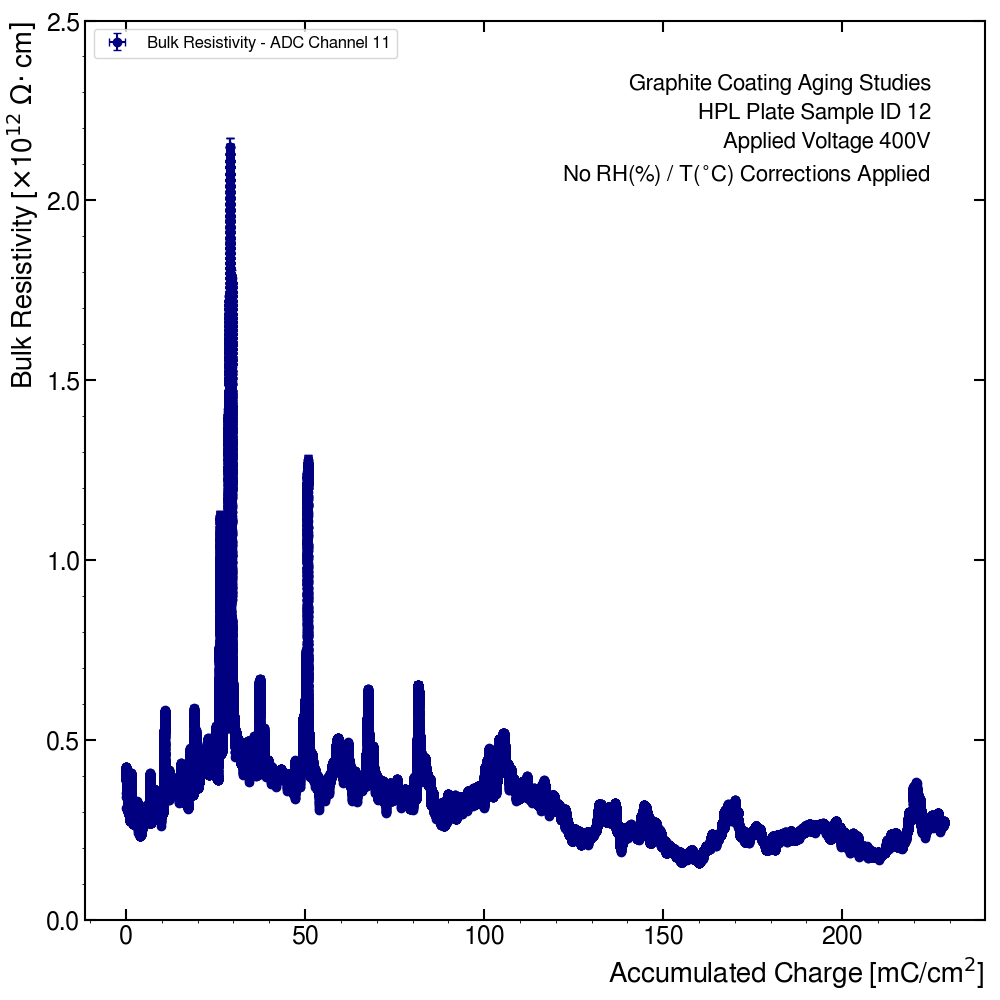}  
    \caption{}             
  \end{subfigure}
    
  \caption{Current (left) and the corresponding bulk resistivity (right) of the HPL test samples as functions of integrated charge at an operating voltage of 400 V. No corrections for temperature or relative‑humidity variations have been applied.}
  \label{fig:appendix_test_sample_400V_aabbccdd}
\end{figure}

Figure \ref{fig:appendix_test_sample_800V_abcd} - \ref{fig:appendix_test_sample_400V_aabbccdd} present a comparison of the current and the corresponding bulk resistivity of the HPL test samples as a function of the integrated charge for two different bias voltages: 800 V [Figure \ref{fig:appendix_test_sample_800V_abcd} - \ref{fig:appendix_test_sample_800V_aabbccdd}] and 400 V [Figure \ref{fig:appendix_test_sample_400V_abcd} - \ref{fig:appendix_test_sample_400V_aabbccdd}].

\section{Irradiation Test Campaign Summary Results}
\label{sec:appendix_B}
\begin{figure}[h!]
    \centering
    \begin{subfigure}[b]{0.49\textwidth}
        \centering
        \includegraphics[width=\textwidth]{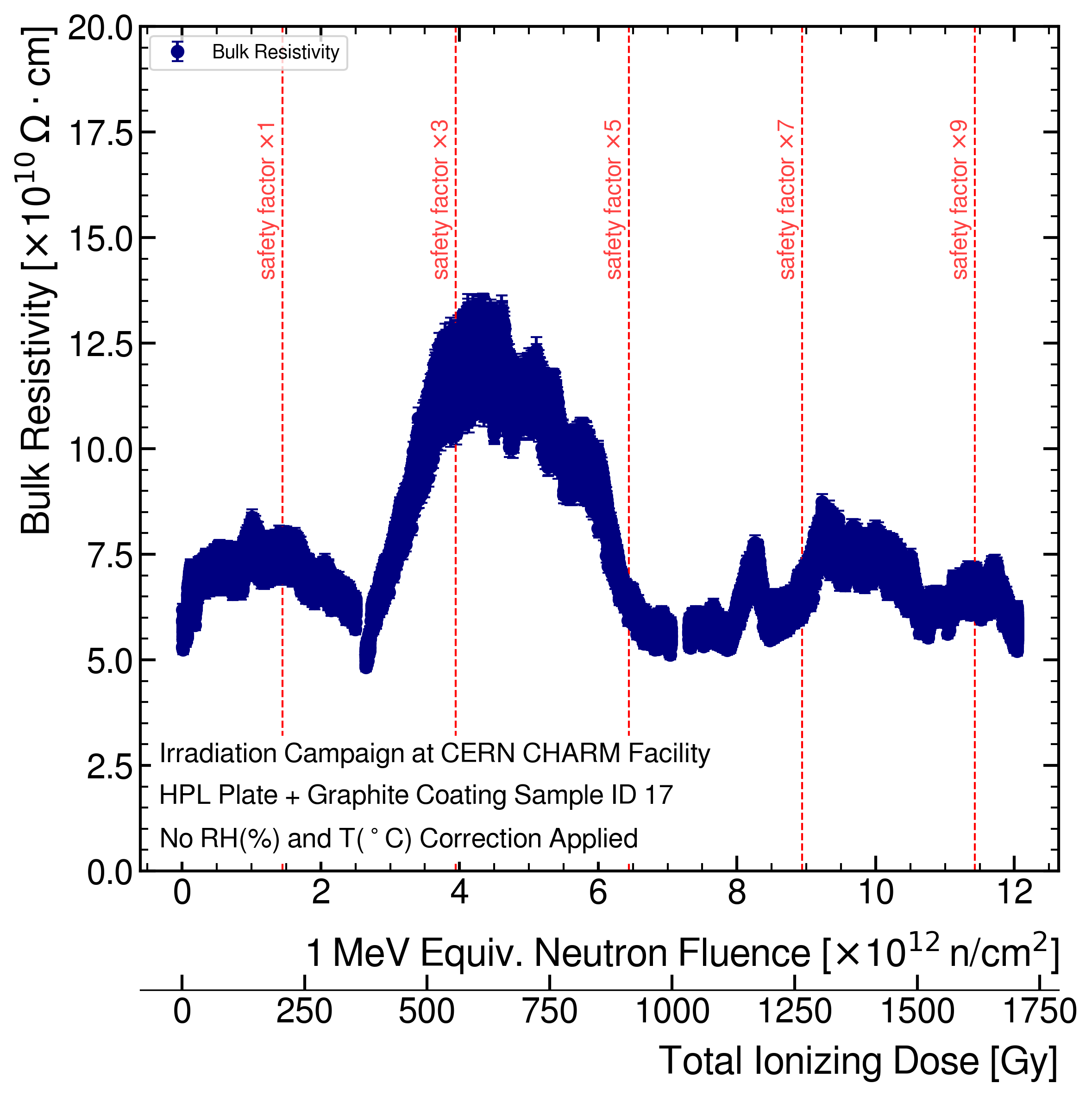}
        \caption{ }
        \label{fig:Bulk_Resistivity_Results_Sample17}
    \end{subfigure}
    \hfill
    \begin{subfigure}[b]{0.49\textwidth}
        \centering
        \includegraphics[width=\textwidth]{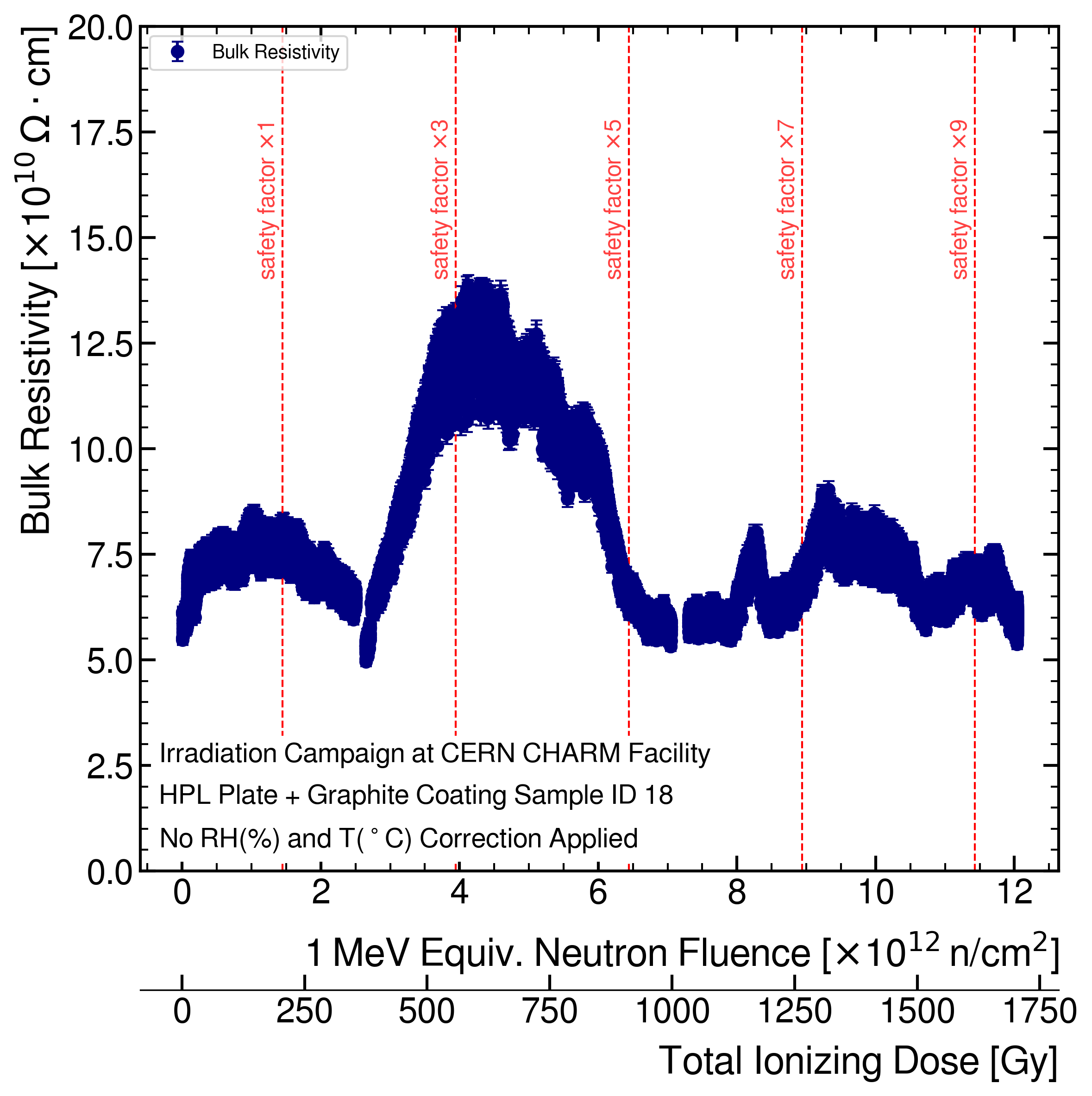}
        \caption{ }
        \label{fig:Bulk_Resistivity_Results_Sample18}
    \end{subfigure}

    \vspace{0.5cm} 

    \begin{subfigure}[b]{0.49\textwidth}
        \centering
        \includegraphics[width=\textwidth]{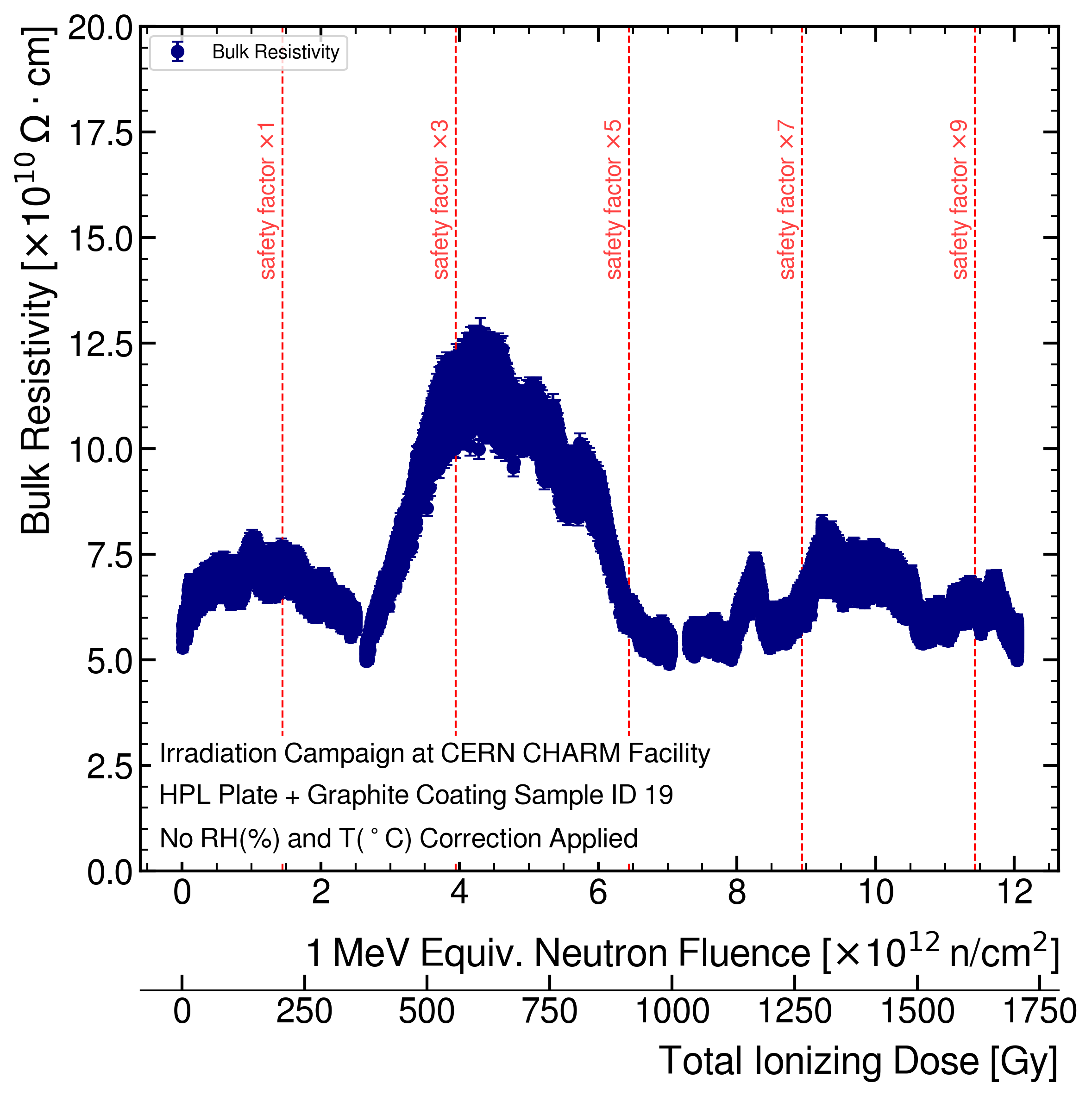}
        \caption{ }
        \label{fig:Bulk_Resistivity_Results_Sample19}
    \end{subfigure}
    \hfill
    \begin{subfigure}[b]{0.49\textwidth}
        \centering
        \includegraphics[width=\textwidth]{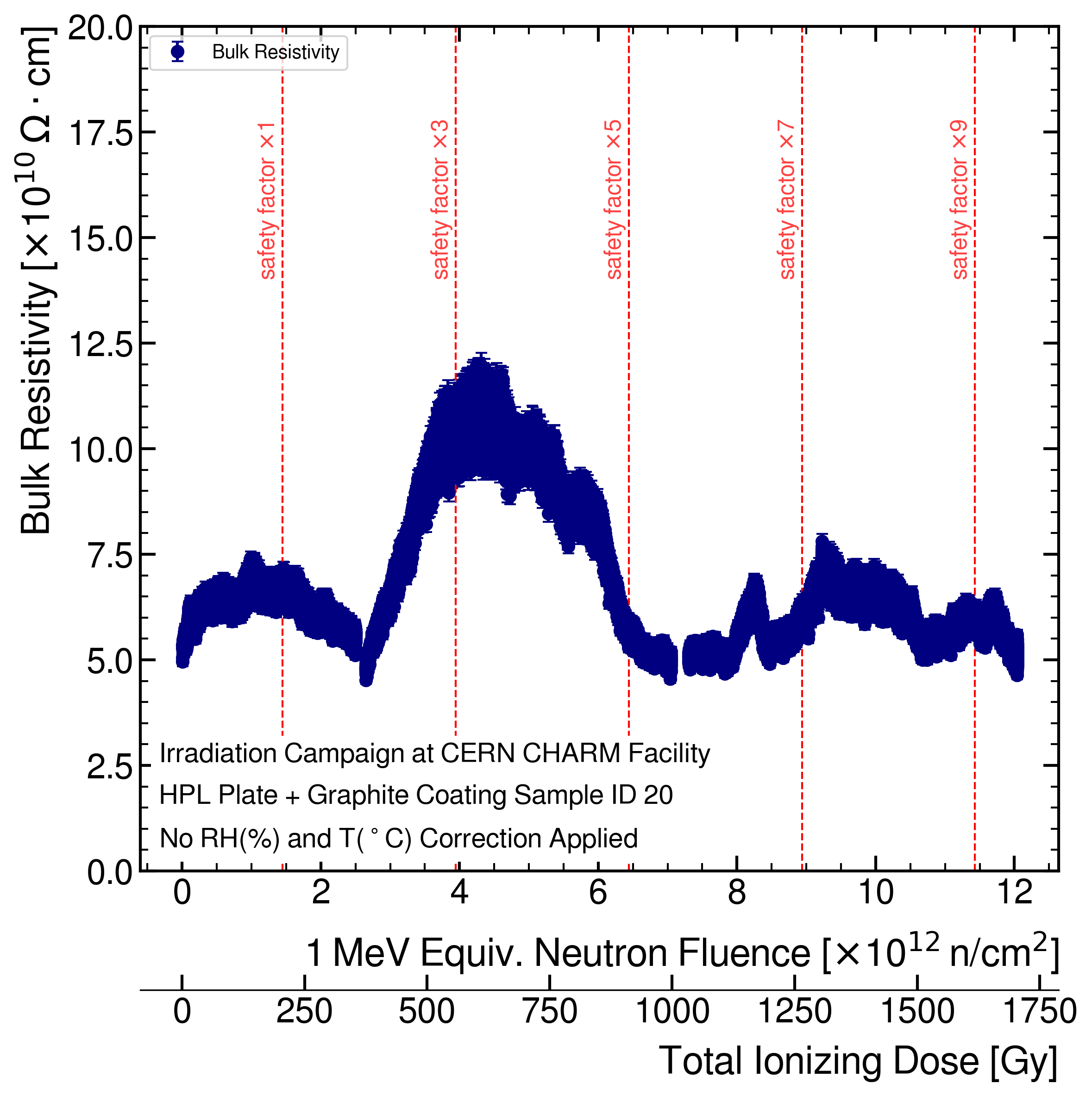}
        \caption{ }
        \label{fig:Bulk_Resistivity_Results_Sample20}
    \end{subfigure}
    
    \caption{ Bulk resistivity of the high-pressure-laminate (HPL) electrode sample as a function of the accumulated 1 MeV Si-eq. neutron fluence and TID.}
    \label{fig:CERN_CHARM_Bulk_Resistivity_Results_Summary}
\end{figure}

\begin{figure}[h!]
    \centering
    \begin{subfigure}[b]{0.49\textwidth}
        \centering
        \includegraphics[width=\textwidth]{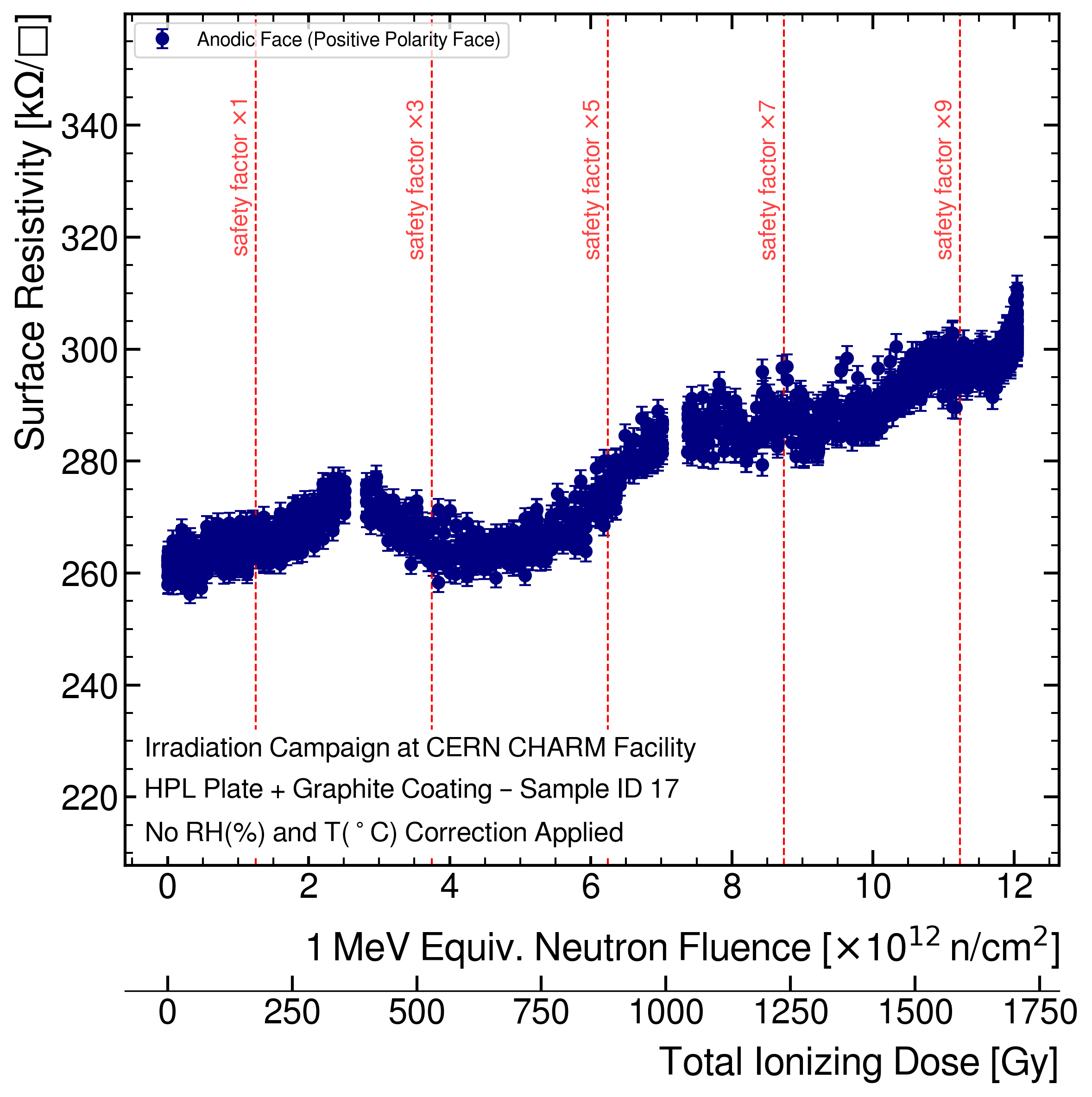}
        \caption{ }
        \label{fig:Surface_Resistivity_Results_Sample_17_F1}
    \end{subfigure}
    \hfill
    \begin{subfigure}[b]{0.49\textwidth}
        \centering
        \includegraphics[width=\textwidth]{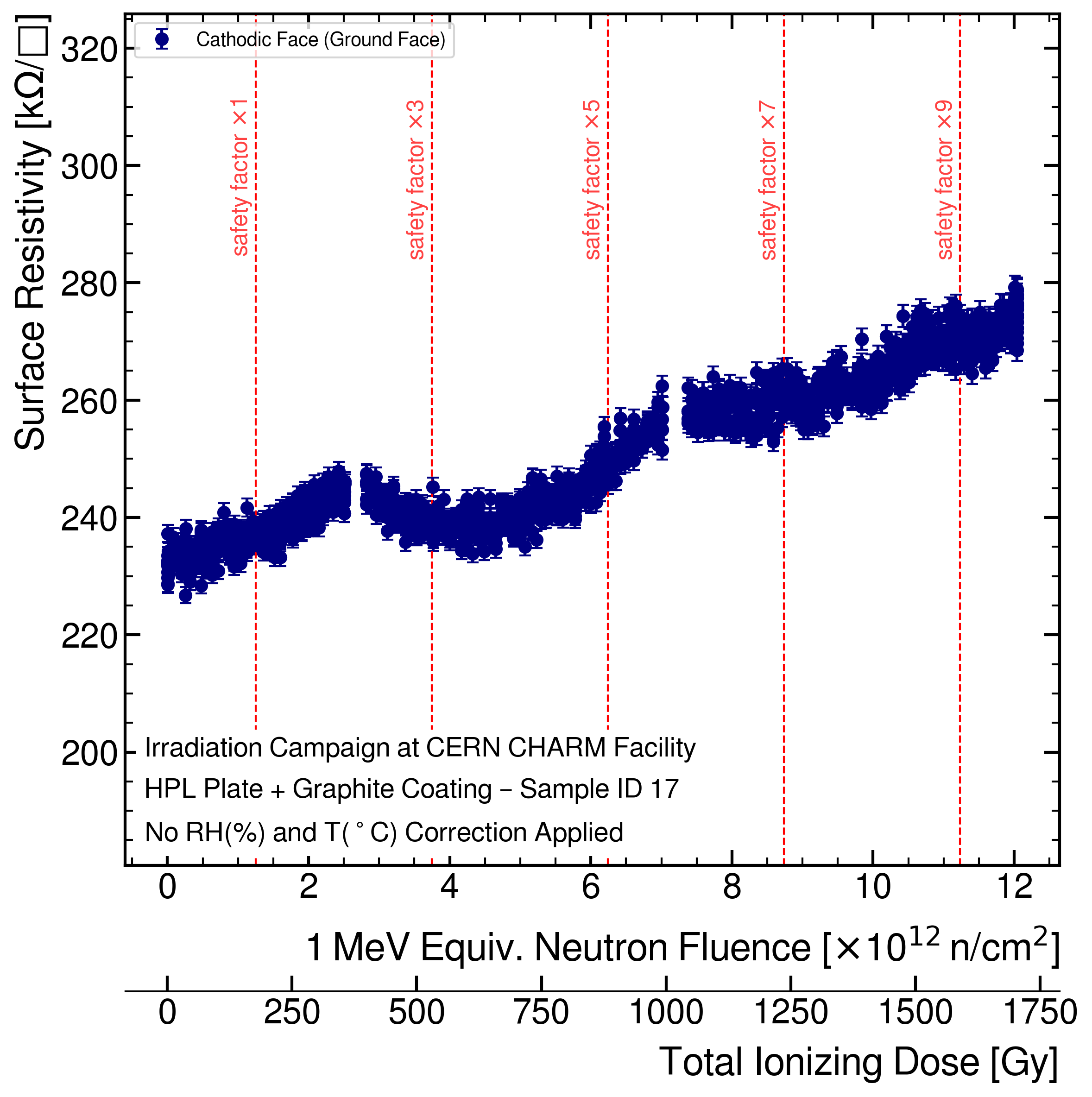}
        \caption{ }
        \label{fig:Surface_Resistivity_Results_Sample_17_F2}
    \end{subfigure}

    \vspace{0.5cm} 

    \begin{subfigure}[b]{0.49\textwidth}
        \centering
        \includegraphics[width=\textwidth]{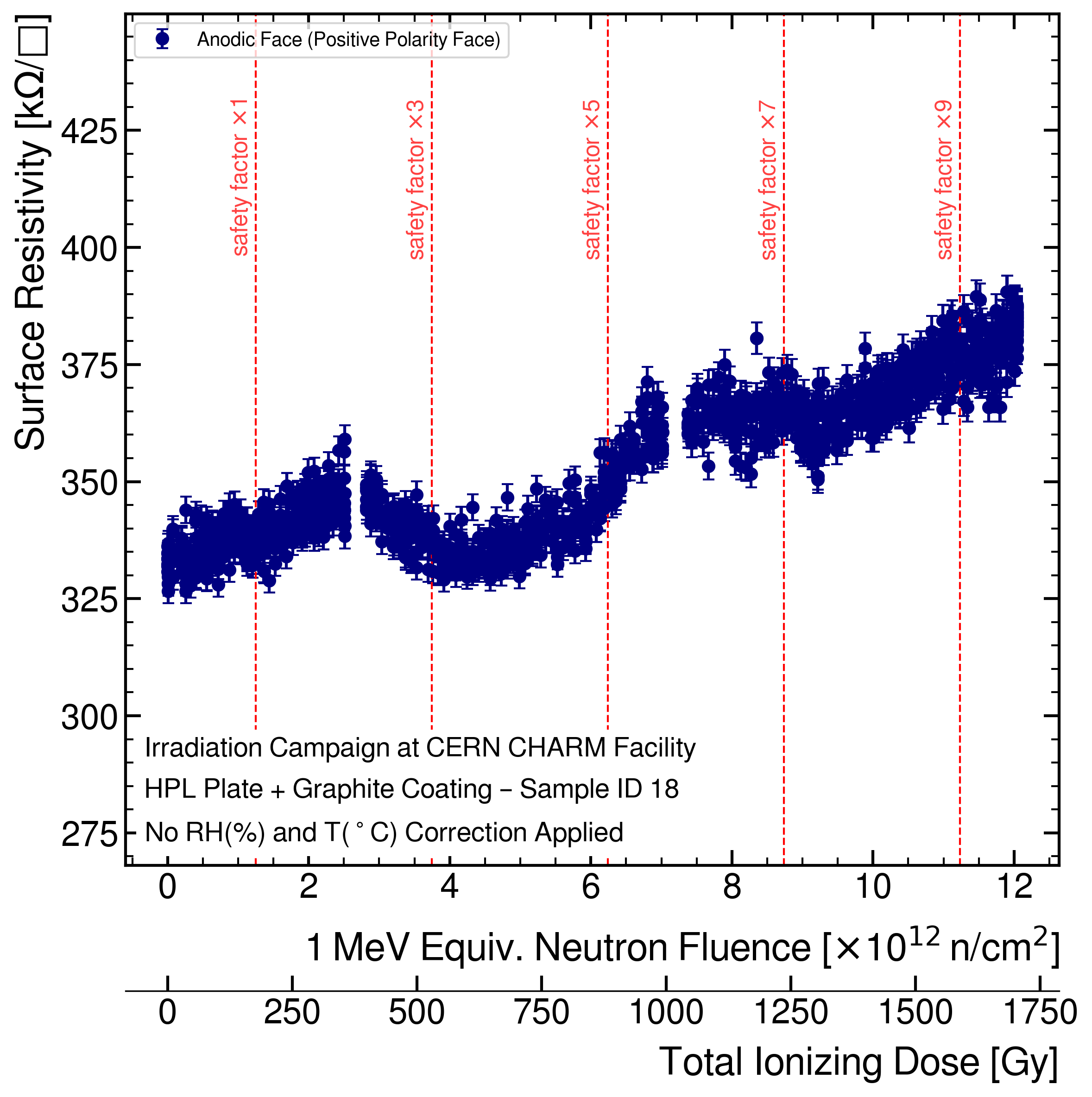}
        \caption{ }
        \label{fig:Surface_Resistivity_Results_Sample_18_F1}
    \end{subfigure}
    \hfill
    \begin{subfigure}[b]{0.49\textwidth}
        \centering
        \includegraphics[width=\textwidth]{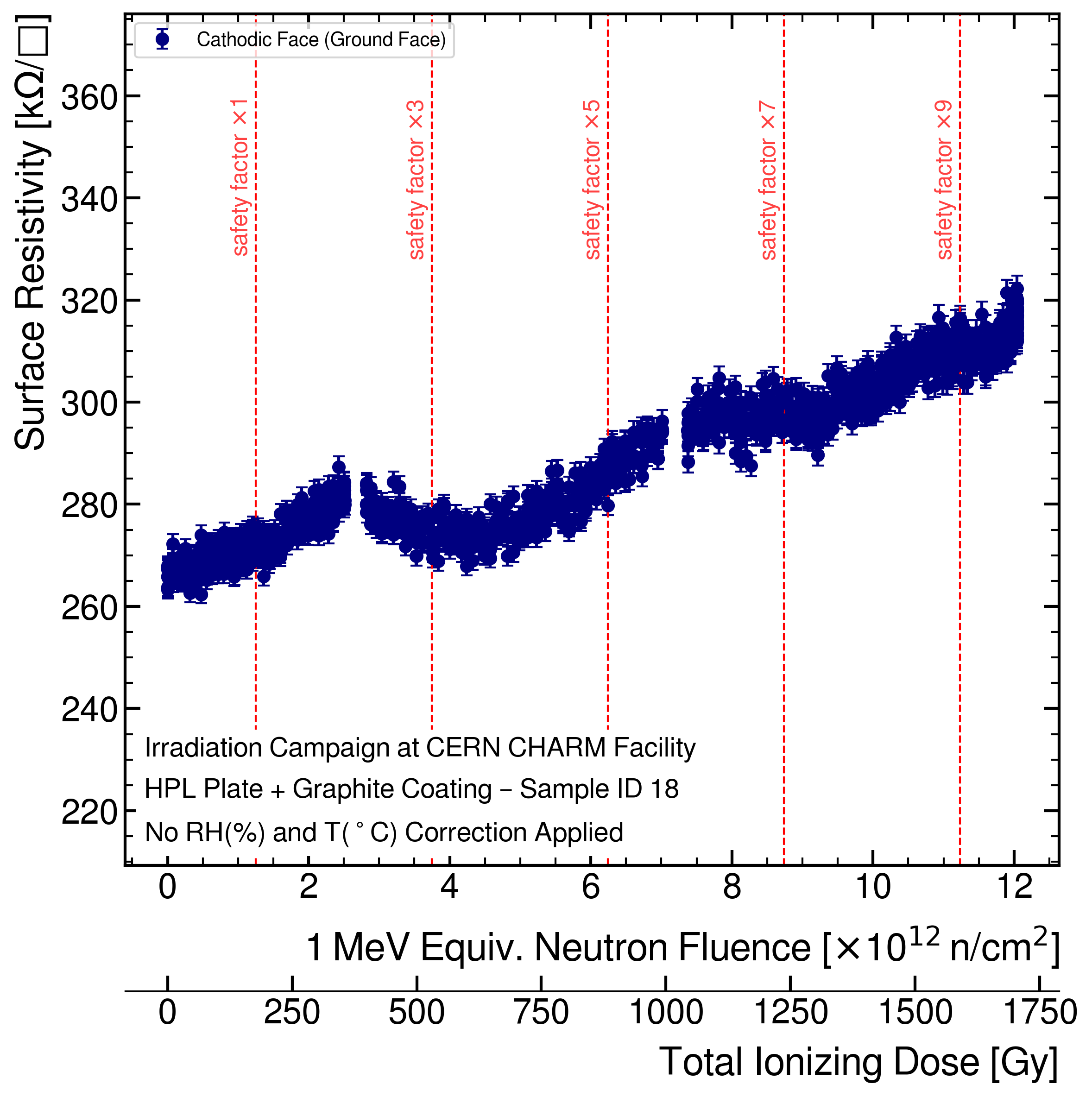}
        \caption{ }
        \label{fig:Surface_Resistivity_Results_Sample_18_F2}
    \end{subfigure}
    
    \caption{Surface resistivity of the carbon-based resistive-ink coating on the high-pressure-laminate (HPL) electrode sample as a function of accumulated 1 MeV Si-equivalent neutron fluence and total ionizing dose (TID): (a, c) anode face; (b, d) cathode face.}
    \label{fig:CERN_CHARM_Surface_Resistivity_Results_Summary_Sample_17_18}
\end{figure}

\begin{figure}[h!]
    \centering
    \begin{subfigure}[b]{0.49\textwidth}
        \centering
        \includegraphics[width=\textwidth]{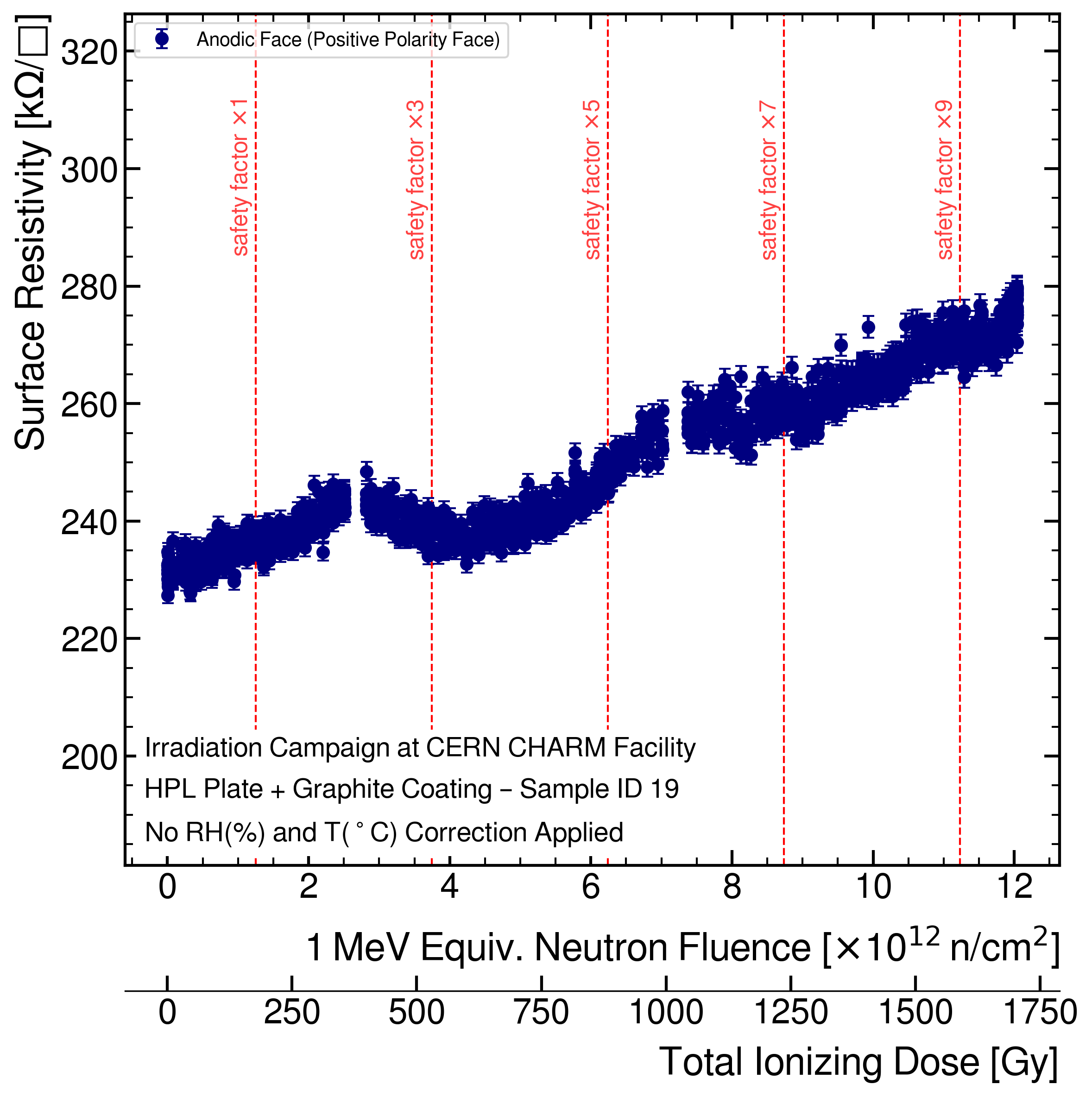}
        \caption{ }
        \label{fig:Surface_Resistivity_Results_Sample_19_F1}
    \end{subfigure}
    \hfill
    \begin{subfigure}[b]{0.49\textwidth}
        \centering
        \includegraphics[width=\textwidth]{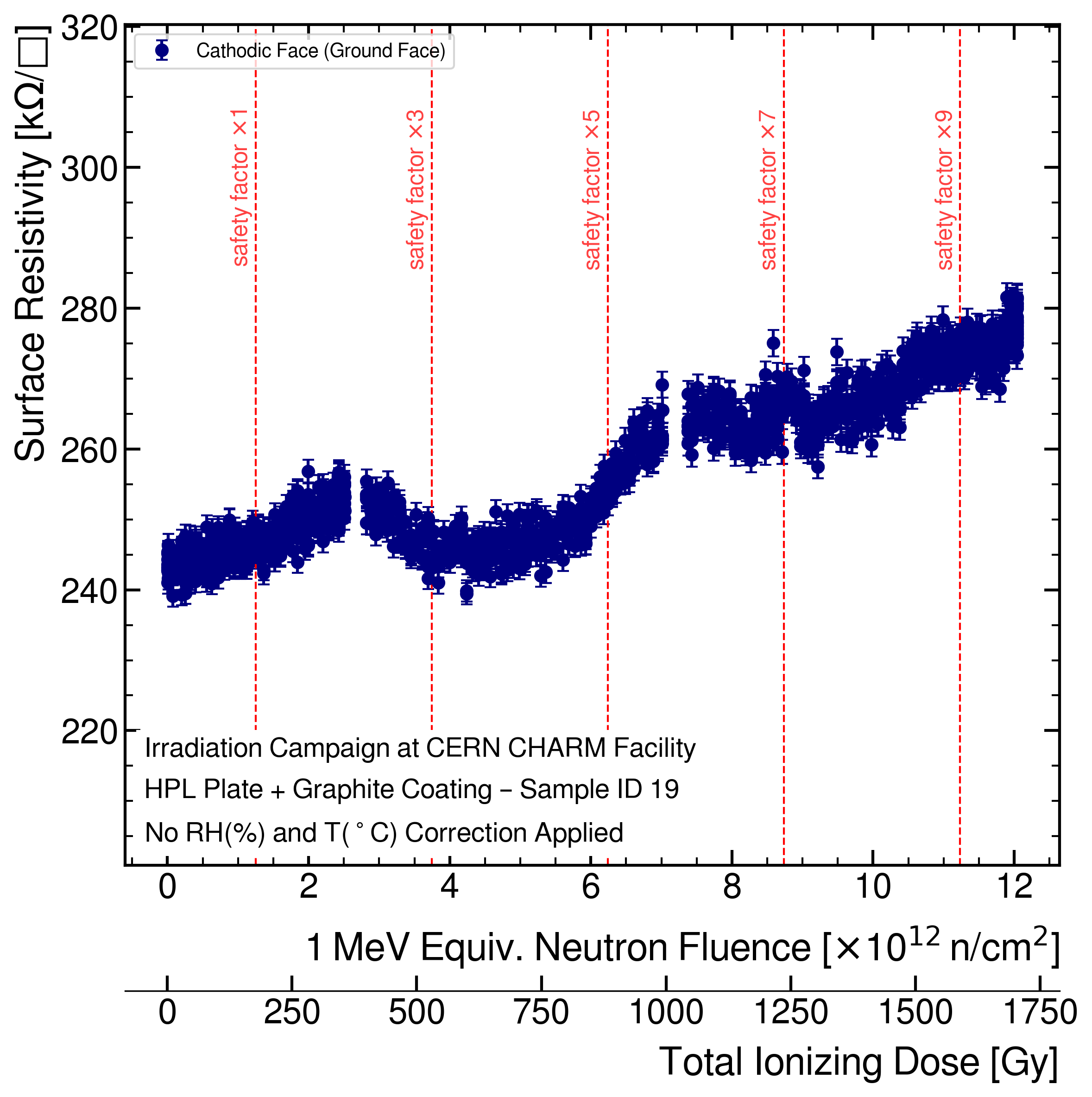}
        \caption{ }
        \label{fig:Surface_Resistivity_Results_Sample_19_F2}
    \end{subfigure}

    \vspace{0.5cm} 

    \begin{subfigure}[b]{0.49\textwidth}
        \centering
        \includegraphics[width=\textwidth]{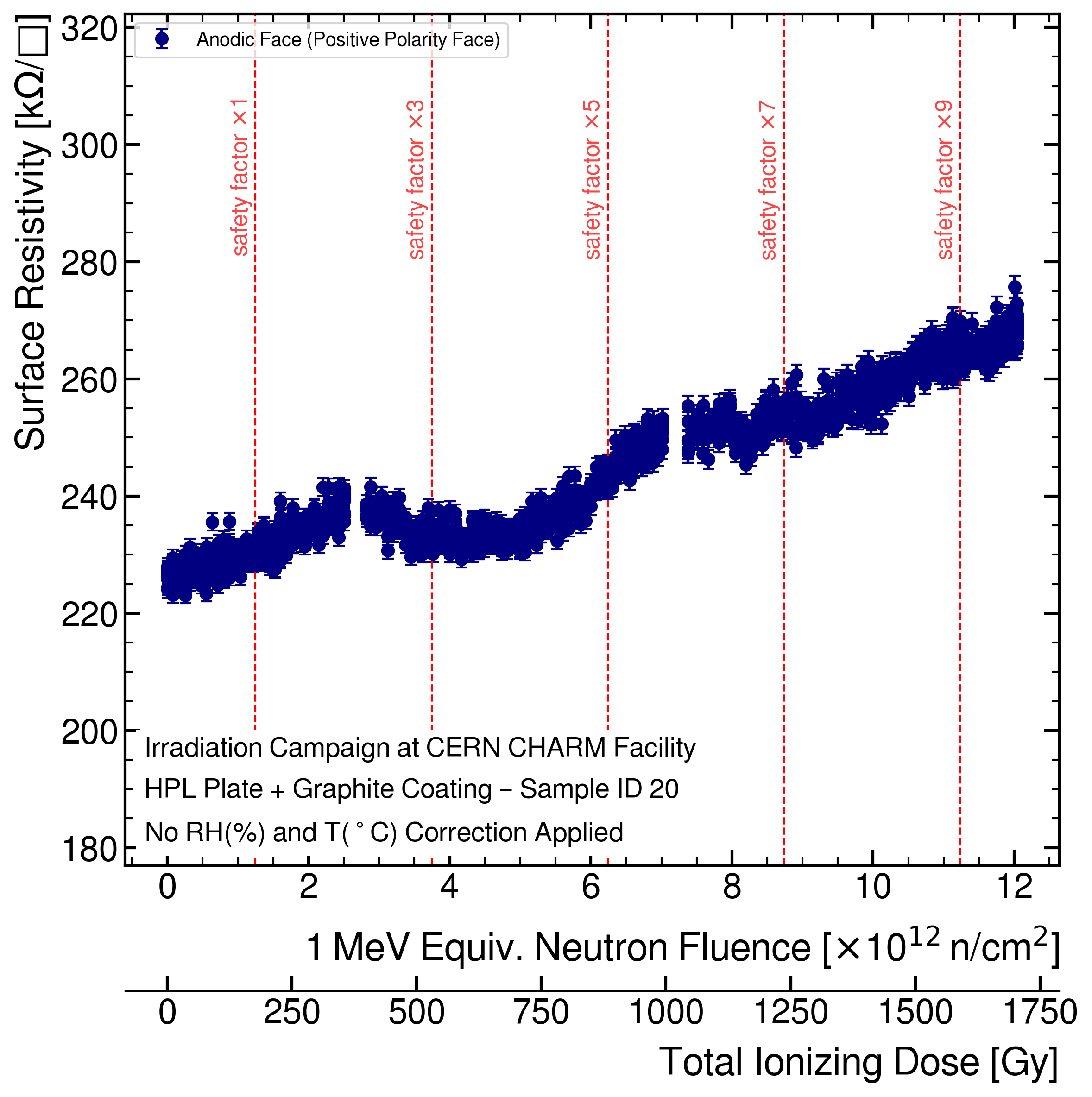}
        \caption{ }
        \label{fig:Surface_Resistivity_Results_Sample_20_F1}
    \end{subfigure}
    \hfill
    \begin{subfigure}[b]{0.49\textwidth}
        \centering
        \includegraphics[width=\textwidth]{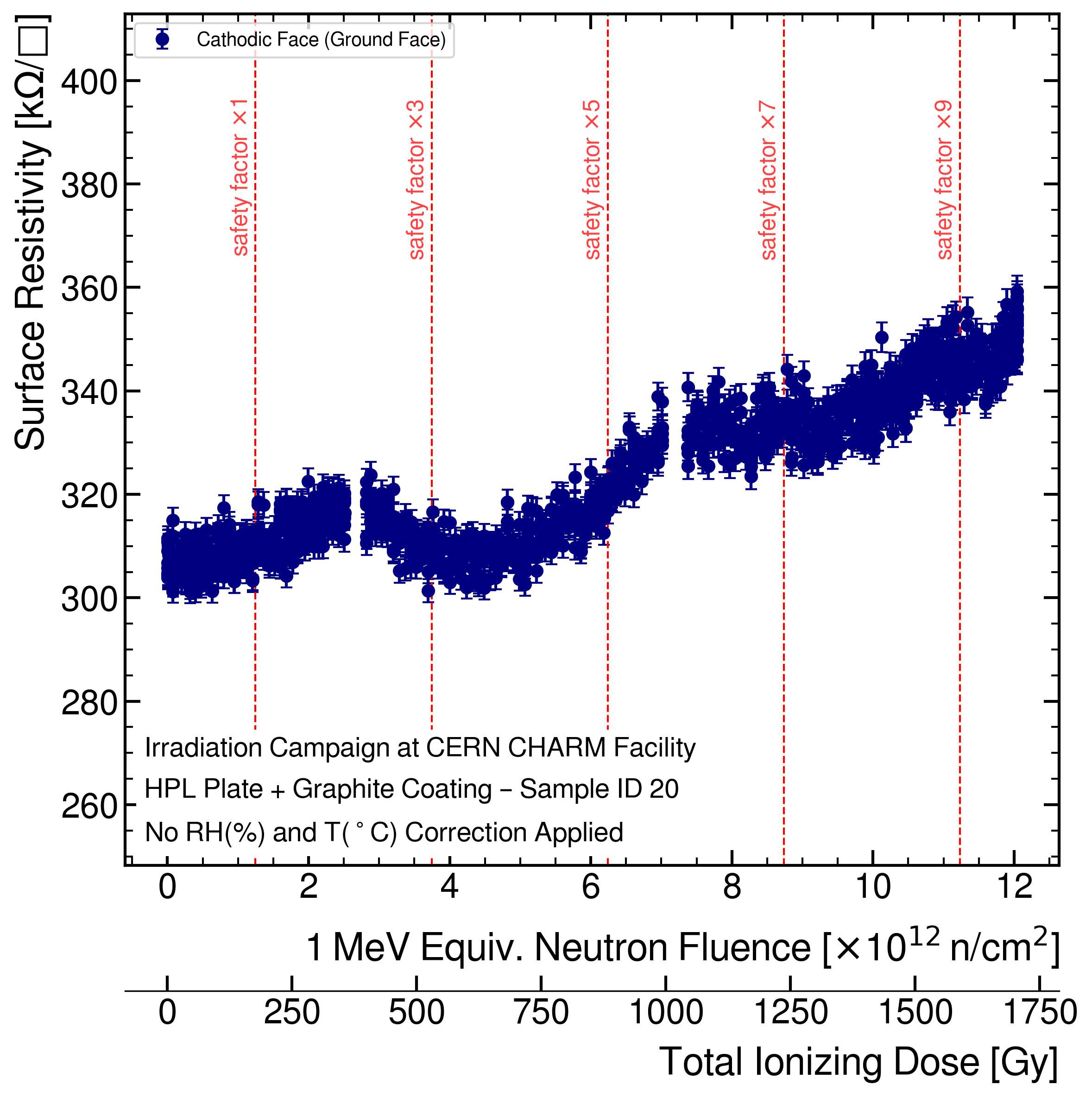}
        \caption{ }
        \label{fig:Surface_Resistivity_Results_Sample_20_F2}
    \end{subfigure}
    
    \caption{Surface resistivity of the carbon-based resistive-ink coating on the high-pressure-laminate (HPL) electrode sample as a function of accumulated 1 MeV Si-equivalent neutron fluence and total ionizing dose (TID): (a, c) anode face; (b, d) cathode face.}
    \label{fig:CERN_CHARM_Surface_Resistivity_Results_Summary_Sample_19_20}
\end{figure}

Figure \ref{fig:CERN_CHARM_Bulk_Resistivity_Results_Summary} illustrates the evolution of the bulk resistivity of the HPL electrode sample as a function of both the 1 MeV Si-equivalent neutron fluence and the accumulated total ionizing dose. Figure \ref{fig:CERN_CHARM_Surface_Resistivity_Results_Summary_Sample_17_18} and \ref{fig:CERN_CHARM_Surface_Resistivity_Results_Summary_Sample_19_20} compare the surface resistivity of the carbon-based resistive ink coating on the anode [Figure \ref{fig:Surface_Resistivity_Results_Sample_17_F1}, \ref{fig:Surface_Resistivity_Results_Sample_18_F1}, \ref{fig:Surface_Resistivity_Results_Sample_19_F1}, \ref{fig:Surface_Resistivity_Results_Sample_20_F1}] and cathode [Figure \ref{fig:Surface_Resistivity_Results_Sample_17_F2}, \ref{fig:Surface_Resistivity_Results_Sample_18_F2}, \ref{fig:Surface_Resistivity_Results_Sample_19_F2}, \ref{fig:Surface_Resistivity_Results_Sample_20_F2} ] faces of the HPL electrode sample as a function of both the 1 MeV Si-equivalent neutron fluence and the accumulated total ionizing dose.

\end{document}